\let\ifarxiv=\iftrue     
\pdfoutput=1
\documentclass[12pt,a4paper]{article}

\ifarxiv\ifnum\pdfoutput=1\else
\PassOptionsToPackage{hypertex}{hyperref}
\PassOptionsToPackage{draft}{graphicx}
\fi\fi

\usepackage{amsmath,amssymb}
\usepackage{graphicx}
\usepackage[nosort]{cite}
\usepackage[bulletsep]{collref}
\usepackage{feynmp}
\usepackage{multirow}
\usepackage{listings}
\usepackage{slashed}
\usepackage{longtable}
\usepackage{multicol}
\usepackage{hhline}
\usepackage{array} 
\usepackage{pdflscape}  
\usepackage{nicefrac}
\usepackage[percent]{overpic}

\usepackage[a4paper,text={173mm,216mm},centering]{geometry}

\allowdisplaybreaks[3]

\usepackage[font=small,labelfont=bf,width=0.85\textwidth]{caption}

\makeatletter
\let\old@startsection=\@startsection
\renewcommand{\@startsection}[6]{\old@startsection{#1}{#2}{#3}{#4}{#5}{#6\mathversion{bold}}}
\makeatother

\makeatletter
\newlength{\apb@width}
\newcommand{\autoparbox}[2][c]{\settowidth{\apb@width}{#2}\parbox[#1]{\apb@width}{#2}}

\makeatother

\let\oldPhi=\Phi
\let\oldPsi=\Psi
\let\oldGamma=\Gamma
\let\oldDelta=\Delta
\let\oldSigma=\Sigma
\let\oldLambda=\Lambda
\let\oldTheta=\Theta
\let\oldPi=\Pi
\let\oldXi=\Xi
\let\oldUpsilon=\Upsilon
\let\oldOmega=\Omega
\renewcommand{\Phi}{\mathnormal{\oldPhi}}
\renewcommand{\Psi}{\mathnormal{\oldPsi}}
\renewcommand{\Gamma}{\mathnormal{\oldGamma}}
\renewcommand{\Sigma}{\mathnormal{\oldSigma}}
\renewcommand{\Delta}{\mathnormal{\oldDelta}}
\renewcommand{\Theta}{\mathnormal{\oldTheta}}
\renewcommand{\Lambda}{\mathnormal{\oldLambda}}
\renewcommand{\Pi}{\mathnormal{\oldPi}}
\renewcommand{\Xi}{\mathnormal{\oldXi}}
\renewcommand{\Upsilon}{\mathnormal{\oldUpsilon}}
\renewcommand{\Omega}{\mathnormal{\oldOmega}}

\ifx\genfrac\sdflkaj\else\fi
\newcommand{\sfrac}[2]{{\textstyle\frac{#1}{#2}}}
\newcommand{\half}{\sfrac{1}{2}}

\newcommand{\vev}[1]{\langle#1\rangle}

\newcommand{\abs}[1]{|#1|}

\newcommand{\be}{\begin{equation}}
\newcommand{\ee}{\end{equation}}

\newcommand{\Tr}{\mathop{\mathrm{Tr}}}

\newcommand{\nn}{\nonumber}

\def\[{\begin{equation}}
\def\]{\end{equation}}
\def\<{\begin{eqnarray}}
\def\>{\end{eqnarray}}

\makeatletter
\def\mr@ignsp#1 {\ifx\:#1\@empty\else #1\expandafter\mr@ignsp\fi}%
\newcommand{\multiref}[1]{\begingroup
\xdef\mr@no@sparg{\expandafter\mr@ignsp#1 \: }%
\def\mr@comma{}%
\@for\mr@refs:=\mr@no@sparg\do{\mr@comma\def\mr@comma{,}\ref{\mr@refs}}%
\endgroup}
\makeatother

\newcommand{\eqn}[1]{(\ref{#1})}

\ifx\href\asklfhas\newcommand{\href}[2]{#2}\fi

\newcommand{\hypref}[2]{\ifx\href\asklfhas #2\else\href{#1}{#2}\fi}

\renewcommand{\eqref}[1]{(\multiref{#1})}

\newcommand{\hl}{\hhline{*{3}{|-}||*{3}{-|}||*{3}-|}}

\newcommand{\hld}{\hhline{*{3}{:=}::*{3}{=:}::*{3}=:}}

\newcommand{\e}{\ensuremath{\varepsilon}}

\newcommand{\soll}{\stackrel{!}{=}}

\newcommand{\tr}[1]{\mathrm{Tr}\left(#1\right)}

\newcommand{\bleq}{\ensuremath{\mathrel{\phantom{=}}}}

\newcommand{\kla}[1]{\left( #1 \right)}

\newcommand{\ekla}[1]{\left[ #1 \right]}
\newcommand{\skla}[1]{\left\langle #1 \right\rangle}

\newcommand{\dv}[1]{\ensuremath{\text{d}^4#1} \,}

\newcommand{\indexfett}[2][f]{\if#1f{\index{#2|textbf}}\else{\index{#2}}\fi}

\newcommand{\lnk}[1]{\ln \frac{\e^2}{x_{#1}^2}}
\newcommand{\lnl}[3]{\ln \frac{\e^2 x_{#1}^2}{x_{#2}^2 x_{#3}^2}}

\newcommand{\Op}{\mathcal{O}}
\newcommand{\cO}{\mathcal{O}}
\newcommand{\K}{\mathcal{K}}
\newcommand{\D}{\mathcal{D}}

\newcommand{\Otil}{\widetilde{\mathcal{O}}}
\newcommand{\gYM}{g_{\text{YM}}}
\newcommand{\Dn}{\Delta^{(0)}}
\newcommand{\Df}{\Delta}

\newcommand{\nnl}{\nonumber\\}

\newcommand{\ps}{$\hspace{-6pt}\phantom{\Big]}$}
\newcommand{\oc}[1]{\raisebox{-1pt}{#1}}
\newcommand{\psx}{\\[4pt]}
\newcommand{\ocx}[1]{\raisebox{-3pt}{#1}}

\renewcommand{\fmfdot}[1]{\fmfv{decor.shape=circle,decor.filled=full,
decor.size=1.5thick}{#1}}

\ifx\hypersetup\sadfkjashdfkxja\newcommand\hypersetup[1]{}\fi

\hypersetup{plainpages=false}
\hypersetup{pdfpagemode=UseNone}
\hypersetup{bookmarksnumbered=true}
\hypersetup{pdfstartview=FitH}
\hypersetup{colorlinks=false}
\hypersetup{citebordercolor={.5 1 .5}}
\hypersetup{urlbordercolor={.5 1 1}}
\hypersetup{linkbordercolor={1 .7 .7}}

\newcommand\oplength{\Delta^{(0)}}

\newcommand{\ii}{\text{i}}

\newcommand\spinchain[2]{\left \{ \begin{matrix} #1 \cr #2 \end{matrix}\right \} }
\newcommand{\ft}[2]{{\textstyle\frac{#1}{#2}}}

\newcommand{\gmZB}{0} 
\newcommand{\gmDB}{4} 
\newcommand{\gmDC}{0} 
\newcommand{\gmVA}{$\frac{13+ \sqrt{41}}{2}$} 
\newcommand{\gmVB}{$ 5 + \sqrt{5}$} 
\newcommand{\gmVE}{$\frac{13 - \sqrt{41}}{2}$} 
\newcommand{\gmVF}{$5 - \sqrt{5}$} 
\newcommand{\gmVG}{0} 

\begin{document}
\thispagestyle{empty}
\vspace*{-20mm}

\begingroup\raggedleft\footnotesize\ttfamily
HU-EP-12/02\\
QMUL-PH-11-23\\
\vspace{15mm}
\endgroup

\begingroup\centering
{\Large\bfseries\mathversion{bold}
Three-point functions in planar $\mathcal{N}=4$ super Yang-Mills Theory for scalar operators
up to length five at the one-loop order\par}%
\hypersetup{pdfsubject={}}%
\hypersetup{pdfkeywords={}}%
\ifarxiv\vspace{15mm}\else\vspace{15mm}\fi

\hypersetup{pdfauthor={}}%
\begingroup\scshape\large
 George Georgiou${}^{1}$,  Valeria Gili${}^{2}$, Andr\'e Gro{\ss}ardt${}^{3,4}$ \\
and Jan Plef\/ka${}^{4}$
\endgroup
\vspace{5mm}

\begingroup\ifarxiv\small\fi
\textit{
${}^{1}$Demokritos National Research Center, Institute of Nuclear Physics\\
Ag.~Paraskevi, GR-15310 Athens, Greece\\[0.5cm]
${}^{2}$Centre for Research in String Theory,
School of Physics,\\
Queen Mary University of London \\
Mile End Road, London, E14NS, United Kingdom \\[0.5cm]
${}^{3}$ ZARM Bremen, Am Fallturm 1, D-28359 Bremen and\\ Institute of Theoretical Physics,
University of Hannover,\\ Appelstrasse 2 D-30167 Hannover, Germany  \\[0.5cm]
${}^{4}$Institut f\"ur Physik, Humboldt-Universit\"at zu Berlin,\\
Newtonstra{\ss}e 15, D-12489 Berlin, Germany\\[0.6cm]}
\texttt{georgiou@inp.demokritos.gr,
v.gili@qmul.ac.uk,\{andre.grossardt,jan.plefka\}@physik.hu-berlin.de}
\endgroup
\vspace{1cm}

\textbf{Abstract}\vspace{5mm}\par
\begin{minipage}{14.7cm}
We report on a systematic perturbative study of  three-point functions in
planar $SU(N)$ $\mathcal{N}=4$ super Yang-Mills theory at the one-loop level
involving scalar field operators up to length five.
For this we have computed
a sample of 40 structure constants   involving
primary operators  of up to and including length five which are
 built entirely from scalar fields.
A combinatorial dressing technique has been developed to promote tree-level
correlators to one-loop level. In addition we have resolved the mixing up to the
order $g_{\text{YM}}^{2}$ level of the operators involved, which amounts to mixings
with bi-fermions, with bi-derivative insertions as well as self-mixing contributions in the
scalar sector. This work supersedes a preprint by two of the authors from 2010
which had neglected
the mixing contributions.
\end{minipage}\par
\endgroup
\newpage


\setcounter{tocdepth}{2}
\hrule height 0.75pt
\tableofcontents
\vspace{0.8cm}
\hrule height 0.75pt
\vspace{1cm}

\setcounter{tocdepth}{2}

\section{Introduction and Conclusions}

Following the discovery of integrable structures \cite{Minahan:2002ve, Beisert:2003tq, Bena:2003wd,Beisert:2003yb} in the AdS/CFT correspondence
\cite{Maldacena:1998re,Witten:1998qj,Gubser:1998bc}
our understanding of
${\mathcal N}=4$ supersymmetric Yang-Mills (SYM) theory \cite{Brink:1976bc, Gliozzi:1976qd}
and the dual $AdS_5\times S^5$ superstring theory
has greatly advanced. To a large extent this progress
occurred in the problem of finding the exact all-loop form of the
anomalous scaling dimensions of local gauge
invariant operators of the gauge theory alias
the spectrum of string excitations in the string model.
The key was a mapping of the problem to an integrable spin chain which emerged from a one-loop
perturbative study of the diagramatics involved by Minahan and Zarembo
\cite{Minahan:2002ve}. Moving on to higher loops
the spectral problem was mapped to the diagonalization of a long-range spin chain model,
whose precise microscopic form remains unknown
\cite{Beisert:2003yb,Beisert:2003ys}. Nevertheless,
assuming integrability the spin-chain S-matrix
could be algebraically constructed and the spectral problem was rephrased for asymptotically long
operators to the solution of a set of nested Bethe equation
\cite{Staudacher:2004tk,Beisert:2005tm} (for reviews see
\cite{Tseytlin:2004cj,Belitsky:2004cz,Zarembo:2004hp,Plefka:2005bk,Minahan:2006sk,Arutyunov:2009ga,Beisert:2004yq,Beisert:2004ry, Beisert:2010jr}).
The central remaining problem is now the understanding of wrapping interactions,
which affect short operators at lower loop orders \cite{Ambjorn:2005wa},
\cite{Bajnok:2009vm,Fiamberti:2009jw}. From the algebraic
viewpoint
important progress was made by thermodynamic Bethe ansatz techniques
\cite{Gromov:2009bc,Bombardelli:2009ns,Arutyunov:2009ur}
which
also lead to a conjecture for the exact numerical scaling dimensions of the
Konishi operator, the shortest unprotected operator in the theory \cite{Gromov:2009zb,Frolov:2010wt}.

Next to the scaling dimensions there also exist remarkable all-order results in planar
${\mathcal N}=4$ SYM for supersymmetric
Wilson-loops of special geometries \cite{Erickson:2000af,Drukker:2000rr}
as well as for scattering amplitudes of four and five external
particles \cite{Bern:2005iz,Drummond:2008vq}, being closely related to light-like Wilson lines
\cite{Alday:2007hr}, see \cite{Alday:2008yw,Henn:2009bd} for reviews.

Given these advances in finding exact results it is natural
to ask if one can make similar statements for three-point functions of local gauge invariant
operators. Due to conformal symmetry the new data appearing are the structure constants
which have a nontrivial coupling constant $\lambda=g^{2}N$ dependence and also appear in the
associated operator product expansion. In detail we have for renormalized operators
\be
 \skla{\Otil_\alpha(x_1) \, \Otil_\beta(x_2) \, \Otil_\gamma(x_3)} = \frac{C_{\alpha \beta \gamma}}{\abs{x_{12}}^{\Df_\alpha + \Df_\beta - \Df_\gamma} \abs{x_{23}}^{\Df_\beta + \Df_\gamma - \Df_\alpha} \abs{x_{13}}^{\Df_\alpha + \Df_\gamma - \Df_\beta} \abs{\mu}^{\gamma_\alpha + \gamma_\beta + \gamma_\gamma} } \, ,
\label{3ptconvn}
\ee
where $\Delta_{\alpha}=\Dn_{\alpha}+\lambda\,\gamma_{\alpha}$
denotes the scaling dimensions of the operators involved
with $\Dn$ the engineering and $\gamma$ the anomalous scaling dimensions, $\mu$ the renormalization
scale and
\be
 C_{\alpha \beta \gamma} = C^{(0)}_{\alpha \beta \gamma} + \lambda \,
 C^{(1)}_{\alpha \beta \gamma} + O(\lambda^2)\,
\label{cdef}
\ee
is the scheme independent structure constant representing the new observable arising in three-point
functions one would like to find. Similar to the case of two-point functions there are non-renormalization
theorems for three-point correlation functions of chiral primary (or 1/2 BPS)
operators, whose structure constants do not receive radiative corrections
\cite{Eden:1999gh,Arutyunov:2001qw,Heslop:2001gp,Lee:1998bxa,Basu:2004nt}.

The study of three-point functions involving non-protected operators allowing for
a non-trivial coupling constant dependence of the structure constants is still largely in its infancy.
Direct computations of three-point functions are
\cite{Bianchi:2001cm,Beisert:2002bb,Roiban:2004va,Okuyama:2004bd,Alday:2005nd,Alday:2005kq,Georgiou:2009tp} while
\cite{Arutyunov:2000im} analyzed the problem indirectly through an OPE decompostition
of four-point functions of chiral primaries. The works
\cite{Beisert:2002bb,Georgiou:2009tp,Chu:2002pd} focused on non-extremal correlators involving
scalar two-impurity operators which are particularly relevant in the BMN limit.
The mixing problem of these operators with fermion and derivative impurities was analyzed in
\cite{Georgiou:2008vk}.
\cite{Roiban:2004va} considered extremal correlators of a very special class of operators
allowing an interesting map to spin-chain correlation functions, while
\cite{Casteill:2007td} addresses similar questions from the perspective of the non-planar
contribution of the dilatation operator.

In the past year important advances in our understanding
of the strong coupling behaviour of three-point functions were made based on the semi-classical analysis
of the dual string theory.
This approach to the calculation of n-point correlators
involving non-BPS states was initiated in
\cite{Yoneya:2006td,Dobashi:2004nm,Tsuji:2006zn,Janik:2010gc,Buchbinder:2010vw,Zarembo:2010rr,Costa:2010rz}.
More recently, the authors of \cite{Janik:2010gc} argued that it should be possible to
obtain the correlation functions of local operators corresponding to classical
spinning string states, at strong coupling, by evaluating the string action on
a classical solution with appropriate boundary conditions
after convoluting  with the classical states wavefunctions.
In \cite{Buchbinder:2010vw,Roiban:2010fe,Ryang:2010bn,Klose:2011rm,Ryang:2011tk}, 2-point and 3-point correlators of vertex operators representing
classical string states with large  spins were calculated.
Moreover, in a series of papers
\cite{Zarembo:2010rr,Costa:2010rz,Hernandez:2010tg,Russo:2010bt,Georgiou:2010an,Park:2010vs,Bak:2011yy,
Bissi:2011dc,Hernandez:2011up,Ahn:2011zg,Arnaudov:2011wq,Ahn:2011dq} the 3-point function coefficients
of correlators involving a massive string state, its conjugate and a third "light" state
state were computed for a variety of massive string states. The result takes the
form of a fattened Witten diagram with the vertex operator of the light state being integrated over the world-sheet
of the classical solution describing the 2-point correlator of the heavy operator.

Furthermore, three-point functions of single trace operators were studied in \cite{Escobedo:2010xs,Escobedo:2011xw,Gromov:2011jh}
from the perspective of integrability. In particular,
the protected $SU(2)$ scalar
subsector of the ${\mathcal N}=4$ theory involving two holomorphic scalar fields was studied
and upon exploting the underlying
integrable spin chain structure analytic expressions for the tree-level piece
$C^{(0)}_{\alpha \beta \gamma}$ could be established \cite{Escobedo:2010xs}.
Recently these results
were extended to the one-loop closed $SU(3)$ sector involving three holomorphic scalar
fields \cite{Escobedo:2011xw}. In both cases limits of one short and two long operators led to
serious simplifications. The one loop structure, however, has only been
started to be explored in the $SU(2)$ subsector (see appendix E of \cite{Escobedo:2010xs}).
Also an intriguing weak/strong coupling match of correlators
involving operators in the $SU(3)$ sector was observed \cite{Escobedo:2011xw} \footnote{The authors of \cite{Bissi:2011ha} have calculated the one-loop
correction to the structure constants of operators in the $SU(2)$ sub-sector to find that this agreement is spoiled. However, there is the subtlety of
the two "heavy" operators being {\it roughly} the conjugate of each other.}.
This match was found to hold for correlators of two non-protected operators in the Frolov-Tseytlin limit and one short BPS operator.
Subsequently, this weak/strong coupling match was extended for correlation functions involving operators in the $SL(2,R)$
closed subsector of $N=4$ SYM theory \cite{Georgiou:2011qk}.
Finally, by performing Pohlmeyer reduction for classical
solutions living in $AdS_2$ but with a prescribed nonzero energy-momentum tensor the authors of \cite{Janik:2011bd}
calculated the AdS contribution to the three-point coefficient of three heavy states  rotating purely
in $S^5$ \footnote{The authors of \cite{Buchbinder:2011jr} questioned this result arguing that string solutions with no $AdS_5$ charges
should be point-like in the $AdS_5$ space.}.
In the same spirit, part of the three-point fusion coefficient of three GKP strings
\cite{Gubser:2002tv} was calculated
in \cite{Kazama:2011cp}\footnote{The contribution coming from the exact form of the vertex operators is still to be found.} .

The two works \cite{Okuyama:2004bd,Alday:2005nd} considered the general problem of finding the structure constants
of scalar field primary operators discussing important aspects of scheme independence for the
determination of $C^{(1)}_{\alpha\beta\gamma}$.

In this paper we shall continue this work and report on a
systematic one-loop study of short single trace conformal primary operators built from the six
real scalar fields of the theory in the planar limit at the leading order. For this we developed a combinatorial
dressing technique to promote tree-level non-extremal three-point correlation functions to
the one-loop level which is similar to the results reported in \cite{Alday:2005nd}.
This is then used to compute a total of 40 structure constants at the one-loop
level involving 11 different scalar field conformal primary operators up to
length five.
The restriction to this particular set of operators arose from the necessity to lift the
operator degeneracy in the pure scalar $SO(6)$ sector by resolving the operator mixing problem
arising from two-point correlators up to the two-loop.
Indeed a large portion of our work is devoted to resolving the mixing problem of the
scalar operators considered at the \emph{two-loop} level. For this the results
of \cite{Georgiou:2008vk,Georgiou:2009tp} for the mixing with bi-fermion and bi-derivative mixings have been extended to
the two-loop self mixing sector as well as to two singlet operators of length four.
In chapter 5 we spell out the form of the operators including all mixing corrections
up to and including order $\cO(\lambda^{2})$. The main results of our work are
collected in the tables \ref{244corr}, \ref{334corr}, \ref{444corr} and
\ref{255corr} of section \ref{sec:corresults}.

The main motivation for this spectroscopic study is to provide data to
test and develop future conjectures on the form of the three-point structure constants most likely making use of integrability.
It is important to stress
that both results only apply for non-extremal correlation functions. Extremal correlation functions
are such that $\Dn_{\gamma}=\Dn_{\alpha}+\Dn_{\beta}$ i.e.~the length of the longest operator
is equal to the sum of the two shorter ones. Here there also exists a proposed one-loop
formula due to Okuyama and Tseng \cite{Okuyama:2004bd} see equation
\eqn{eqn:form of structure constants for extremal correlators}.

It would be very interesting to see whether these simple structures are stable at higher loop-order
and also for non-purely scalar field primary operators such as the twist $J$ operators for example.

\section{General structure and scheme dependence of two and three-point functions}

We want to compute planar two- and three-point functions of local scalar operators
at the one-loop order.
For this it is important to identify the regularization scheme independent information.

To begin with a scalar two-point function of bare local operators $\Op_{\alpha}^{B}(x)$
in a random basis can be brought into diagonal form under a suitable linear transformation
$\Op_\alpha = M_{\alpha \beta} \Op^B_\beta$ with a coupling constant
$\lambda=g^{2}\,N$ independent mixing matrix $M_{\alpha\beta}$ as we are working at the one-loop
level\footnote{Note that the two-loop diagonalization will involve a mixing matrix proportional to $\lambda$.}
\begin{equation}
 \skla{ \Op_\alpha(x_1) \, \Op_\beta(x_2) } = \frac{\delta_{\alpha \beta}}{x_{12}^{2 \Dn_\alpha}} \kla{ 1 + \lambda \, g_\alpha - \lambda\, \gamma_\alpha \,\ln \abs{x_{12} \epsilon^{-1}}^2 }\, ,
\qquad x_{12}^{2}:=(x_{1}-x_{2})^{2}\, ,\end{equation}
where $\epsilon$ represents a space-time UV-cutoff and $\Delta^{(0)}_{\alpha}$ the engineering
scaling dimension of $\cO_{\alpha}$. Clearly the finite contribution
to the one-loop normalization $g_{\alpha}$ is scheme dependent \cite{Okuyama:2004bd,Alday:2005nd}
as a shift in the cutoff parameter
$\epsilon\to e^{c}\,\epsilon$ changes
\be
\label{5}
g_{\alpha}\to g_{\alpha}+ 2\, c\, \gamma_\alpha\, .
\ee

One may now define the renormalized operators via
\begin{equation}
 \Otil_\alpha = \Op_\alpha \kla{ 1 - \frac{\lambda}{2} g_\alpha - \lambda \gamma_\alpha \ln \abs{\mu\, \epsilon} + O(\lambda^2) } \label{eqn:Renormalized ops}
\end{equation}
with a renormalization momentum scale $\mu$ to
obtain finite canonical two-point correlation
functions
\begin{equation}
\label{2ptren}
\skla{ \Otil_\alpha(x_1) \, \Otil_\beta(x_2) } = \frac{\delta_{\alpha \beta}}{\abs{x_{12}}^{2 \Dn_\alpha}} \kla{1 - \lambda \gamma_\alpha \ln \abs{x_{12} \mu}^2 + O(\lambda^2)}
= \frac{\delta_{\alpha \beta}}{\abs{x_{12}}^{2 \Dn_\alpha} \abs{x_{12} \mu}^{2 \lambda \gamma_\alpha}}\, ,
\end{equation}
allowing one to extract the scheme independent
one-loop scaling dimensions $\Df_\alpha = \Dn_\alpha + \lambda \gamma_\alpha$.

Moving on to three-point functions of the un-renormalized diagonal operators $\cO_{\alpha}$
one obtains to the one-loop order in $\lambda$
\begin{align}
\label{3ptfct}
\langle \, \cO_\alpha(x_1)\, &\cO_\beta(x_2) \, \cO_\gamma(x_3)\,\rangle =
\frac{1}{|x_{12}|^{\Dn_\alpha+\Dn_\beta-\Dn_\gamma}\,|x_{23}|^{\Dn_\beta+\Dn_\gamma-\Dn_\alpha}\,
|x_{31}|^{\Dn_\gamma+\Dn_\alpha-\Dn_\beta} }\, \nn\\
&\times \left [ C^{(0)}_{\alpha\beta\gamma} \left ( 1+ \frac{1}{2}\, \lambda\, \left \{ \gamma_\alpha\,
\ln \frac{\epsilon^2\, x_{23}^2}{x_{12}^2\, x_{31}^2} +  \gamma_\beta\,
\ln \frac{\epsilon^2\, x_{31}^2}{x_{12}^2\, x_{23}^2} +  \gamma_\gamma\,
\ln \frac{\epsilon^2\, x_{12}^2}{x_{23}^2\, x_{31}^2} \,\right \}\,
\right ) + \lambda\,  \tilde C^{(1)}_{\alpha\beta\gamma}\,
 \right ]
\end{align}
Now again the finite one-loop contribution to the structure
constant $\tilde C^{(1)}_{\alpha\beta\gamma}$ is scheme dependent
\cite{Okuyama:2004bd,Alday:2005nd} as it changes under
$\epsilon\to\epsilon\, e^{c}$ as
\be
\tilde C^{(1)}_{\alpha\beta\gamma}\to \tilde C^{(1)}_{\alpha\beta\gamma}
+c\, (\gamma_\alpha+\gamma_\beta+\gamma_\gamma)\, C^{(0)}_{\alpha\beta\gamma}\, ,
\qquad \mbox{(no sums on the indices)}\, .
\ee
However, the following combination of the unrenormalized three-point function
structure constant and the normalization is scheme independent
\be
C^{(1)}_{\alpha\beta\gamma}:= \tilde C^{(1)}_{\alpha\beta\gamma} -\frac{1}{2}\, (
g_{\alpha} \,C^{(0)}_{\alpha\beta\gamma} + g_{\beta} \,C^{(0)}_{\alpha\beta\gamma}
+ g_{\gamma} \,C^{(0)}_{\alpha\beta\gamma}\,
) \, .
\label{schemeindepc}
\ee
This is the only datum to be extracted from three-point functions. It also
directly arises as the structure constant in the three-point function of the
renormalized operators $\Otil_{\alpha}$
\be
 \skla{\Otil_\alpha(x_1) \, \Otil_\beta(x_2) \, \Otil_\gamma(x_3)} = \frac{C_{\alpha \beta \gamma}}{\abs{x_{12}}^{\Df_\alpha + \Df_\beta - \Df_\gamma} \abs{x_{23}}^{\Df_\beta + \Df_\gamma - \Df_\alpha} \abs{x_{13}}^{\Df_\alpha + \Df_\gamma - \Df_\beta} \abs{\mu}^{\lambda (\gamma_\alpha + \gamma_\beta + \gamma_\gamma)} } \, ,
\label{3ptconv}
\ee
where
$
 C_{\alpha \beta \gamma} = C^{(0)}_{\alpha \beta \gamma} + \lambda \,
 C^{(1)}_{\alpha \beta \gamma} + O(\lambda^2)\,
$
is the scheme independent structure constant of \eqn{schemeindepc}.

An important further point is the following. If one wishes to compute the one-loop piece
$C^{(1)}_{\alpha\beta\gamma}$ starting from a generic basis of operators
one has to resolve the mixing problem at the two-loop order. This is so as
the above discussed mixing matrix $M_{\alpha\beta}$ will receive $O(\lambda)$ terms
once one computes the two-point function out to the order $O(\lambda^{2})$. These
mixing terms will contribute to the final $C^{(1)}_{\alpha\beta\gamma}$
through tree-level contractions.
In this work we shall be interested in three-point
functions of single-trace operators which are given by purely scalar operators at
leading order $\cO_{\alpha}$. The mixing effects then induce a correction pattern
of the schematic form
\be
{\hat\cO}_{\alpha}(g_{\text{YM}},N) \sim  \Tr(\phi^{L}) + g_{\text{YM}}\, N\,
 \Tr(\psi\psi\phi^{L-3}) + g_{\text{YM}}^{2}\, N^{2}\, \Tr(D_{\mu}D_{\mu}\phi^{L-2}) +
 g^{2}_{\text{YM}}\, N\,  \Tr(\phi^{L})+ \ldots \, ,
\ee
where $L$ is the length or engineering scaling dimension of $\cO_{\alpha}$. Note that in the
above schematic formula each trace-operator stands as a representative of a particular
weighted combination of permutations of the same field content under the trace.
Of course all operators transform in the same representation of the R-symmetry
group $su(4)$ and are space-time scalars. We refer to the
three-contributions as the fermionic $\cO_{\psi\psi}$, the derivative $\cO_{DD}$
and the self-mixing $\cO_{\text{self}}$ contributions. Clearly the tree-level insertions
of these mixings, such as $\vev{\cO_{\alpha,0}\, \cO_{\beta,\psi\psi}\,
\cO_{\beta,\psi\psi}}$ or $\vev{\cO_{\alpha,0}\, \cO_{\beta,0}\,
\cO_{\beta,DD}}$ or $\vev{\cO_{\alpha,0}\, \cO_{\beta,0}\,
\cO_{\beta,\text{self}}}$ contribute to the one-loop structure constants $C^{(1)}_{\alpha\beta\gamma}$
next to the radiative corrections discussed above. In addition the correlator
$\vev{\cO_{\alpha,0}\, \cO_{\beta,0}\,
\cO_{\beta,\psi\psi}}$ with a Yukawa-vertex insertion will also contrbute potentially.

In our work we evaluate both these contributions -- the radiative and mixing ones --
and state the final result for the
scheme independent structure constant
$C^{(1)}_{\alpha\beta\gamma}$ for a large number of three-point functions.
For this the mixing of scalar operators up to and including engineering length 5 has
been determined.

\section{The one-loop planar dressing formulae}
\label{three}

\subsection{Derivation}

In this section we derive an efficient set of combinatorial dressing formulae to
dress up tree-level graphs to one-loop. Similar formulae appeared in \cite{Drukker:2008pi}.

Following \cite{Beisert:2002bb} we introduce the 4d  propagator and the relevant
one-loop integrals in configuration space
\begin{align}
 I_{12} &= \frac{1}{(2\pi)^2 x_{12}^2}\, ,\nn\\
 Y_{123} &= \int \dv{w} I_{1w} I_{2w} I_{3w}\, ,\nn \\
 X_{1234} &= \int \dv{w} I_{1w} I_{2w} I_{3w} I_{4w}\, ,\nn\\
 H_{12,34} &= \int \dv{v} \dv{w} I_{1v} I_{2v} I_{vw} I_{3w} I_{4w}\, ,\nn \\
 F_{12,34} &= \frac{(\partial_1 - \partial_2) \cdot (\partial_3 - \partial_4) H_{12,34}}{I_{12} I_{34}}\, .
\end{align}
We have put the space-time points as indices to the function to make the expressions
more compact. These functions are all finite except in certain limits. For example
$Y_{123}$ , $X_{1234}$ and $H_{12,34}$ diverge logarithmically when $x_{1}\to x_{2}$. In point
splitting regularization one has the limiting formulae ($\lim_{i\to j}x^{2}_{ij}=\epsilon^{2}$)
\begin{align}
 X_{1123}  &= -\frac{1}{16 \pi^2}\, I_{12} I_{13} \kla{\ln \frac{x_{23}^2 \e^2}{x_{12}^2 x_{13}^2} -2}, \label{eqn:PointSplittingLimits1:X1123}\\
 Y_{112}   &= -\frac{1}{16 \pi^2}\, I_{12} \kla{\ln \frac{\e^2}{x_{12}^2} -2} = Y_{122},\label{eqn:PointSplittingLimits2:Y112}\\
 F_{12,13} &= -\frac{1}{16 \pi^2} \, \kla{\ln \frac{\e^2}{x_{23}^2} -2} + Y_{123} \kla{ \frac{1}{I_{12}} + \frac{1}{I_{13}} - \frac{2}{I_{23}} },\label{eqn:PointSplittingLimits3:F1213}\\
 X_{1122}  &= -\frac{1}{8 \pi^2} I_{12}^2 \kla{\ln \frac{\e^2}{x_{12}^2} -1},\label{eqn:PointSplittingLimits4:X1122}\\
 F_{12,12} &= -\frac{1}{8 \pi^2} \,\kla{\ln \frac{\e^2}{x_{12}^2} -3}.\label{eqn:PointSplittingLimits5:F1212}
\end{align}

We introduce a graphical symbol for the scalar propagators and work in a normalization
where
\begin{equation}
 \skla{\phi^{I}(x_1) \phi^{J}(x_2)}_{\text{tree}} u_1^{I} u_2^{J} =
 \begin{minipage}[h]{10mm}\begin{center}\footnotesize

\begin{fmffile}{ScalarPropagator}
\begin{fmfgraph*}(7,7)
\fmfpen{thick}
\fmftop{a1}
\fmfbottom{a2}
\fmfdot{a1,a2}
\fmf{plain}{a1,a2}
\fmflabel{$u_1$}{a1}
\fmflabel{$u_2$}{a2}
\end{fmfgraph*}
\end{fmffile}

\end{center}\end{minipage} = (u_1 \cdot u_2) \, I_{12}\, , \phantom{\Bigg(}
\end{equation}
here the $SO(6)$-indices of the scalar fields are contracted with dummy
six-vectors $u_1^{I}$ and $u_2^{J}$ for bookmarking purposes.

The one-loop corrections are then built of the following three components
\begin{alignat}{2}
\begin{minipage}[h]{18mm}\begin{center}\footnotesize

\begin{fmffile}{LoopSelfEnergy}
\begin{fmfgraph*}(7,7)
\fmfpen{thick}
\fmfleft{a1}
\fmfright{a2}
\fmf{plain}{a1,v,a2}
\fmfdot{a1,a2}
\fmfblob{6}{v}
\fmfdot{a1,a2}
\fmflabel{$u_1$}{a1}
\fmflabel{$u_2$}{a2}
\end{fmfgraph*}
\end{fmffile}

\end{center}\end{minipage} &= - \lambda (u_1 \cdot u_2) \, I_{12} \, \frac{Y_{112} + Y_{122}}{I_{12}}\label{eqn:BasicInteractionsLoopSelfEnergy} &&\quad \quad \text{(self-energy),}\phantom{\Bigg(}\\
\begin{minipage}[h]{18mm}\begin{center}\footnotesize

\begin{fmffile}{LoopGluon}
\begin{fmfgraph*}(7,7)
\fmfpen{thick}
\fmfright{a4,a2}
\fmfleft{a3,a1}
\fmf{plain}{a1,v1,a2}
\fmf{plain}{a3,v2,a4}
\fmffreeze
\fmf{wiggly,width=1}{v1,v2}
\fmfdot{a1,a2,a3,a4}
\fmfv{label=$u_1$,label.dist=1.5}{a1}
\fmfv{label=$u_2$,label.dist=1.5}{a2}
\fmfv{label=$u_3$,label.dist=1.5}{a3}
\fmfv{label=$u_4$,label.dist=1.5}{a4}
\end{fmfgraph*}
\end{fmffile}

\end{center}\end{minipage} &= \frac{\lambda}{2} (u_1 \cdot u_2) (u_3 \cdot u_4) \, I_{12} \, I_{34} \, F_{12,34}\label{eqn:BasicInteractionsLoopGluon} &&\quad \quad \text{(gluon),}\phantom{\Bigg(}\\
\begin{minipage}[h]{18mm}\begin{center}\footnotesize

\begin{fmffile}{LoopVertex}
\begin{fmfgraph*}(7,7)
\fmfpen{thick}
\fmfright{a4,a2}
\fmfleft{a3,a1}
\fmf{plain}{a1,v,a4}
\fmf{plain}{a3,v,a2}
\fmfdot{a1,a2,a3,a4,v}
\fmfv{label=$u_1$,label.dist=1.5}{a1}
\fmfv{label=$u_2$,label.dist=1.5}{a2}
\fmfv{label=$u_3$,label.dist=1.5}{a3}
\fmfv{label=$u_4$,label.dist=1.5}{a4}
\end{fmfgraph*}
\end{fmffile}

\end{center}\end{minipage} &= \frac{\lambda}{2} \big[2 (u_2 \cdot u_3) (u_1 \cdot u_4) - (u_2 \cdot u_4) (u_1 \cdot u_3) \phantom{\Bigg(} \hspace{-5cm} &&\nn\\ &\qquad
 - (u_1 \cdot u_2) (u_3 \cdot u_4)\big] \, X_{1234}\label{eqn:BasicInteractionsLoopVertex} &&\quad \quad  \text{(vertex).}
\end{alignat}
With these basic interactions we can now diagrammatically dress up the tree-level
two- and three-point correlation functions to the one-loop level.
To do so we note that a generic planar three-point function will be made of two-gon and three-gon
sub-graphs which need to be dressed, see figure 1.

\begin{figure}
 \centering
 \raisebox{0.5cm}{\includegraphics[width=4cm]{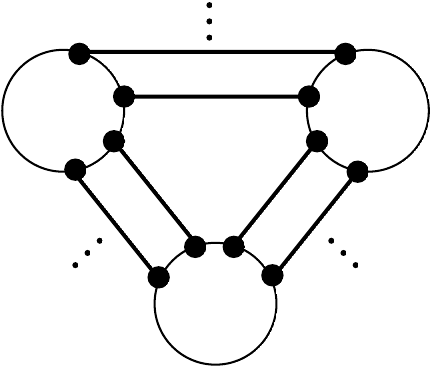}}
 \caption{The generic tree-level three-point function.}
 \label{fig:Ord3pt}
\end{figure}

For the two-gon dressing one finds the basic dressing formula
\begin{align}
\skla{\begin{minipage}[h]{12mm}\begin{center}\footnotesize

\begin{fmffile}{2gon}
\begin{fmfgraph*}(5,14)
\fmfstraight
\fmfleft{dummy1,p1,a2,a1,p2,dummy2}
\fmfright{dummy3,p3,b1,b2,p4,dummy4}
\fmfpen{thin}
\fmf{plain}{dummy1,p1,a2,a1,p2,dummy2}
\fmf{plain}{dummy3,p3,b1,b2,p4,dummy4}
\fmffreeze
\fmf{plain,width=2}{a1,b2}
\fmf{plain,width=2}{a2,b1}
\fmf{dots}{p1,p3}
\fmf{dots}{p2,p4}
\fmfdot{a1,a2,b1,b2}
\fmfv{label=$u_1$,label.dist=1.5thick}{a1}
\fmfv{label=$u_2$,label.dist=1.5thick}{a2}
\fmfv{label=$v_1$,label.dist=1.5thick}{b1}
\fmfv{label=$v_2$,label.dist=1.5thick}{b2}
\fmflabel{$x_1$}{dummy1}
\fmflabel{$x_2$}{dummy3}
\end{fmfgraph*}
\end{fmffile}

\end{center}\end{minipage}}_\text{1-loop} &= \begin{minipage}[h]{10mm}\begin{center}\footnotesize

\begin{fmffile}{2gonVertex}
\begin{fmfgraph*}(5,14)
\fmfstraight
\fmfleft{dummy1,p1,a2,a1,p2,dummy2}
\fmfright{dummy3,p3,b1,b2,p4,dummy4}
\fmfpen{thin}
\fmf{plain}{dummy1,p1,a2,a1,p2,dummy2}
\fmf{plain}{dummy3,p3,b1,b2,p4,dummy4}
\fmffreeze
\fmf{plain,width=2}{a1,v,b1}
\fmf{plain,width=2}{a2,v,b2}
\fmf{dots}{p1,p3}
\fmf{dots}{p2,p4}
\fmfdot{a1,a2,b1,b2,v}
\end{fmfgraph*}
\end{fmffile}

\end{center}\end{minipage} + \begin{minipage}[h]{10mm}\begin{center}\footnotesize

\begin{fmffile}{2gonGluon}
\begin{fmfgraph*}(5,14)
\fmfstraight
\fmfleft{dummy1,p1,a2,a1,p2,dummy2}
\fmfright{dummy3,p3,b1,b2,p4,dummy4}
\fmfpen{thin}
\fmf{plain}{dummy1,p1,a2,a1,p2,dummy2}
\fmf{plain}{dummy3,p3,b1,b2,p4,dummy4}
\fmffreeze
\fmf{plain,width=2}{a1,v1,b2}
\fmf{plain,width=2}{a2,v2,b1}
\fmffreeze
\fmf{wiggly}{v1,v2}
\fmf{dots}{p1,p3}
\fmf{dots}{p2,p4}
\fmfdot{a1,a2,b1,b2}
\end{fmfgraph*}
\end{fmffile}

\end{center}\end{minipage} + \frac{1}{2} \; \begin{minipage}[h]{10mm}\begin{center}\footnotesize

\begin{fmffile}{2gonSelfEnergyTop}
\begin{fmfgraph*}(5,14)
\fmfstraight
\fmfleft{dummy1,p1,a2,a1,p2,dummy2}
\fmfright{dummy3,p3,b1,b2,p4,dummy4}
\fmfpen{thin}
\fmf{plain}{dummy1,p1,a2,a1,p2,dummy2}
\fmf{plain}{dummy3,p3,b1,b2,p4,dummy4}
\fmffreeze
\fmf{plain,width=2}{a1,v,b2}
\fmf{plain,width=2}{a2,b1}
\fmfblob{6}{v}
\fmf{dots}{p1,p3}
\fmf{dots}{p2,p4}
\fmfdot{a1,a2,b1,b2}
\end{fmfgraph*}
\end{fmffile}

\end{center}\end{minipage} + \frac{1}{2} \; \begin{minipage}[h]{10mm}\begin{center}\footnotesize

\begin{fmffile}{2gonSelfEnergyBottom}
\begin{fmfgraph*}(5,14)
\fmfstraight
\fmfleft{dummy1,p1,a2,a1,p2,dummy2}
\fmfright{dummy3,p3,b1,b2,p4,dummy4}
\fmfpen{thin}
\fmf{plain}{dummy1,p1,a2,a1,p2,dummy2}
\fmf{plain}{dummy3,p3,b1,b2,p4,dummy4}
\fmffreeze
\fmf{plain,width=2}{a1,b2}
\fmf{plain,width=2}{a2,v,b1}
\fmfblob{6}{v}
\fmf{dots}{p1,p3}
\fmf{dots}{p2,p4}
\fmfdot{a1,a2,b1,b2}
\end{fmfgraph*}
\end{fmffile}

\end{center}\end{minipage} \nnl
& = I_{12}^{2}\,\, \frac{\lambda}{8\pi^2}\, \Bigl (\ln\frac{\epsilon^2}{x_{12}^2}-1\Bigr )\,
\Bigl ({u_1\cdot v_2\, v_1\cdot u_2}-{u_1\cdot u_2\, v_1\cdot v_2 - \frac{1}{2}\, u_1\cdot v_1\, u_2\cdot v_2 } \Bigr )\,  \nn \\
&=  I_{12}^2 \, \frac{\lambda}{8 \pi^2} \Bigl (\ln \frac{\e^2}{x_{12}^2} - 1\Bigr ) \Bigg( \begin{minipage}[h]{8mm}\begin{center}\footnotesize

\begin{fmffile}{2gonBlank}
\begin{fmfgraph*}(5,14)
\fmfstraight
\fmfleft{dummy1,p1,a2,a1,p2,dummy2}
\fmfright{dummy3,p3,b1,b2,p4,dummy4}
\fmfpen{thin}
\fmf{plain}{dummy1,p1,a2,a1,p2,dummy2}
\fmf{plain}{dummy3,p3,b1,b2,p4,dummy4}
\fmffreeze
\fmf{plain,width=2}{a1,b2}
\fmf{plain,width=2}{a2,b1}
\fmf{dots}{p1,p3}
\fmf{dots}{p2,p4}
\fmfdot{a1,a2,b1,b2}
\end{fmfgraph*}
\end{fmffile}

\end{center}\end{minipage} - \begin{minipage}[h]{8mm}\begin{center}\footnotesize

\begin{fmffile}{2gonCrossed}
\begin{fmfgraph*}(5,14)
\fmfstraight
\fmfleft{dummy1,p1,a2,a1,p2,dummy2}
\fmfright{dummy3,p3,b1,b2,p4,dummy4}
\fmfpen{thin}
\fmf{plain}{dummy1,p1,a2,a1,p2,dummy2}
\fmf{plain}{dummy3,p3,b1,b2,p4,dummy4}
\fmffreeze
\fmf{plain,width=2}{a1,b1}
\fmf{plain,width=2}{a2,b2}
\fmf{dots}{p1,p3}
\fmf{dots}{p2,p4}
\fmfdot{a1,a2,b1,b2}
\end{fmfgraph*}
\end{fmffile}

\end{center}\end{minipage} + \frac{1}{2} \begin{minipage}[h]{8mm}\begin{center}\footnotesize

\begin{fmffile}{2gonSelf}
\begin{fmfgraph*}(5,14)
\fmfstraight
\fmfleft{dummy1,p1,a2,a1,p2,dummy2}
\fmfright{dummy3,p3,b1,b2,p4,dummy4}
\fmfpen{thin}
\fmf{plain}{dummy1,p1,a2,a1,p2,dummy2}
\fmf{plain}{dummy3,p3,b1,b2,p4,dummy4}
\fmfcurved
\fmf{plain,left,width=2,tension=0}{a1,a2}
\fmf{plain,left,width=2,tension=0}{b1,b2}
\fmf{dots}{p1,p3}
\fmf{dots}{p2,p4}
\fmfdot{a1,a2,b1,b2}
\end{fmfgraph*}
\end{fmffile}

\end{center}\end{minipage} \Bigg)\, ,
\label{eqn:2-Gon Dressing}
\end{align}
where the diagrams in the last line only stand for the index contractions not for
propagators. This contraction structure is of course that of an integrable
nearest neighbor $SO(6)$ vector spin-chain Hamiltonian as was first noted
in \cite{Minahan:2002ve}.

Analogously, for the three-gon we find
\begin{align}
\skla{\begin{minipage}[h]{20mm}\begin{center}\footnotesize

\begin{fmffile}{3gon}
\begin{fmfgraph*}(14,14)
\fmfpen{thin}
\fmfsurroundn{d}{6}
\fmffreeze
\fmf{plain}{d1,b1,b2,d2}
\fmf{plain}{d3,a1,a2,d4}
\fmf{plain}{d5,c1,c2,d6}
\fmf{dots}{d2,d3}
\fmf{dots}{d4,d5}
\fmf{dots}{d6,d1}
\fmffreeze
\fmf{plain,width=2}{a1,b2}
\fmf{plain,width=2}{a2,c1}
\fmf{plain,width=2}{b1,c2}
\fmfdot{a1,a2,b1,b2,c1,c2}
\fmfv{label=$u_1$,label.dist=1.5thick}{a1}
\fmfv{label=$u_2$,label.dist=1.5thick}{a2}
\fmfv{label=$v_1$,label.dist=1.5thick}{b1}
\fmfv{label=$v_2$,label.dist=1.5thick}{b2}
\fmfv{label=$w_1~$}{c1}
\fmfv{label=$~w_2$}{c2}
\end{fmfgraph*}
\end{fmffile}

\end{center}\end{minipage}}_\text{1-loop} &=
\frac{1}{2} \; \begin{minipage}[h]{20mm}\begin{center}\footnotesize

\begin{fmffile}{3gonSelfEnergyA}
\begin{fmfgraph*}(14,14)
\fmfpen{thin}
\fmfsurroundn{d}{6}
\fmffreeze
\fmf{plain}{d1,b1,b2,d2}
\fmf{plain}{d3,a1,a2,d4}
\fmf{plain}{d5,c1,c2,d6}
\fmf{dots}{d2,d3}
\fmf{dots}{d4,d5}
\fmf{dots}{d6,d1}
\fmffreeze
\fmf{plain,width=2}{a1,v,b2}
\fmf{plain,width=2}{a2,c1}
\fmf{plain,width=2}{b1,c2}
\fmfdot{a1,a2,b1,b2,c1,c2}
\fmfblob{6}{v}
\fmfv{label=$u_1$,label.dist=1.5thick}{a1}
\fmfv{label=$u_2$,label.dist=1.5thick}{a2}
\fmfv{label=$v_1$,label.dist=1.5thick}{b1}
\fmfv{label=$v_2$,label.dist=1.5thick}{b2}
\fmfv{label=$w_1~$}{c1}
\fmfv{label=$~w_2$}{c2}
\end{fmfgraph*}
\end{fmffile}

\end{center}\end{minipage} + \begin{minipage}[h]{20mm}\begin{center}\footnotesize

\begin{fmffile}{3gonGluonA}
\begin{fmfgraph*}(14,14)
\fmfpen{thin}
\fmfsurroundn{d}{6}
\fmffreeze
\fmf{plain}{d1,b1,b2,d2}
\fmf{plain}{d3,a1,a2,d4}
\fmf{plain}{d5,c1,c2,d6}
\fmf{dots}{d2,d3}
\fmf{dots}{d4,d5}
\fmf{dots}{d6,d1}
\fmffreeze
\fmf{plain,width=2}{a1,v1,b2}
\fmf{plain,width=2}{a2,v2,c1}
\fmf{plain,width=2}{b1,c2}
\fmffreeze
\fmf{wiggly}{v1,v2}
\fmfdot{a1,a2,b1,b2,c1,c2}
\fmfv{label=$u_1$,label.dist=1.5thick}{a1}
\fmfv{label=$u_2$,label.dist=1.5thick}{a2}
\fmfv{label=$v_1$,label.dist=1.5thick}{b1}
\fmfv{label=$v_2$,label.dist=1.5thick}{b2}
\fmfv{label=$w_1~$}{c1}
\fmfv{label=$~w_2$}{c2}
\end{fmfgraph*}
\end{fmffile}

\end{center}\end{minipage}
+\begin{minipage}[h]{20mm}\begin{center}\footnotesize

\begin{fmffile}{3gonVertexA}
\begin{fmfgraph*}(14,14)
\fmfpen{thin}
\fmfsurroundn{d}{6}
\fmffreeze
\fmf{plain}{d1,b1,b2,d2}
\fmf{plain}{d3,a1,a2,d4}
\fmf{plain}{d5,c1,c2,d6}
\fmf{dots}{d2,d3}
\fmf{dots}{d4,d5}
\fmf{dots}{d6,d1}
\fmffreeze
\fmf{plain,width=2}{a2,v,b2}
\fmf{plain,width=2}{a1,v,c1}
\fmf{plain,width=2}{b1,c2}
\fmffreeze
\fmfdot{a1,a2,b1,b2,c1,c2,v}
\fmfv{label=$u_1$,label.dist=1.5thick}{a1}
\fmfv{label=$u_2$,label.dist=1.5thick}{a2}
\fmfv{label=$v_1$,label.dist=1.5thick}{b1}
\fmfv{label=$v_2$,label.dist=1.5thick}{b2}
\fmfv{label=$w_1~$}{c1}
\fmfv{label=$~w_2$}{c2}
\end{fmfgraph*}
\end{fmffile}

\end{center}\end{minipage} + \mbox{2 permutations} \nn\\
&= I_{12} I_{13} I_{23} \times \frac{\lambda}{16 \pi^2} \nnl
&\times \Bigg[\kla{\lnl{23}{12}{13} - 2} \kla{\begin{minipage}[h]{16mm}\begin{center}\footnotesize

\begin{fmffile}{3gonBlank}
\begin{fmfgraph*}(14,14)
\fmfpen{thin}
\fmfsurroundn{d}{6}
\fmffreeze
\fmf{plain}{d1,b1,b2,d2}
\fmf{plain}{d3,a1,a2,d4}
\fmf{plain}{d5,c1,c2,d6}
\fmf{dots}{d2,d3}
\fmf{dots}{d4,d5}
\fmf{dots}{d6,d1}
\fmffreeze
\fmf{plain,width=2}{a1,b2}
\fmf{plain,width=2}{a2,c1}
\fmf{plain,width=2}{b1,c2}
\fmfdot{a1,a2,b1,b2,c1,c2}
\end{fmfgraph*}
\end{fmffile}

\end{center}\end{minipage} -
\begin{minipage}[h]{16mm}\begin{center}\footnotesize

\begin{fmffile}{3gonCrossedA}
\begin{fmfgraph*}(14,14)
\fmfpen{thin}
\fmfsurroundn{d}{6}
\fmffreeze
\fmf{plain}{d1,b1,b2,d2}
\fmf{plain}{d3,a1,a2,d4}
\fmf{plain}{d5,c1,c2,d6}
\fmf{dots}{d2,d3}
\fmf{dots}{d4,d5}
\fmf{dots}{d6,d1}
\fmffreeze
\fmf{plain,width=2}{a2,b2}
\fmf{plain,width=2}{a1,c1}
\fmf{plain,width=2}{b1,c2}
\fmfdot{a1,a2,b1,b2,c1,c2}
\end{fmfgraph*}
\end{fmffile}

\end{center}\end{minipage} + \frac{1}{2} \begin{minipage}[h]{16mm}\begin{center}\footnotesize

\begin{fmffile}{3gonSelfA}
\begin{fmfgraph*}(14,14)
\fmfpen{thin}
\fmfsurroundn{d}{6}
\fmffreeze
\fmf{plain}{d1,b1,b2,d2}
\fmf{plain}{d3,a1,a2,d4}
\fmf{plain}{d5,c1,c2,d6}
\fmf{dots}{d2,d3}
\fmf{dots}{d4,d5}
\fmf{dots}{d6,d1}
\fmffreeze
\fmfcurved
\fmf{plain,left,width=2}{a1,a2}
\fmf{plain,right,width=2}{b2,c1}
\fmf{plain,width=2}{b1,c2}
\fmfdot{a1,a2,b1,b2,c1,c2}
\end{fmfgraph*}
\end{fmffile}

\end{center}\end{minipage}} \nnl
&+ \kla{\lnl{13}{12}{23} - 2} \kla{\begin{minipage}[h]{16mm}\begin{center}\footnotesize

\begin{fmffile}{3gonBlank}
\begin{fmfgraph*}(14,14)
\fmfpen{thin}
\fmfsurroundn{d}{6}
\fmffreeze
\fmf{plain}{d1,b1,b2,d2}
\fmf{plain}{d3,a1,a2,d4}
\fmf{plain}{d5,c1,c2,d6}
\fmf{dots}{d2,d3}
\fmf{dots}{d4,d5}
\fmf{dots}{d6,d1}
\fmffreeze
\fmf{plain,width=2}{a1,b2}
\fmf{plain,width=2}{a2,c1}
\fmf{plain,width=2}{b1,c2}
\fmfdot{a1,a2,b1,b2,c1,c2}
\end{fmfgraph*}
\end{fmffile}

\end{center}\end{minipage}
- \begin{minipage}[h]{16mm}\begin{center}\footnotesize

\begin{fmffile}{3gonCrossedB}
\begin{fmfgraph*}(14,14)
\fmfpen{thin}
\fmfsurroundn{d}{6}
\fmffreeze
\fmf{plain}{d1,b1,b2,d2}
\fmf{plain}{d3,a1,a2,d4}
\fmf{plain}{d5,c1,c2,d6}
\fmf{dots}{d2,d3}
\fmf{dots}{d4,d5}
\fmf{dots}{d6,d1}
\fmffreeze
\fmf{plain,width=2}{a1,b1}
\fmf{plain,width=2}{a2,c1}
\fmf{plain,width=2}{b2,c2}
\fmfdot{a1,a2,b1,b2,c1,c2}
\end{fmfgraph*}
\end{fmffile}

\end{center}\end{minipage} + \frac{1}{2} \begin{minipage}[h]{16mm}\begin{center}\footnotesize

\begin{fmffile}{3gonSelfB}
\begin{fmfgraph*}(14,14)
\fmfpen{thin}
\fmfsurroundn{d}{6}
\fmffreeze
\fmf{plain}{d1,b1,b2,d2}
\fmf{plain}{d3,a1,a2,d4}
\fmf{plain}{d5,c1,c2,d6}
\fmf{dots}{d2,d3}
\fmf{dots}{d4,d5}
\fmf{dots}{d6,d1}
\fmffreeze
\fmf{plain,left,width=2}{a1,c2}
\fmf{plain,width=2}{a2,c1}
\fmf{plain,left,width=2}{b1,b2}
\fmfdot{a1,a2,b1,b2,c1,c2}
\end{fmfgraph*}
\end{fmffile}

\end{center}\end{minipage}} \nnl
&+ \kla{\lnl{12}{13}{23} - 2} \kla{\begin{minipage}[h]{16mm}\begin{center}\footnotesize

\begin{fmffile}{3gonBlank}
\begin{fmfgraph*}(14,14)
\fmfpen{thin}
\fmfsurroundn{d}{6}
\fmffreeze
\fmf{plain}{d1,b1,b2,d2}
\fmf{plain}{d3,a1,a2,d4}
\fmf{plain}{d5,c1,c2,d6}
\fmf{dots}{d2,d3}
\fmf{dots}{d4,d5}
\fmf{dots}{d6,d1}
\fmffreeze
\fmf{plain,width=2}{a1,b2}
\fmf{plain,width=2}{a2,c1}
\fmf{plain,width=2}{b1,c2}
\fmfdot{a1,a2,b1,b2,c1,c2}
\end{fmfgraph*}
\end{fmffile}

\end{center}\end{minipage} -
\begin{minipage}[h]{16mm}\begin{center}\footnotesize

\begin{fmffile}{3gonCrossedC}
\begin{fmfgraph*}(14,14)
\fmfpen{thin}
\fmfsurroundn{d}{6}
\fmffreeze
\fmf{plain}{d1,b1,b2,d2}
\fmf{plain}{d3,a1,a2,d4}
\fmf{plain}{d5,c1,c2,d6}
\fmf{dots}{d2,d3}
\fmf{dots}{d4,d5}
\fmf{dots}{d6,d1}
\fmffreeze
\fmf{plain,width=2}{a1,b2}
\fmf{plain,width=2}{a2,c2}
\fmf{plain,width=2}{b1,c1}
\fmfdot{a1,a2,b1,b2,c1,c2}
\end{fmfgraph*}
\end{fmffile}

\end{center}\end{minipage} + \frac{1}{2} \begin{minipage}[h]{16mm}\begin{center}\footnotesize

\begin{fmffile}{3gonSelfC}
\begin{fmfgraph*}(14,14)
\fmfpen{thin}
\fmfsurroundn{d}{6}
\fmffreeze
\fmf{plain}{d1,b1,b2,d2}
\fmf{plain}{d3,a1,a2,d4}
\fmf{plain}{d5,c1,c2,d6}
\fmf{dots}{d2,d3}
\fmf{dots}{d4,d5}
\fmf{dots}{d6,d1}
\fmffreeze
\fmf{plain,width=2}{a1,b2}
\fmf{plain,right,width=2}{a2,b1}
\fmf{plain,left,width=2}{c1,c2}
\fmfdot{a1,a2,b1,b2,c1,c2}
\end{fmfgraph*}
\end{fmffile}

\end{center}\end{minipage}} \Bigg].
\label{eqn:3-Gon Dressing}
\end{align}
Again the graphs in the last three lines only represent the index contractions.
Interestingly a similar structure to the integrable spin-chain Hamiltonian
of \eqn{eqn:2-Gon Dressing} emerges also for the one-loop three-gon interactions.

\subsection{Gauge invariance and Wilson line contributions}

There is one important point we have not addressed so far. The point splitting regularization method
that we employed violates gauge invariance as the space-time locations of the
two neighboring operators in the trace are no longer coincident. The natural way to recover
gauge invariance is to connect the two split points through a straight Wilson line. This, however, gives rise
to new diagrams not yet accounted for in which a gluon is radiated off the Wilson line.
Luckily we are able to show that this
contribution vanishes entirely at the one-loop level for $|\epsilon|\to 0$.

Setting $\epsilon^{\mu}=x^{\mu}_{13}$
the Wilson line is parametrized by
\be
x^{\mu}(\tau) = x_{3}^{\mu}+\epsilon^{\mu}\, \tau\, , \quad \tau\in[0,1]\, .
\ee
We then have the contribution
\begin{align}
\begin{minipage}[h]{20mm}\begin{center}\footnotesize

\begin{fmffile}{WilsonLine}
\begin{fmfgraph*}(16,16)
\fmfstraight
\fmfpen{thick}
\fmfright{d4,a4,tau2,a2,d2}
\fmfleft{d3,a3,tau,a1,d1}
\fmf{plain}{a1,v1,a2}
\fmf{plain}{a3,v2,a4}
\fmffreeze
\fmf{dbl_plain,width=1}{a3,tau,a1}
\fmf{wiggly,width=1}{v1,tau}
\fmf{dots}{d1,a1}
\fmf{dots}{d2,d4}
\fmf{dots}{d3,a3}
\fmfdot{a1,a2,a3,a4,tau,v1}
\fmfv{label=$1$}{a1}
\fmfv{label=$2$}{a2}
\fmfv{label=$3$}{a3}
\fmfv{label=$4$}{a4}
\fmfv{label=$\tau$}{tau}
\fmfv{label=$\omega$}{v1}
\end{fmfgraph*}
\end{fmffile}

\end{center}\end{minipage} \,\,\, &=
\lambda (u_{1}\cdot u_{2})(u_{3}\cdot u_{4})\,
\int_{0}^{1}d\tau \, \epsilon\cdot(\partial_{1}-\partial_{2})\, Y_{12\tau} \nn\\
&=
-\frac{2\lambda (u_{1}\cdot u_{2})(u_{3}\cdot u_{4})}{(2\pi)^{6}}\,
\int_{0}^{1}d\tau \int d^{4}\omega \, \frac{\epsilon\cdot x_{1\omega}}
{(x_{1\omega}^{2})^{2}\, x_{2\omega}^{2}\, x_{\tau\omega}^{2}}\, .
\end{align}
This five dimensional integral is by power-counting logarithmically divergent for
coincident points $x_{3},x(\tau)\to x_{1}$ i.e.~$|\epsilon|\to 0$ and one has
\be
\lim_{|\epsilon|\to 0}\int_{0}^{1} d\tau\int d^{4}\omega \, \frac{\epsilon\cdot x_{1\omega}}
{(x_{1\omega}^{2})^{2}\, x_{2\omega}^{2}\, x_{\tau\omega}^{2}}
\sim \lim_{|\epsilon|\to 0}\, \epsilon\cdot x_{12}\, \left (
\ln\frac{\epsilon^{2}}{x_{12}^{2}} +\text{finite} + O(\epsilon)\, \right )
\to 0\, .
\ee
There is also a novel ladder-diagram in which a gluon is exchanged between two Wilson lines
extending from $x_{1}$ to $x_{3}$ and from $x_{2}$ to $x_{4}$. This ladder-graph is
manifestly finite and vanishes
as $\epsilon^{2}$.
Therefore all the Wilson line contributions to the point splitting regularization vanish at this
order of perturbation theory.

\subsection{Extremal three-point functions}

\begin{figure}
 \centering
 \raisebox{0.5cm}{\includegraphics[width=4cm]{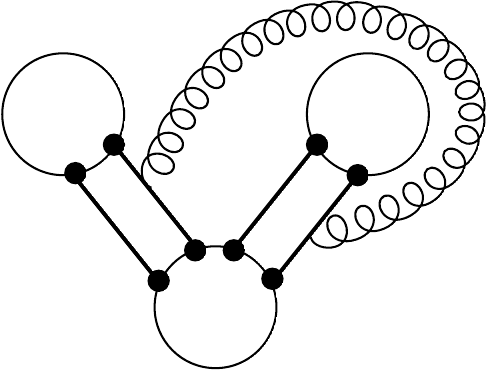}}
 \caption{Additional Feynman-Graphs for extremal three-point functions.}
 \label{fig:Extremal3pt}
\end{figure}

Three-point functions of operators with lengths $\Dn_\alpha$, $\Dn_\beta$ and
$\Dn_\gamma$ where
$\Dn_\alpha + \Dn_\beta = \Dn_\gamma$ are called extremal. For these extremal functions the dressing formulae above do not hold any longer for two reasons: First, there appear additional diagrams with a gluon exchange or a vertex between non-neighboring propagators as the one in figure \ref{fig:Extremal3pt}. These non-nearest neighbor interactions lead to additional terms in the dressing formulae. Second, unlike non-extremal ones extremal three-point functions with double-trace operators contain the same factor of $N$ as those with single-trace operators. This results in an operator mixing \index{operator!mixing} of single-trace with double-trace operators already at tree-level. This is described in detail in \cite{D'Hoker:1999ea,Okuyama:2004bd}.

We will refrain from studying these extremal three-point correlators in the following.
In any case the one-loop structure constants follow a simple pattern:
They are a given by a linear function of the anomalous scaling dimensions
of the operators involved \cite{Okuyama:2004bd}
\begin{equation}
 C^{(1)}_{\alpha \beta \gamma, \,\text{extremal}} \Bigr |_{\text{loop}} = \frac{1}{2} \,
 C^{(0)}_{\alpha \beta \gamma, \, \text{extremal}} \,
 \kla{\gamma_\alpha + \gamma_\beta - \gamma_\gamma}\, ,
 \label{eqn:form of structure constants for extremal correlators}
\end{equation}
hence the three-point problem has been reduced to the two-point one. In particular
structure constants of protected operators are free of radiative corrections.
However, the crucial mixing contributions are not taken into account in this formula
which will also involve mixings with multi-trace operators.

\subsection{Two convenient regularization schemes}

We have seen in \eqn{schemeindepc} how to extract the regularization scheme independent
structure constant from a combination of the bare structure constant and the one-loop
finite normalization shifts. As the latter arises from the finite contribution to
the two-gon dressing \eqn{eqn:2-Gon Dressing} one may pick a regularization to simply
cancel these contributions. I.e.~making the transformation on the point-splitting
parameter
\be
\epsilon \to \sqrt{e}\, \epsilon
\ee
transforms
\be
\lnk{ij}-1 \to \lnk{ij}\, , \qquad \text{and}\qquad
\lnl{ij}{ik}{jk}-2  \to \lnl{ij}{ik}{jk}-1\ .
\ee
Hence in this scheme the finite part of the two-gon dressing vanishes resulting in
a vanishing finite correction to the two-point functions
\be
g_{\alpha}=0\, ,
\ee
which in turn implies that the bare and the renormalized structure functions
coincide in this scheme
\be
\widetilde{C}^{(1)}_{\alpha \beta \gamma} = C^{(1)}_{\alpha \beta \gamma}\, .
\ee
This implies that the structure function may be read off solely from the three-gon
dressings of the non-extremal correlator, which may be graphically represented by
\begin{align}
 C_{\alpha \beta \gamma}^{(1)} &= -\frac{1}{16 \pi^2} \, \sum_{\substack{\text{cyclic}\\\text{perm.}}} \, \Bigg[ 3 \times \begin{minipage}[h]{16mm}\begin{center}\footnotesize

\begin{fmffile}{3gonBlank}
\begin{fmfgraph*}(14,14)
\fmfpen{thin}
\fmfsurroundn{d}{6}
\fmffreeze
\fmf{plain}{d1,b1,b2,d2}
\fmf{plain}{d3,a1,a2,d4}
\fmf{plain}{d5,c1,c2,d6}
\fmf{dots}{d2,d3}
\fmf{dots}{d4,d5}
\fmf{dots}{d6,d1}
\fmffreeze
\fmf{plain,width=2}{a1,b2}
\fmf{plain,width=2}{a2,c1}
\fmf{plain,width=2}{b1,c2}
\fmfdot{a1,a2,b1,b2,c1,c2}
\end{fmfgraph*}
\end{fmffile}

\end{center}\end{minipage} - \begin{minipage}[h]{16mm}\begin{center}\footnotesize

\begin{fmffile}{3gonCrossedA}
\begin{fmfgraph*}(14,14)
\fmfpen{thin}
\fmfsurroundn{d}{6}
\fmffreeze
\fmf{plain}{d1,b1,b2,d2}
\fmf{plain}{d3,a1,a2,d4}
\fmf{plain}{d5,c1,c2,d6}
\fmf{dots}{d2,d3}
\fmf{dots}{d4,d5}
\fmf{dots}{d6,d1}
\fmffreeze
\fmf{plain,width=2}{a2,b2}
\fmf{plain,width=2}{a1,c1}
\fmf{plain,width=2}{b1,c2}
\fmfdot{a1,a2,b1,b2,c1,c2}
\end{fmfgraph*}
\end{fmffile}

\end{center}\end{minipage}
 + \frac{1}{2} \times \begin{minipage}[h]{16mm}\begin{center}\footnotesize

\begin{fmffile}{3gonSelfA}
\begin{fmfgraph*}(14,14)
\fmfpen{thin}
\fmfsurroundn{d}{6}
\fmffreeze
\fmf{plain}{d1,b1,b2,d2}
\fmf{plain}{d3,a1,a2,d4}
\fmf{plain}{d5,c1,c2,d6}
\fmf{dots}{d2,d3}
\fmf{dots}{d4,d5}
\fmf{dots}{d6,d1}
\fmffreeze
\fmfcurved
\fmf{plain,left,width=2}{a1,a2}
\fmf{plain,right,width=2}{b2,c1}
\fmf{plain,width=2}{b1,c2}
\fmfdot{a1,a2,b1,b2,c1,c2}
\end{fmfgraph*}
\end{fmffile}

\end{center}\end{minipage} \nnl
 &\bleq - \begin{minipage}[h]{16mm}\begin{center}\footnotesize

\begin{fmffile}{3gonCrossedB}
\begin{fmfgraph*}(14,14)
\fmfpen{thin}
\fmfsurroundn{d}{6}
\fmffreeze
\fmf{plain}{d1,b1,b2,d2}
\fmf{plain}{d3,a1,a2,d4}
\fmf{plain}{d5,c1,c2,d6}
\fmf{dots}{d2,d3}
\fmf{dots}{d4,d5}
\fmf{dots}{d6,d1}
\fmffreeze
\fmf{plain,width=2}{a1,b1}
\fmf{plain,width=2}{a2,c1}
\fmf{plain,width=2}{b2,c2}
\fmfdot{a1,a2,b1,b2,c1,c2}
\end{fmfgraph*}
\end{fmffile}

\end{center}\end{minipage} + \frac{1}{2} \times \begin{minipage}[h]{16mm}\begin{center}\footnotesize

\begin{fmffile}{3gonSelfB}
\begin{fmfgraph*}(14,14)
\fmfpen{thin}
\fmfsurroundn{d}{6}
\fmffreeze
\fmf{plain}{d1,b1,b2,d2}
\fmf{plain}{d3,a1,a2,d4}
\fmf{plain}{d5,c1,c2,d6}
\fmf{dots}{d2,d3}
\fmf{dots}{d4,d5}
\fmf{dots}{d6,d1}
\fmffreeze
\fmf{plain,left,width=2}{a1,c2}
\fmf{plain,width=2}{a2,c1}
\fmf{plain,left,width=2}{b1,b2}
\fmfdot{a1,a2,b1,b2,c1,c2}
\end{fmfgraph*}
\end{fmffile}

\end{center}\end{minipage}
 - \begin{minipage}[h]{16mm}\begin{center}\footnotesize

\begin{fmffile}{3gonCrossedC}
\begin{fmfgraph*}(14,14)
\fmfpen{thin}
\fmfsurroundn{d}{6}
\fmffreeze
\fmf{plain}{d1,b1,b2,d2}
\fmf{plain}{d3,a1,a2,d4}
\fmf{plain}{d5,c1,c2,d6}
\fmf{dots}{d2,d3}
\fmf{dots}{d4,d5}
\fmf{dots}{d6,d1}
\fmffreeze
\fmf{plain,width=2}{a1,b2}
\fmf{plain,width=2}{a2,c2}
\fmf{plain,width=2}{b1,c1}
\fmfdot{a1,a2,b1,b2,c1,c2}
\end{fmfgraph*}
\end{fmffile}

\end{center}\end{minipage} + \frac{1}{2} \times \begin{minipage}[h]{16mm}\begin{center}\footnotesize

\begin{fmffile}{3gonSelfC}
\begin{fmfgraph*}(14,14)
\fmfpen{thin}
\fmfsurroundn{d}{6}
\fmffreeze
\fmf{plain}{d1,b1,b2,d2}
\fmf{plain}{d3,a1,a2,d4}
\fmf{plain}{d5,c1,c2,d6}
\fmf{dots}{d2,d3}
\fmf{dots}{d4,d5}
\fmf{dots}{d6,d1}
\fmffreeze
\fmf{plain,width=2}{a1,b2}
\fmf{plain,right,width=2}{a2,b1}
\fmf{plain,left,width=2}{c1,c2}
\fmfdot{a1,a2,b1,b2,c1,c2}
\end{fmfgraph*}
\end{fmffile}

\end{center}\end{minipage} \Bigg]\, .
\label{eqn:Structure Constant as 3-Gon Dressing}
\end{align}
Alternatively one may apply the transformation
\be
\epsilon \to e\, \epsilon
\ee
yielding
\be
\lnk{ij}-1 \to \lnk{ij} +1\, , \qquad \text{and}\qquad
\lnl{ij}{ik}{jk}-2  \to \lnl{ij}{ik}{jk}\, .
\label{here}
\ee

Now the finite contributions to the three-gon dressings vanish and the bare structure
constant may be computed from only dressing the two-gons in the tree-level
correlator
\be
 \widetilde{C}^{(1)}_{\alpha \beta \gamma} = \frac{1}{8 \pi^2} \sum_{\substack{\text{cyclic}\\\text{perm.}}} \; \sum_{\substack{\text{all}\\\text{2-gons}}} \Bigg( \begin{minipage}[h]{8mm}\begin{center}\footnotesize

\begin{fmffile}{2gonBlank}
\begin{fmfgraph*}(5,14)
\fmfstraight
\fmfleft{dummy1,p1,a2,a1,p2,dummy2}
\fmfright{dummy3,p3,b1,b2,p4,dummy4}
\fmfpen{thin}
\fmf{plain}{dummy1,p1,a2,a1,p2,dummy2}
\fmf{plain}{dummy3,p3,b1,b2,p4,dummy4}
\fmffreeze
\fmf{plain,width=2}{a1,b2}
\fmf{plain,width=2}{a2,b1}
\fmf{dots}{p1,p3}
\fmf{dots}{p2,p4}
\fmfdot{a1,a2,b1,b2}
\end{fmfgraph*}
\end{fmffile}

\end{center}\end{minipage} - \begin{minipage}[h]{8mm}\begin{center}\footnotesize

\begin{fmffile}{2gonCrossed}
\begin{fmfgraph*}(5,14)
\fmfstraight
\fmfleft{dummy1,p1,a2,a1,p2,dummy2}
\fmfright{dummy3,p3,b1,b2,p4,dummy4}
\fmfpen{thin}
\fmf{plain}{dummy1,p1,a2,a1,p2,dummy2}
\fmf{plain}{dummy3,p3,b1,b2,p4,dummy4}
\fmffreeze
\fmf{plain,width=2}{a1,b1}
\fmf{plain,width=2}{a2,b2}
\fmf{dots}{p1,p3}
\fmf{dots}{p2,p4}
\fmfdot{a1,a2,b1,b2}
\end{fmfgraph*}
\end{fmffile}

\end{center}\end{minipage} + \frac{1}{2} \begin{minipage}[h]{8mm}\begin{center}\footnotesize

\begin{fmffile}{2gonSelf}
\begin{fmfgraph*}(5,14)
\fmfstraight
\fmfleft{dummy1,p1,a2,a1,p2,dummy2}
\fmfright{dummy3,p3,b1,b2,p4,dummy4}
\fmfpen{thin}
\fmf{plain}{dummy1,p1,a2,a1,p2,dummy2}
\fmf{plain}{dummy3,p3,b1,b2,p4,dummy4}
\fmfcurved
\fmf{plain,left,width=2,tension=0}{a1,a2}
\fmf{plain,left,width=2,tension=0}{b1,b2}
\fmf{dots}{p1,p3}
\fmf{dots}{p2,p4}
\fmfdot{a1,a2,b1,b2}
\end{fmfgraph*}
\end{fmffile}

\end{center}\end{minipage} \Bigg)\, .
 \ee
The scheme independent structure constants can then be calculated using
 \eqn{schemeindepc} with $g_\alpha=\gamma_\alpha$ by virtue of \eqn{here}, i.e.
\be
 C^{(1)}_{\alpha \beta \gamma} = \widetilde{C}^{(1)}_{\alpha \beta \gamma} - \frac{1}{2} \, C^{(0)}_{\alpha \beta \gamma} \, \kla{\gamma_\alpha  + \gamma_\beta + \gamma_\gamma}. \label{eqn:structure constants as 2-gon dressing}
\ee
In our actual computations we have used both schemes depending on the problem at hand.

\section{The considered operators and their mixing}
\label{sect:mixing}

As discussed above we also have to face the subtle problem of operator mixing.
In this paper we aim at computing the structure constants of space-time scalar operators
constructed exclusively from the $SO(6)$ scalar fields of $\mathcal{N}=4$ super Yang-Mills
\be
\cO_{\alpha}=\sum c_{i_{1}\ldots i_{L}}\, \Tr(\phi_{i_{1}}\ldots \phi_{i_{L}}) \, .
\ee
Due to operator mixing these will be corrected perturbatively by operators with bi-fermion
and bi-derivative insertions, as well as 'self-mixings' in the purely scalar sector, i.e. we are facing
the mixing structure
\begin{align}
\hat{\cO}_{\alpha} &=  \cO_{\alpha} + \gYM\, N\,\cO_{\alpha,\psi\psi} + \gYM^{2}
\,N^{2}\, \cO_{\alpha,DD} +
\gYM^{2}\, N\, \cO_{\alpha,\text{self}} +\cO(\gYM^{3}) \nn \\ & \text{with}\nn \\
\cO_{\alpha,\psi\psi} &=   \sum d_{i_{1}\ldots i_{L-3}}\, \Tr(
\psi^{\alpha}\phi_{i_{1}}\ldots\phi_{i_{\ast-1}}\psi_{\alpha}\phi_{i_{\ast}}
 \ldots \phi_{i_{L-3}}) \nn\\
\cO_{\alpha,DD} &=   \sum e_{i_{1}\ldots i_{L-3};kl}\, \Tr(D^{\mu}\phi_{i_{k}}
\phi_{i_{1}}\ldots \phi_{i_{\ast-1}} D_{\mu}\phi_{i_{l}}\phi_{i_{\ast}}
 \ldots \phi_{i_{L-4}}) \nn\\
\cO_{\alpha,\text{self}}&=
\sum f_{i_{1}\ldots i_{L}}\, \Tr(\phi_{i_{1}}\ldots \phi_{i_{L}}) \, .
\label{mixingpattern}
\end{align}
These mixings will lead to contributions to the one-loop structure constant
$C^{(1)}_{\alpha\beta\gamma}$ beyond the radiative ones discussed in the previous chapter
through the tree-level correlators
\be
\vev{\cO_{\alpha,\psi\psi}\, \cO_{\beta,\psi\psi}\, \cO_{\gamma}}_{0}\, ,\quad
\vev{\cO_{\alpha,DD}\, \cO_{\beta}\, \cO_{\gamma}}_{0}\, ,\quad
\vev{\cO_{\alpha,\text{self}}\, \cO_{\beta}\, \cO_{\gamma}}_{0}\, ,
\ee
and their permutations, as well as the insertion of a Yukawa-vertex into the correlators
\be
\vev{\cO_{\alpha,\psi\psi}\, \cO_{\beta}\, \cO_{\gamma}}_{\text{Yukawa}}\, ,
\ee
which are all of order $\cO(\gYM^{2})$.

The operators we shall be considering in the non-protected sector  are at leading order
($\Phi_{{AB}}=\frac{1}{2}\,\epsilon_{ABCD}\, \Phi^{CD}$ and $Z=\frac{1}{\sqrt{2}}\,(\phi_{5}+i\phi_{6})$ are the complexified scalars,
see appendix \ref{App:A} for conventions)
\begin{align}
\cO_{2A}&=  \mathcal{K} = \frac{8 \pi^2}{\sqrt{3} N} \tr{\Phi_{AB}\Phi^{AB}} \\
{\mathcal{O}}^J_n &=
\sqrt{\frac{(8\pi^{2})^{J+2}}{N^{J+2}(J+3)}}
\Big\{
\sum_{p=0}^J\cos{\frac{\pi n (2p+3)}{J+3}}
\tr{\Phi_{AB} Z^p \Phi^{AB} Z^{J-p}}\Big \} \nn\\
&\qquad J=1,2,3 \qquad n=1,\ldots,[{J}/{2}]
\label{OJn}\\
\hat{\mathcal{O}}_{4A/E} &=
     \Big(
     \frac{8\pi^2}{N}
     \Big)^2
     \frac{4}{\sqrt{738\mp102\sqrt{41}}}
     \Big \{
       \tr{\Phi_{AB}\Phi^{AB}\Phi_{CD}\Phi^{CD}} +
       \frac{5 \mp \sqrt{41}}{4}
       \tr{\Phi_{AB}\Phi_{CD}\Phi^{AB}\Phi^{CD}}\Big \}
       \label{O4AE}
\end{align}
From the mixing pattern in \eqn{mixingpattern} we see that the length two Konishi operator
$\cO_{2A}=\mathcal{K}$
is protected from mixing. The operators ${\mathcal{O}}^J_n$ constitute a whole
family of $SO(4)\times SO(4)$ single trace singlets
with classical conformal dimension
$\Delta^{(0)}=J+2$, where $J$ is the charge under a $U(1)\in
SO(6)$ under which the $Z$ fields carry charge 1.
These states belong to the $[0,J,0]$ representation of
$SU(4)_R$ and they are the highest weights of a long representation of
the $PSU(2,2|4)$ superconformal algebra.
For each value of $J$, there are
$E\big[\frac{J+2}{2}\big]$ such eigenstates of the planar two-loop
conformal dimension, labeled by an integer number $1\leq n \leq
E\big[\frac{J+2}{2}\big]$.
Indeed, the bi-fermion and bi-derivative mixings of these operators
have been worked out in \cite{Georgiou:2008vk,Georgiou:2009tp}:
\begin{align}
\label{hws}
\hat{\mathcal{O}}^J_n =
&\mathcal{N}\left\{
\sum_{p=0}^J\cos{\frac{\pi n (2p+3)}{J+3}}
\tr{\Phi_{AB} Z^p \Phi^{AB} Z^{J-p}}
\nonumber\right.\\ &\left.
g\frac{N}{8\sqrt{2}\pi^2}
\sin{\frac{\pi n}{J+3}}\sum_{p=0}^{J-1}\sin{\frac{\pi n(2p+4)}{J+3}}
\tr{\psi^{1 \alpha} Z^p \psi^2_\alpha Z^{J-p-1}}
\nonumber\right.\\ &\left.
-g\frac{N}{8\sqrt{2}\pi^2}
\sin{\frac{\pi n}{J+3}}\sum_{p=0}^{J-1}\sin{\frac{\pi n(2p+4)}{J+3}}
\tr{\bar{\psi}_{3 \dot{\alpha}} Z^p \bar{\psi}^{\dot{\alpha}}_4 Z^{J-p-1}}
\nonumber\right.\\ &\left.
+g^2\frac{N^2}{(8\sqrt{2}\pi^2)^2}
\sin^2{\frac{\pi n}{J+3}}\sum_{p=0}^{J-2}\cos{\frac{\pi n(2p+5)}{J+3}}
\tr{D_\mu Z Z^p D^\mu Z Z^{J-p-2}}
\nonumber\right.\\ &\left.
+g^2\sum_{\substack{m=1\\m\neq n}}^{E\big[\frac{J+2}{2}\big]}\mathcal{C}_{n,m}^{\text{self}}
 \sum_{p=0}^J\cos{\frac{\pi m (2p+3)}{J+3}}
\tr{\Phi_{AB} Z^p \Phi^{AB} Z^{J-p}}
\right\},
\end{align}
where the self-mixing coefficient $\mathcal{C}_{n,m}^{\text{self}}$ is still undetermined and
the normalization up to one loop is (see the computation in appendix
\ref{App:norm})
\begin{align}
  \label{eq:norm}
  \mathcal{N} = \sqrt{\frac{N_0^{-J-2}}{J+3}}
  \Bigl[ &
    1 + g_{YM}^2N\frac{ \sin^2{\frac{\pi n}{J+3}}}{2 \pi^2 (J+3)}
    \left(
      \frac{J-1}{2} + 2 \cos^2{\frac{2\pi n}{J+3}}
    \right)
    -\frac{g_{YM}^2N}{2} g_{\mathcal{O}^J_n} \nn\\ 
    &+
    g_{YM}^2N \gamma_{\mathcal{O}^J_n}
    \ln{
      \left\vert
        \frac{\Lambda}{\mu}
      \right\vert}
   + \mathcal{O}(g_{YM}^4)
 \Bigr ],
\end{align}
where $N_0=\frac{N}{8\pi^2}$ and $g_{\mathcal{O}^J_n}$ is the scheme dependent finite one-loop
contribution discussed in section 2.  The mixing with the terms containing
fermionic and derivative impurities in \eqref{hws} has been computed
in \cite{Georgiou:2008vk,Georgiou:2009tp} by requiring that
the operator is annihilated by the superconformal charges up to
one-loop. In the next section we shall compute the coefficient of the self-mixing in the last
line of \eqref{hws}, by requiring that the full operator
$\hat{\mathcal{O}}^J_n$ is an eigenstate of the dilatation operator at
two loops constructed in \cite{Georgiou:2011xj}. As we shall be studying only correlators
involving operators up to engineering lenth 5, we shall be studying the set of
operators
\be
\{\,
\mathcal{O}_{4F}:=\mathcal{O}^2_1, \,
\mathcal{O}_{4B}:=\mathcal{O}^2_2, \,
\mathcal{O}_{5J}:=\mathcal{O}^3_1, \,
\mathcal{O}_{5E}:=\mathcal{O}^3_2\,
\}
\label{0j0ops}
\ee
in this $\hat\cO^{J}_{n}$ operator family. We also often use the alternative
(historic) nomenclature pattern introduced in the preprint \cite{Grossardt:2010xq}. Finally we note
that the one-loop scaling dimension of $\hat{\mathcal{O}}^J_{n}$ reads
$\gamma_{Jn}=4 \lambda \sin^2{\frac{\pi n}{J+3}}$.

\subsection{Self-mixing contributions from $H_{4}$}

With $a,b,c,d,e,f, i, j ,k=1,\ldots,6$ denoting the $SO(6)$ indices of the scalar fields
$\Phi_{i}$ one may write the pure $SO(6)$ piece of the one and two-loop hamiltonians
$H_{2}$ and $H_{4}$ as follows \cite{Minahan:2002ve,Beisert:2003ys,Georgiou:2011xj}
\begin{align}
\label{H2andH4}
H_{2}&= \frac{\lambda}{4}\, \Bigl ( 2 \spinchain{ab}{ab} - 2  \spinchain{ba}{ab} +
\spinchain{bb}{aa}  \Bigr)\, ,
\qquad \lambda := \frac{\gYM^{2}\, N}{4\pi^{2}}\, ,\\
H_{4}&= \frac{\lambda^{2}}{4}\, \Bigl (
-2\, \spinchain{abc}{abc}
+ \frac{3}{2}\, \Bigl [ \spinchain{bac}{abc} +  \spinchain{acb}{abc}\Bigr ]
-\frac{1}{2} \, \Bigl [ \spinchain{bca}{abc} +  \spinchain{cab}{abc}\Bigr ] \nn\\
& \qquad -\frac{11}{16} \Bigl [ \spinchain{bbc}{aac} +  \spinchain{cbb}{caa}\Bigr ]
+ \frac{1}{4}\, \spinchain{bcb}{aca}
-\frac{1}{16}\, \Bigl [ \spinchain{bbc}{caa} +  \spinchain{cbb}{aac}\Bigr ]\nn\\
& \qquad +\frac{1}{8}\, \Bigl [ \spinchain{bbc}{aca} +  \spinchain{cbb}{aca}
+\spinchain{bcb}{aac} +  \spinchain{bcb}{caa} \Bigr ]\, .
\end{align}
Where the action of theses operators is defined by
\begin{equation}
\spinchain{abc}{def}\, |\Phi_{i}\,\Phi_{j}\Phi_{k}\rangle =
\delta_{di}\, \delta_{ej}\, \delta_{fk}\, |\Phi_{a}\,\Phi_{b}\Phi_{c}\rangle\, ,
\end{equation}
and repeated indices are summed over. Note that we are now employing a spin-chain
language representing a local gauge invariant operator by a state.

We now use this result in order to determine the scalar contributions to the
self-mixing coefficients in (\ref{hws}).
Upon acting on $H_{4}$ one finds for the operators of \eqn{OJn} and \eqn{O4AE}
\begin{align}
H_{4}\circ {\mathcal{O}}^2_1 & = \frac{-19+7\sqrt{5}}{8}\lambda^2\, {\mathcal{O}}^2_1
+ \frac{-2}{4}\lambda^2\, {\mathcal{O}}^2_2 \, , \nn\\
H_{4}\circ {\mathcal{O}}^2_2 &= \frac{-2}{4}\lambda^2\, {\mathcal{O}}^2_1
+ \frac{-19-7\sqrt{5}}{8}\lambda^2\, {\mathcal{O}}^2_2 \, , \nn\\
H_{4}\circ {\mathcal{O}}^3_1  &= -\frac{1}{8}\lambda^2\, {\mathcal{O}}^3_1
- \frac{\sqrt{3}}{8}\lambda^2\, {\mathcal{O}}^3_2 \, , \nn\\
H_{4}\circ {\mathcal{O}}^3_2  &= -\frac{\sqrt{3}}{8}\lambda^2\, {\mathcal{O}}^3_1
- \frac{27}{8}\lambda^2\, {\mathcal{O}}^3_2 \, , \nn\\
H_{4}\circ {\mathcal{O}}_{4A}  &= (-\frac{25}{8}-\frac{181}{8\sqrt{41}})\lambda^2\, {\mathcal{O}}_{4A}
 - \frac{\sqrt{5}}{2\sqrt{41}}\lambda^2\,  {\mathcal{O}}_{4E} \, , \nn\\
H_{4}\circ {\mathcal{O}}_{4E}  &=  - \frac{\sqrt{5}}{2\sqrt{41}}\lambda^2\, {\mathcal{O}}_{4A}
+(-\frac{25}{8}+\frac{181}{8\sqrt{41}})\lambda^2\, {\mathcal{O}}_{4E} \, .
\end{align}
From this one deduces the following corrected two-loop eigenstates using standard non-degenerate
perturbation theory
\begin{align}
\hat{\mathcal{O}}^2_1 & = {\mathcal{O}}^2_1  + \lambda \, \frac{1}{2 \sqrt{5} }\,
{\mathcal{O}}^2_2  = \hat{\mathcal{O}}_{4F}\, , \\
\hat{\mathcal{O}}^2_2 & = {\mathcal{O}}^2_2  - \lambda \, \frac{1}{2 \sqrt{5} }\,
{\mathcal{O}}^2_1  = \hat{\mathcal{O}}_{4B}\, , \\
\hat{\mathcal{O}}^3_1 & = {\mathcal{O}}^3_1  + \lambda \, \frac{\sqrt{3}}{16}\,
{\mathcal{O}}^3_2  = \hat{\mathcal{O}}_{5J}\, , \\
\hat{\mathcal{O}}^3_2 & = {\mathcal{O}}^3_2  - \lambda \, \frac{\sqrt{3}}{16}\,
{\mathcal{O}}^3_1  = \hat{\mathcal{O}}_{5E}\, ,\\
\hat{\mathcal{O}}_{4A} & = {\mathcal{O}}_{4A}  - \lambda \,  \frac{\sqrt{5}}{41}\,
{\mathcal{O}}_{4E}  \, ,\\
\hat{\mathcal{O}}_{4E} & = {\mathcal{O}}_{4E}  + \lambda \, \frac{\sqrt{5}}{41}\,
{\mathcal{O}}_{4A} \, .
\end{align}
Note, however, that the self-mixing coefficients appearing in (\ref{hws}) receive
additional contributions from the interactions with fermions appearing in $H_{3}$.

\subsection{Self mixing contributions originating from $H_{3}$}

We now evaluate the additional contributions to the
self-mixing coefficients appearing in (\ref{hws}) originating
from the interactions with fermions.

As discussed in \cite{Georgiou:2011xj} the self mixing has two sources. One is the 2-loop purely scalar Hamiltonian
$H_4$ and the other is $H_{3}$. The total contributrion to the self-mixing is given by
\begin{eqnarray}\label{eigenstatefinal}
{\tilde{\cal O}}_{\alpha}={\cal O}_{\alpha}+ \gYM\,{\cal O}_{\alpha,\psi\psi}+
\gYM^{2}\,{\cal O}_{\alpha,DD}+ \gYM^{2\,}
 \sum_{\beta\neq \alpha}\frac{\langle{\cal O}_{\beta}|H_3|{\cal O}_{\alpha,\psi\psi}\rangle
+\langle{\cal O}_{\beta}|H_4|{\cal O}_{\alpha}\rangle}
{E_{2.\alpha}-E_{2,\beta}}\,{\cal O}_{\beta} \, .
\end{eqnarray}
The 2-loop energy of the state \eqref{eigenstatefinal} is
\begin{eqnarray}\label{eigenenergy}
E_{4,\alpha}=\langle{\cal O}_{\alpha}|H_4|{\cal O}_{\alpha}\rangle+\langle{\cal O}_{\alpha}|H_3|{\cal O}_{\alpha,\psi\psi}\rangle\qquad
\text{where}\quad E_{\alpha}=\lambda\, E_{2,\alpha}+ \lambda^{2}\, E_{4,\alpha}+\ldots\, .
\end{eqnarray}
Before resolving the mixing for the specific states we are interested in, let us
calculate via \eqref{eigenenergy} the 2-loop eigenvalue of the operator $O^2_1$.
The form of this  class of operators is given up to order $\gYM$ follows from
the first three lines of \eqn{hws}.
The corresponding spin chain state is obtained by performing the following
substitutions \cite{Georgiou:2004ty}
\begin{eqnarray}\label{substitutions}
\Big(\frac{8 \pi^2}{N}\Big)^{\frac{1}{2}}Z_{YM}\rightarrow Z_{sp}, \qquad
\Big(\frac{8 \pi^2}{N}\Big)^{\frac{1}{2}}\frac{1}{\sqrt{2}}\psi^A_{YM}\rightarrow \psi^A_{sp}
\end{eqnarray}
This correspondence is such that $\langle{\bar Z}_i|Z_i\rangle=1$ and similarly for the fermions.
To simplify notation we will drop the subscript 'sp'. Consequently,
the first three lines of eqn. \eqref{hws} can be written in the spin chain language as
the state
\begin{eqnarray}
\label{spinOJng}
|\hat{\mathcal{O}}^J_n\rangle=|\mathcal{O}^J_n\rangle+\gYM\,|\mathcal{O}^J_{n,\psi\psi}\rangle=\frac{1}{\sqrt{J+3}}\sum_{p=0}^J\cos{\frac{\pi n (2p+3)}{J+3}}
 |Z_i Z^p {\bar Z}_i Z^{J-p}\rangle \nonumber\\
 +\gYM\frac{\sqrt{N}}{4\pi}\frac{1}{\sqrt{J+3}}
\sin{\frac{\pi n}{J+3}}\sum_{p=0}^{J-1}\sin{\frac{\pi n(2p+4)}{J+3}}\,
2\,|\psi^{1 \alpha} Z^p \psi^2_\alpha Z^{J-p-1}\rangle
\nonumber\\
-\gYM\frac{\sqrt{N}}{4\pi}\frac{1}{\sqrt{J+3}}
\sin{\frac{\pi n}{J+3}}\sum_{p=0}^{J-1}\sin{\frac{\pi n(2p+4)}{J+3}}\,
2\,|\bar{\psi}_{3 \dot{\alpha}} Z^p \bar{\psi}^{\dot{\alpha}}_4 Z^{J-p-1}\rangle.
\end{eqnarray}
We should mention that the tree-level operator in the first line of \eqref{spinOJng}
is normalized in such a way that its two point function is 1.
Now we use the equation
\begin{eqnarray}\label{H3intermsH2}
\langle\mathcal{O}_{\beta,\psi\psi}|H_3|\mathcal{O}_{\alpha }\rangle=
E_{2,\alpha}\langle\mathcal{O}_{\beta, \psi\psi}|\mathcal{O}_{\alpha, \psi\psi}\rangle-
\langle\mathcal{O}_{\beta, \psi\psi}|H_2|\mathcal{O}_{\alpha, \psi\psi}\rangle \, ,
\end{eqnarray}
being a direct consequence of the full eigenvalue equation for
$|\hat{\mathcal O}_{\alpha}\rangle$
and the absence of
$\mathcal{O}(\gYM^{3})$ contributions to the scaling dimensions,
to express the matrix element of $H_3$ between the leading bosonic term
of the correlator and the subleading one involving fermions in terms
of matrix elements of the 1-loop Hamiltonian $H_2$.
In the sector we are interested in (e.g. scalar fermion pair of the type
$\psi_1 \psi_2$, ${\bar \psi}_3 {\bar \psi}_4$ and the complex field $Z$)
this can be written as
\begin{eqnarray}\label{H2full}
H_2=\frac{\lambda}{2}\sum_{i=1}^L (1-\Pi_{i,i+1}),
\end{eqnarray}
where $\Pi_{i,i+1}$ is the graded permutation operator.

Specialized to the case where $J=2,\, n=1$ we get
\begin{eqnarray}\label{inner}
\langle\mathcal{O}^2_{1,\psi\psi}|\mathcal{O}^2_{1,\psi\psi}\rangle=
4\times 2 \Big( \gYM\, \frac{\sqrt{N}}{2\pi}\frac{1}{\sqrt{5}}
\sin{\frac{\pi }{5}}\sin{\frac{4 \pi }{5}}\Big)^2
\end{eqnarray}
Here the factor of 4 appearing in \eqref{H2O21} is due to the inner
product of $|\psi^{1 \alpha} \psi^2_\alpha Z\rangle-|\psi^{2 \alpha} \psi^1_\alpha Z\rangle$,
while the factor of 2 is due to the fact that there is a similar term with the
fermions being ${\bar \psi}_3$ and ${\bar \psi}_4$. Finally the $\sin{\frac{4 \pi }{5}}$
is due to the phase factor inside the sum of the second line of \eqref{spinOJng}.
Similarly, the matrix element of $H_2$  is given by
\begin{eqnarray}\label{H2O21}
\langle\mathcal{O}^2_{1, \psi\psi}|H_2|\mathcal{O}^2_{1, \psi\psi}\rangle=
4\times 2 \Big( g\frac{\sqrt{N}}{2\pi}\frac{1}{\sqrt{5}}
\sin{\frac{\pi }{5}}\sin{\frac{4 \pi }{5}}\Big)^2 3\lambda
\end{eqnarray}
Putting everything together we obtain for the two-loop energy of $\hat{O}^2_1$
the value
\begin{align}\label{energyO21}
E_4&=\langle\mathcal{O}^2_{1}|H_4|\mathcal{O}^2_{1}\rangle+
E_{2}\langle\mathcal{O}^2_{1, \psi\psi}|\mathcal{O}^2_{1, \psi\psi}\rangle
=
\frac{\lambda^2}{8}(-17+5 \sqrt{5}).
\end{align}
This value is in complete agreement with the two-loop anomalous dimension of a level 4 descendant of the $\mathcal{O}^2_{1}$ primary state
which belongs to an $SU(2)$ sub-sector.
The precise form of this descendant operator is
\begin{eqnarray}\label{desc.}
\mathcal{O}_{desc.}=\frac{1}{J+1}\sum_{p=0}^J \cos{\frac{\pi n(2 p+1)}{J+1}} \tr{Z_1 Z^p Z_1 Z^{J-p}},\,\,\,J=4,\, n=1,
\end{eqnarray}
and its two-loop anomalous dimension was found to be \cite{Beisert:2004ry}
\begin{eqnarray}
E_4=16 \lambda^2 \sin^4{\frac{\pi n}{J+1}}(-\frac{1}{4}-\frac{\cos^2{\frac{\pi n}{J+1}}}{J+1}),
\end{eqnarray}
which for $J=4$ and $n=1$ gives precisely \eqref{energyO21}.

We now proceed to the resolution of the mixing for the operators $\mathcal{O}^2_{n},\, n=1,2$.
The subleading terms of order $\gYM$ are given by
\begin{eqnarray}\label{}
|\mathcal{O}^2_{n,\psi\psi}\rangle=\sqrt{\lambda}\frac{1}{\sqrt{5}}\sin{\frac{\pi n}{5}}\sin{\frac{4\pi n}{5}}
(|\psi^{1 \alpha} \psi^2_\alpha Z\rangle-|\psi^{2 \alpha} \psi^1_\alpha Z\rangle-
|{\bar \psi}_{3 \dot \alpha} {\bar \psi}_4^{\dot \alpha} Z\rangle+
|{\bar \psi}_{4 \dot \alpha} {\bar \psi}_3^{\dot \alpha} Z\rangle).
\end{eqnarray}
This gives
\begin{align}\label{inner12}
\langle\mathcal{O}^2_{2, \psi\psi}|\mathcal{O}^2_{1,\psi\psi}\rangle &=
 -\frac{8}{5}\lambda
\sin^2{\frac{\pi }{5}}\sin^2{\frac{2 \pi }{5}},
\\
\label{HERE}
\langle\mathcal{O}^2_{2, \psi\psi}|H_2|\mathcal{O}^2_{1, \psi\psi}\rangle &=
-\frac{8}{5}\lambda
\sin^2{\frac{\pi }{5}}\sin^2{\frac{2 \pi }{5}} 3\lambda
\end{align}
From  \eqref{inner12} and \eqref{HERE} one obtains
\begin{align}\label{H3 12}
\langle\mathcal{O}^2_{2, \psi\psi}|H_3|\mathcal{O}^2_{1}\rangle=
-\frac{8}{5}\lambda^2
\sin^2{\frac{\pi }{5}}\sin^2{\frac{2 \pi }{5}}(4\sin^2{\frac{\pi }{5}}-  3) \\
\label{H3 21}
\langle\mathcal{O}^2_{1, \psi\psi}|H_3|\mathcal{O}^2_{2}\rangle=
-\frac{8}{5}\lambda^2
\sin^2{\frac{\pi }{5}}\sin^2{\frac{2 \pi }{5}}(4\sin^2{\frac{2 \pi }{5}}-  3).
\end{align}
Finally, then the self-mixing contributions from the $H_{3}$ sector are given by
\begin{align}
\hat{\mathcal{O}}^2_1 & = {\mathcal{O}}^2_1  +
\frac{\langle\mathcal{O}^2_{2}|H_3|\mathcal{O}^2_{1, \psi\psi}\rangle}{4\lambda(\sin^2{\frac{\pi }{5}}-\sin^2{\frac{2 \pi }{5}})} \,
{\mathcal{O}}^2_2={\mathcal{O}}^2_1  +\lambda\frac{5-\sqrt{5}}{20}{\mathcal{O}}^2_2
= \hat{\mathcal{O}}_{4F}\, , \\
\hat{\mathcal{O}}^2_2 & = \mathcal{O}^2_{2}+\frac{\langle\mathcal{O}^2_{1}|H_3|\mathcal{O}^2_{2, \psi\psi}\rangle}{4\lambda(\sin^2{\frac{2 \pi }{5}}-\sin^2{\frac{ \pi }{5}})} \,
{\mathcal{O}}^2_1={\mathcal{O}}^2_2  +\lambda\frac{5+\sqrt{5}}{20}{\mathcal{O}}^2_1 = \hat{\mathcal{O}}_{4B}\, ,
\end{align}
In a similar way, one can resolve the self-mixing for the operators $\mathcal{O}^3_{n},\, n=1,2$.
\begin{align}
\hat{\mathcal{O}}^3_1 & = {\mathcal{O}}^3_1  +\lambda\frac{\sqrt{3}}{8}{\mathcal{O}}^3_2
= \hat{\mathcal{O}}_{5J}\, , \\
\hat{\mathcal{O}}^3_2 & = {\mathcal{O}}^3_2  +\lambda\frac{\sqrt{3}}{8}{\mathcal{O}}^3_1 = \hat{\mathcal{O}}_{5E}\, ,
\end{align}

Lastly, we focus on the length 4 primary operators ${\mathcal{O}}_{4A},\, {\mathcal{O}}_{4E}$.
In order to resolve the self-mixing here the most general form for the 1-loop Hamiltonian
$H_2$ acting on one scalar and one fermion is needed. It is given by
\begin{eqnarray}\label{H2scalarfermion}
H_2|{\bar \psi}_A \Phi^{BC}\rangle=\frac{\lambda}{2}(|{\bar \psi}_A \Phi^{BC}\rangle-|\Phi^{BC} {\bar \psi}_A \rangle)
+\frac{\lambda}{4}\big( \delta^B_A |{\bar \psi}_K \Phi^{KC}\rangle+ \delta^B_A |\Phi^{KC}{\bar \psi}_K \rangle+
\delta^C_A |{\bar \psi}_K \Phi^{BK}\rangle\nn \\
+ \delta^C_A |\Phi^{BK}{\bar \psi}_K \rangle\big)+c \delta^B_A \sigma^{\mu}_{\alpha \dot \beta}\, |D_{\mu}\psi^C\rangle +
c'\delta^C_A \sigma^{\mu}_{\alpha \dot \beta}\, |D_{\mu}\psi^B\rangle \, .
\end{eqnarray}
The  two last terms with the covariant derivatives acting on the fermion in the fundamental will not be needed
in what follows. The action of $H_2$ on states where the scalar and the fermion are swaped or on states like
$| \psi^A \Phi_{BC}\rangle$ should be obvious from \eqref{H2scalarfermion}.
Let us write the two operators as
\begin{eqnarray}\label{4A4E}
| \tilde{\mathcal{O}}_{4A}\rangle & =\frac{4}{\sqrt{738-102\sqrt{41}}}\big( |\Phi_{AB}\Phi^{AB} \Phi_{CD}\Phi^{CD} \rangle
+ \frac{5-\sqrt{41}}{4}|\Phi_{AB} \Phi_{CD}\Phi^{AB}\Phi^{CD} \rangle \big) +|\mathcal{O}_{4A, \psi\psi}\rangle \nn \\
| \tilde{\mathcal{O}}_{4E}\rangle & =\frac{4}{\sqrt{738+102\sqrt{41}}}\big( |\Phi_{AB}\Phi^{AB} \Phi_{CD}\Phi^{CD} \rangle
+ \frac{5+\sqrt{41}}{4}|\Phi_{AB} \Phi_{CD}\Phi^{AB}\Phi^{CD} \rangle \big) +|\mathcal{O}_{4E, \psi\psi}\rangle \, ,
\end{eqnarray}
where
\begin{eqnarray}\label{4A4Epsi}
|\mathcal{O}_{4A, \psi\psi}\rangle=-\gYM \, c_{A}\,2\,(|\Phi_{AB}\psi^{A\alpha}\psi^B_{\alpha}\rangle-
|\Phi^{AB}{\bar \psi}_{A \dot\alpha}{\bar \psi}_B^{\dot\alpha}\rangle)\frac{4}{\sqrt{738-102\sqrt{41}}}\nn \\
|\mathcal{O}_{4E, \psi\psi}\rangle=-\gYM  \,c_{E}\,2\, (|\Phi_{AB}\psi^{A\alpha}\psi^B_{\alpha}\rangle-
|\Phi^{AB}{\bar \psi}_{A \dot\alpha}{\bar \psi}_B^{\dot\alpha}\rangle)\frac{4}{\sqrt{738+102\sqrt{41}}}
\end{eqnarray}
The constants $c_A$ and $c_E$ are given by
\be
c_{A/E}= -\frac{\sqrt{N}}{8\pi}(\frac{1}{2}-\frac{5\mp\sqrt{41}}{4})\, .
\ee
Furthermore, the normalizations
appearing in \eqref{4A4E} are such that the the leading scalar terms are normalized to 1.
As above one can evaluate the following quantities
\begin{eqnarray}\label{inner4A4E}
\langle\mathcal{O}_{4E, \psi\psi}|\mathcal{O}_{4A, \psi\psi}\rangle=48 \gYM^2
c_A c_E \frac{2}{3 \sqrt{205}}
\end{eqnarray}
and
\begin{eqnarray}\label{H24A4E}
\langle\mathcal{O}_{4E, \psi\psi}|H_2|\mathcal{O}_{4A, \psi\psi}\rangle=48 g^2 c_A c_E \frac{2}{3 \sqrt{205}}(3 \lambda),
\end{eqnarray}
using \eqref{H2scalarfermion}.
From this we deduce the matrix elements
\begin{align}\label{H24A4E}
\langle\mathcal{O}_{4E, \psi\psi}|H_3|\mathcal{O}_{4A}\rangle&=48 \gYM^2 c_A c_E \frac{2}{3 \sqrt{205}}(\frac{\lambda}{4}(13+\sqrt{41})-3 \lambda)\, ,\\
\label{H24E4A}
\langle\mathcal{O}_{4A, \psi\psi}|H_3|\mathcal{O}_{4E}\rangle&=48 \gYM^2 c_A c_E \frac{2}{3 \sqrt{205}} (\frac{\lambda}{4}(13-\sqrt{41})-3 \lambda)\, .
\end{align}
Using these we may now also write down the self-mixing contributions originating from $H_{3}$
\begin{align}
\hat{\mathcal{O}}_{4A} & = {\mathcal{O}}_{4A}  + \,  48 g^2 c_A c_E\, \frac{2}{3 \sqrt{205}}
\frac{1-\sqrt{41}}{2\sqrt{41}}{\mathcal{O}}_{4E}={\mathcal{O}}_{4A}+\lambda\frac{2(-1+\sqrt{41})}{41\sqrt{5}} \,
{\mathcal{O}}_{4E}  \, ,\\
\hat{\mathcal{O}}_{4E} & = {\mathcal{O}}_{4E}  + \,48 g^2 c_A c_E \, \frac{2}{3 \sqrt{205}}
 \frac{-1-\sqrt{41}}{2\sqrt{41}}{\mathcal{O}}_{4A}={\mathcal{O}}_{4E}  +\lambda\frac{2(1+\sqrt{41})}{41\sqrt{5}}\,
{\mathcal{O}}_{4A} \, ,
\end{align}

Finally, we may now state the complete self-mixing contributions to the operators that we are considering
by combining the results of this and the previous subsection. We find
\begin{align}
\label{first}
\hat{\mathcal{O}}^2_1 & = {\mathcal{O}}^2_1  + \lambda \, \frac{1}{10}\frac{5+\sqrt{5}}{2}\,
{\mathcal{O}}^2_2 ={\mathcal{O}}^2_1  +  \frac{1}{10}\gamma_{22}\,{\mathcal{O}}^2_2 = \hat{\mathcal{O}}_{4F}\, , \\
\hat{\mathcal{O}}^2_2 & = {\mathcal{O}}^2_2  + \lambda \, \frac{1}{10}\frac{5-\sqrt{5}}{2}\,
{\mathcal{O}}^2_1 ={\mathcal{O}}^2_2  +  \frac{1}{10}\gamma_{21}\,{\mathcal{O}}^2_1= \hat{\mathcal{O}}_{4B}\, , \\
\hat{\mathcal{O}}^3_1 & = {\mathcal{O}}^3_1  + \lambda \, \frac{3\sqrt{3}}{16}\,
{\mathcal{O}}^3_2  = {\mathcal{O}}^3_1  +   \frac{\sqrt{3}}{16}\,\gamma_{32}
{\mathcal{O}}^3_2 =\hat{\mathcal{O}}_{5J}\, , \\
\hat{\mathcal{O}}^3_2 & = {\mathcal{O}}^3_2  + \lambda \, \frac{\sqrt{3}}{16}\,
{\mathcal{O}}^3_1  = {\mathcal{O}}^3_2  +   \frac{\sqrt{3}}{16}\,\gamma_{31}
{\mathcal{O}}^3_1= \hat{\mathcal{O}}_{5E}\, ,\\
\hat{\mathcal{O}}_{4A} & = {\mathcal{O}}_{4A}  + \lambda \,  \frac{-7+2\sqrt{41}}{41\sqrt{5}}\,
{\mathcal{O}}_{4E}  \, ,\\
\hat{\mathcal{O}}_{4E} & = {\mathcal{O}}_{4E}  + \lambda \, \frac{7+2\sqrt{41}}{41\sqrt{5}}\,
{\mathcal{O}}_{4A} \, ,
\end{align}
where $\gamma_{Jn}=4 \lambda \sin^2{\frac{\pi n}{J+3}}$ is the 1-loop anomalous dimension of the operator ${\mathcal{O}}^J_n$.

\subsection{Fermionic and derivative mixing terms for the $\mathcal{O}_{4A}$ and
$\mathcal{O}_{4E}$  operators}
What remains to be found are the bi-fermionic and bi-derivative mixing contributions
for the scalar $\mathcal{O}_{4A}$ and
$\mathcal{O}_{4E}$ operators.
Let us rewrite these operators
in complex notation\footnote{Our conventions are stated in
appendix \ref{App:A}.}:
\begin{equation}
  \label{eq:o4ae}
\mathcal{O}_{4A/E}^{(0)} =
\tr{\Phi_{AB}\Phi^{AB}\Phi_{CD}\Phi^{CD}} +
\alpha_{A/E}
\tr{\Phi_{AB}\Phi_{CD}\Phi^{AB}\Phi^{CD}}
\end{equation}
where $\alpha_{A/E} =\frac{5\mp\sqrt{41}}{4}$.

At orders $\gYM$ and $\gYM^2$ we expect to find subleading mixing terms,
singlets under the $SU(4)$ R-symmetry group and with naive scaling
dimension four, containing respectively two fermions and two
derivative impurities. The natural candidates are then:
\begin{equation}
  \label{eq:phipsipsi}
  c_1 \,\gYM\,
  \tr{\Phi_{AB} \psi^A_\alpha \psi^B_\beta}
  \epsilon^{\alpha \beta} +
  c_2 \,\gYM \,
  \tr{\Phi^{AB} \bar{\psi}_A^{\dot{\alpha}} \bar{\psi}_B^{\dot{\beta}}}
  \epsilon_{\dot{\alpha} \dot{\beta}}
\end{equation}
and
\begin{equation}
  \label{eq:DD}
 d \,\gYM^2\,
  D_\mu \Phi_{AB} D^\mu\Phi^{AB}.
\end{equation}

We can compute the mixing coefficients by requiring that this operator - being a
highest
weight state - is annihilated by the action of the superconformal
charges $\bar{S}^{\dot{\alpha}}_A$ and $S_\alpha^A$ for any
$A=1,\ldots,4$ at any order in perturbation theory. As it is composed by
scalar fields only, the leading term is trivially annihilated by the
superconformal charges at order $\gYM^0$. This is no longer true at
higher orders as the superconformal charges receive quantum corrections,
thus their action on the leading term should be compensated by the
action of $S$ and $\bar{S}$ at lower orders on \eqref{eq:phipsipsi}
and \eqref{eq:DD}.

Let us first resolve the mixing at order $\gYM$, demanding that the
action $S$ and $\bar{S}$ at this order on \eqref{eq:o4ae} is cancelled
by the tree-level action of the same supercharges on the two terms in
\eqref{eq:phipsipsi}. These are determined by the
contraction of the relevant term of the supercurrents in eqs \eqref{supc}.

Since these terms have opposite sign, we can write $c_1 = -c_2 = c$.

Let us focus on the variation generated by the charge $\bar{S}^{\dot{\alpha}}_{A=1}$ at order $g_{YM}$, and let us first compute the coefficient $c$ by
specializing the action of $\bar{S}^{\dot{\alpha}}_1$
on $\cO^{(0)}_{4A/E}$ from \eqref{eq:o4ae}. From equation \eqref{eq:spp}, we get
\begin{equation}
  \label{eq:sphiphi}
  \bar{S}^{\dot{\alpha}}_1\Phi_{AB}\Phi_{CD} =
  \ii\frac{\gYM N}{32\pi^2}
  \left(
    \epsilon_{1AB[C}\bar{\psi}^{\dot{\alpha}}_{D]} -
    \epsilon_{1CD[A}\bar{\psi}^{\dot{\alpha}}_{B]}
  \right),
 \end{equation}
 where $\epsilon_{1AB[C}\bar{\psi}_{D]} =
\frac{1}{2}(\epsilon_{1ABC}\bar{\psi}_D - \epsilon_{1ABD}\bar{\psi}_C )$.

We can act with $\bar{S}^{\dot{\alpha}}_1$ on
$\tr{\Phi_{AB}\Phi_{A'B'}\Phi_{CD}\Phi_{C'D'}}$. Then, we must
contract the result with $\epsilon^{ABA'B'}\epsilon^{CDC'D'}$ and
$\epsilon^{ABCD}\epsilon^{A'B'C'D'}$ to obtain the action of $\bar{S}$
at order-$\gYM$ on the first and second terms of \eqref{eq:o4ae}
respectively.
We obtain
\begin{align}
  \bar{S}^{\dot{\alpha}}_1\tr{\Phi_{AB}\Phi_{A'B'}\Phi_{CD}\Phi_{C'D'}}
\epsilon^{ABA'B'}\epsilon^{CDC'D'} & =
-\ii\frac{\gYM N}{2\pi^2}
\tr{\Phi^{AB} [\bar{\psi}_B^{\dot{\alpha}},\Phi_{1A} ]}
\\
  \bar{S}^{\dot{\alpha}}_1\tr{\Phi_{AB}\Phi_{A'B'}\Phi_{CD}\Phi_{C'D'}}
\epsilon^{ABCD}\epsilon^{A'B'C'D'} & =
\ii\frac{\gYM N}{\pi^2}
\tr{\Phi^{AB} [\bar{\psi}_B^{\dot{\alpha}},\Phi_{1A} ]}.
\end{align}
Thus
\begin{equation}
  \label{eq:lt}
\bar{S}^{\dot{\alpha}}_1 \mathcal{O}_{4A/E}^{(0)} =
\ii\frac{\gYM N}{4 \pi^2}
\left(
-\frac{1}{2} + \alpha_{A/E}
\right)
\tr{\Phi^{AB} [\bar{\psi}_B^{\dot{\alpha}},\Phi_{1A} ]}.
\end{equation}
Now let us consider the tree-level action of $\bar{S}_1$ on the
subleading term in \eqref{eq:phipsipsi}. Since (see eq. \eqref{eq:sf})
\begin{equation}
  \label{eq:spsi}
  \bar{S}^{\dot{\alpha}}_1 \bar{\psi}_{B\,\dot{\beta}}
  = 4\sqrt{2} \, \ii\,  \Phi_{1B}\delta^{\dot{\alpha}}_{\dot{\beta}},
\end{equation}
we have
\begin{equation}
  \label{eq:slt}
  \bar{S}^{\dot{\alpha}}_1
  \tr{\Phi^{AB} \bar{\psi}_A^{\dot{\gamma}} \bar{\psi}_B^{\dot{\beta}}}
  \epsilon_{\dot{\gamma} \dot{\beta}}=
  4 \sqrt{2}\, \ii \,
  \tr{\Phi^{AB}  [\bar{\psi}_B^{\dot{\alpha}}, \Phi_{1A}]}.
\end{equation}
Recalling that $c_2 = - c_1 = - c$ one has
\begin{equation}
  \label{eq:c}
  c =  -\frac{N}{16\sqrt{2}\pi^2}
  \left(
    \frac{1}{2} - \alpha_{A/E}
  \right).
\end{equation}

We now proceed to calculate the mixing of the operators 4A/4E with operators involving derivative terms.
There is a single candidate consistent with the dimension of 4A/4E which is four and with the fact that
the 4A/4E operators are singlets under both the $SU(4)$ and the Lorentz group.
We denote this mixing term by $d \tr{D_{\mu}\Phi_{AB}D^{\mu}\Phi^{AB}}$.
An easy way to determine $d$ is by demanding orthogonality of the 4A/4E operators with the Konishi up to order $g^2$.
It is enough to consider just part of the Konishi operator, namely $\tr{Z \bar{Z}}$. All other terms
can be manipulated in a similar way and give the same result for $d$.

Firstly, we write the relevant Yukawa term as $-4 \sqrt{2} g \,\tr{\Phi_{AB}\psi^{\alpha B} \psi_{\alpha}^A}$.
Then we focus on the 2-point function
\begin{equation}
\langle \tr{{\bar\psi}_{\dot \alpha 1} {\bar \psi}^{\dot \alpha}_ 2 Z}(x) \, \,\,  \tr{Z {\bar Z} }(0) \rangle
\end{equation}
Inserting the Yukawa and making the contractions
 we get
\begin{eqnarray}\label{2-pointfermions}
-2 \sqrt{2}\, g\, i\, N^3 \,\frac{1}{2^4} \,\Delta(x)\,\int d^4z\, \Delta(z)\,(i \,\sigma_{\alpha \dot\alpha}^{\mu} \partial_{\mu}^x\Delta(x-z))\,
(-i \,{\bar \sigma}^{\nu \dot\alpha \alpha} \partial_{\nu}^z\Delta(x-z))=\nonumber\\
2 \sqrt{2}\, g  N^3 \,\frac{1}{2^4}\, \Delta^3(x),
\end{eqnarray}
where to pass from the first to the second line of \eqref{2-pointfermions} we have used the identities
\begin{eqnarray}\label{identities}
\sigma_{\alpha \dot\alpha}^{\mu}{\bar \sigma}^{\nu \dot\alpha \alpha}=2 \eta_{\mu \nu}\\
\int d^4z \Delta(z) \partial_{\mu}^x\Delta(x-z) \partial^{\mu}_x\Delta(x-z)=-\frac{i}{2}\Delta^2(x).
\end{eqnarray}
Furthermore, the factor $N^3$ comes from the fact that the corresponding diagram has 3 loops while the factor of
$1/2^4$ is related to the fact that there are 4 propagators each of which brings a factor of $1/2$.
To get the full contribution of the bi-fermion term for the operators 4A/4E we should multiply
the result of \eqref{2-pointfermions} by 4. The first 2 is to take into account the contribution
when the fermions ${\bar \psi}_{1} {\bar \psi}_ 2$ in the 4A/4E state are substituted by $\psi^3 \psi^4$ while the second
one to take into account the contributions when the Yukawa vertex is contracted with the $\bar Z$ field of the Konishi.
Putting everything together and multiplying by the coefficient of the bi-fermion term which is $g\frac{N}{16\sqrt{2}\pi^2}\frac{3\mp \sqrt{41}}{4}$
for 4A and 4E respectively we get the final contribution from the bi-fermion term to be
\begin{eqnarray}\label{2-pointfermionsfinal}
\frac{g^2 N^4}{16\, 2 \pi^2}\frac{3\mp \sqrt{41}}{4}
\end{eqnarray}
We now turn to the contribution of the bi-derivative insertions
\begin{equation}
d\, \langle \tr{D^{\mu}{\bar Z}D_{\mu}Z}(x) \, \,\,  \tr{Z {\bar Z} }(0) \rangle,
\end{equation}
where $d$ is the coefficient which we need to determine.
These involve only free contractions and read
\begin{eqnarray}\label{2-pointderiv}
d\, N^2\,\frac{1}{2^2}\partial_{\mu}^x\Delta(x) \partial^{\mu}_x\Delta(x)=d\, N^2\,\frac{1}{2^2}(-16 \pi^2)\, \Delta^3(x)
\end{eqnarray}
Demanding that the sum of \eqref{2-pointfermionsfinal} and \eqref{2-pointderiv} is zero we deduce
\begin{equation}
d=\frac{g^2 N^2}{2^7 \pi^4}\frac{3\mp \sqrt{41}}{4}.
\end{equation}
Finally, we should mention that the leading term of the 4A/4E operators and the Konishi have no
overlap at order $g^2$.

\section{Operators up to length $L=5$}
\label{sect:mixresults}

In this short section we now collect the results of the previous chapter and
write down the explicit form of the non-BPS operators we
are going to use in the computation of the three point functions
\begin{itemize}
\item $L=2$
\begin{equation}
  \label{eq:konishi}
  \mathcal{K} = \frac{8 \pi^2}{\sqrt{3} N} \tr{\Phi_{AB}\Phi^{AB}}
\end{equation}
\item $L=3$
\begin{equation}
  \label{eq:O11}
  \hat{\mathcal{O}}^1_1 =
\frac{8\pi^3}{N^{\frac{3}{2}}}\big[
\tr{\Phi^{AB} \Phi_{AB} Z} + \tr{\Phi^{AB} Z \Phi_{AB}}
\big]
\end{equation}
\item  $L=4$
  \begin{align}
    \hat{\mathcal{O}}^2_1 = &
    \frac{(8 \pi^2)^2}{N^2\sqrt{5}}
\Big(
1+\gYM^2\frac{N}{32\pi^2}(3-\sqrt{5})
\Big)
    \sum_{p=0}^2\cos{\frac{\pi(2p+3)}{5}}
    \tr{\Phi^{AB} Z^p \Phi_{AB} Z^{2-p}}  \\\notag
    &+ \gYM \frac{\pi^2}{N}\sqrt{2}(\sqrt{5}-1)
    \Big[
    \tr{\psi^{[1\,\alpha} \psi^{2]}_\alpha Z}
    -\tr{\bar{\psi}_{[3\,\dot{\alpha}} \bar{\psi}_{4]}^{\dot{\alpha}} Z}
    \Big] \\\notag
    &+ \gYM^2 \frac{\sqrt{5}-1}{16} \tr{D_{\mu}Z D^{\mu}Z} \\\notag
    &+ \gYM^2 \frac{4 \pi^2}{5 N}(\sqrt{5}+1)
    \sum_{p=0}^2\cos{\frac{2\pi(2p+3)}{5}}
    \tr{\Phi^{AB} Z^p \Phi_{AB} Z^{2-p}} \\
     \hat{\mathcal{O}}^2_2 = &
     \frac{(8 \pi^2)^2}{N^2\sqrt{5}}
 \Big(
 1+\gYM^2\frac{N}{32\pi^2}(3+\sqrt{5})
 \Big)
     \sum_{p=0}^2\cos{\frac{2\pi(2p+3)}{5}}
     \tr{\Phi^{AB} Z^p \Phi_{AB} Z^{2-p}}  \\\notag
     &- \gYM \frac{\pi^2}{N}\sqrt{2}(1+\sqrt{5})
     \Big[
     \tr{\psi^{[1\,\alpha} \psi^{2]}_\alpha Z} -
     \tr{\bar{\psi}_{[3\,\dot{\alpha}} \bar{\psi}_{4]}^{\dot{\alpha}} Z}
     \Big] \\\notag
     &+\gYM^2 \frac{1+\sqrt{5}}{16} \tr{D_{\mu}Z D^{\mu}Z} + \\\notag
     &+\gYM^2 \frac{4 \pi^2}{5 N}(\sqrt{5}-1)
     \sum_{p=0}^2\cos{\frac{\pi(2p+3)}{5}}
     \tr{\Phi^{AB} Z^p \Phi_{AB} Z^{2-p}} \\
     \hat{\mathcal{O}}_{4A} = &
     \Big(
     \frac{8\pi^2}{N}
     \Big)^2
     \mathcal{N}_A
     \left\{
       \tr{\Phi_{AB}\Phi^{AB}\Phi_{CD}\Phi^{CD}} +
       \frac{5 - \sqrt{41}}{4}
       \tr{\Phi_{AB}\Phi_{CD}\Phi^{AB}\Phi^{CD}}
     \right. \\\notag
     &- \gYM \frac{N}{16 \sqrt{2} \pi^2} \frac{3-\sqrt{41}}{4}
     \Big[
     \tr{\Phi_{AB}\psi^{A\alpha}\psi^B_\alpha} -
     \tr{\Phi^{AB}\bar{\psi}_{A \dot{\alpha}}\psi_B^{\dot{\alpha}}}
     \Big]
     \notag\\ &\left.
       + \gYM^2 \frac{N^2}{2^{7}\pi^4}\frac{3-\sqrt{41}}{4}
       \tr{D_\mu\Phi_{AB} D^\mu\Phi^{AB}}
     \right\} \notag\\\notag &
     - \gYM^2\frac{16 \pi^2}{N}
     \frac{(7-2\sqrt{41})}{41 \sqrt{5}} \mathcal{N}_E
     \Big[
       \tr{\Phi_{AB}\Phi^{AB}\Phi_{CD}\Phi^{CD}} +
       \frac{5+\sqrt{41}}{4}
       \tr{\Phi_{AB}\Phi_{CD}\Phi^{AB}\Phi^{CD}}
     \Big]\\
     \hat{\mathcal{O}}_{4E} = &
     \Big(
     \frac{8\pi^2}{N}
     \Big)^2
     \mathcal{N}_E
     \left\{ \tr{\Phi_{AB}\Phi^{AB}\Phi_{CD}\Phi^{CD}} +
       \frac{5+\sqrt{41}}{4}
       \tr{\Phi_{AB}\Phi_{CD}\Phi^{AB}\Phi^{CD}} \right.\\\notag &
        - \gYM \frac{N}{16 \sqrt{2} \pi^2} \frac{3+\sqrt{41}}{4}
       \Big[
       \tr{\Phi_{AB}\psi^{A\alpha}\psi^B_\alpha} -
       \tr{\Phi^{AB}\bar{\psi}_{A \dot{\alpha}}\psi_B^{\dot{\alpha}}}
       \Big]
     \\\notag &\left.
       + \gYM^2 \frac{N^2}{2^{7}\,\pi^4}\frac{3+\sqrt{41}}{4}
       \tr{D_\mu\Phi_{AB} D^\mu\Phi^{AB}}
     \right\}  \\\notag &
     +\gYM^2\frac{16 \pi^2}{N}
     \frac{(7+2\sqrt{41})}{41 \sqrt{5}} \mathcal{N}_A
\Big[\tr{\Phi_{AB}\Phi^{AB}\Phi_{CD}\Phi^{CD}} +
\frac{5-\sqrt{41}}{4}
\tr{\Phi_{AB}\Phi_{CD}\Phi^{AB}\Phi^{CD}}
\Big]
   \end{align}
   where the antisymmetrization of the indices is defined as
   $\psi^{[1\,\alpha} \psi^{2]}_\alpha = \frac{1}{2}(\psi^{1\,\alpha}
   \psi^2_\alpha - \psi^{2\,\alpha} \psi^1_\alpha)$ and the normalization reads
   $\mathcal{N}_{4A/4E} = \frac{4}{\sqrt{738 \mp 102\sqrt{41}}}
   \Big(
   1 + \frac{\gYM^2 N}{4 \pi^2}\frac{25 \mp 3\sqrt{41}}{246 \mp 34 \sqrt{41}}
+ \mathcal{O}(\gYM^4)
   \Big)$.\\
\item  $L=5$
  \begin{align}
    \hat{\mathcal{O}}^3_1 = &
    \Big(\frac{8 \pi^2}{N}\Big)^{\frac{5}{2}}
\frac{1}{\sqrt{6}}
\Big(
1+\gYM^2\frac{N}{32\pi^2}
\Big)
    \sum_{p=0}^3\cos{\frac{\pi(2p+3)}{6}}
    \tr{\Phi^{AB} Z^p \Phi_{AB} Z^{3-p}}  \\\notag
    &+ \gYM \frac{4\sqrt{2}\pi^3}{\sqrt{3} N^{\frac{3}{2}}}
\sum_{p=0}^2\sin{\frac{\pi(2p+4)}{6}}
 \Big[
    \tr{\psi^{1\,\alpha} Z^p \psi^{2}_\alpha Z^{2-p}} -
    \tr{\bar{\psi}_{3\,\dot{\alpha}} Z^p \bar{\psi}_{4}^{\dot{\alpha}} Z^{2-p}}
    \Big] \notag\\
&- \gYM^2 \frac{\pi}{8 \sqrt{N}}
\big[
\tr{D_{\mu}Z D^{\mu}Z Z} + \tr{D_{\mu}Z Z D^{\mu}Z}
\big]
 \\\notag
    &+\gYM^2
 \frac{6 \pi^3}{N^{\frac{3}{2}}}
    \sum_{p=0}^3\cos{\frac{\pi(2p+3)}{3}}
    \tr{\Phi^{AB} Z^p \Phi_{AB} Z^{3-p}} \\
    \hat{\mathcal{O}}^3_2 = &
    \Big(\frac{8 \pi^2}{N}\Big)^{\frac{5}{2}}
\frac{1}{\sqrt{6}}
\Big(
1+\gYM^2\frac{3 N}{32\pi^2}
\Big)
    \sum_{p=0}^3\cos{\frac{\pi(2p+3)}{3}}
    \tr{\Phi^{AB} Z^p \Phi_{AB} Z^{3-p}}  \\\notag
    &+ \gYM \frac{4\sqrt{2}\pi^3}{N^{\frac{3}{2}}}
\sum_{p=0}^2\sin{\frac{\pi(2p+4)}{3}}
 \Big[
    \tr{\psi^{1\,\alpha} Z^p \psi^{2}_\alpha Z^{2-p}} -
    \tr{\bar{\psi}_{3\,\dot{\alpha}} Z^p \bar{\psi}_{4}^{\dot{\alpha}} Z^{2-p}}
    \Big] \notag\\
&+ \gYM^2 \frac{\sqrt{3}\pi}{8 \sqrt{N}}
\big[
\tr{D_{\mu}Z D^{\mu}Z Z} + \tr{D_{\mu}Z Z D^{\mu}Z}
\big]
 \\\notag
    &+\gYM^2
 \frac{2 \pi^3}{N^{\frac{3}{2}}}
    \sum_{p=0}^3\cos{\frac{\pi(2p+3)}{6}}
    \tr{\Phi^{AB} Z^p \Phi_{AB} Z^{3-p}}
\end{align}
\item {\bf BPS Operators}\\[0.3cm]
We will also need explicit forms of the protected $1/2$ BPS operators beyond the
 lengths three and four. The maximally charged operators are of course simply $\tr Z^{J}$ carrying
$U(1)$ charge $J$. At length three we note the BPS operator
\begin{eqnarray}
   \label{3Cfinal}
\mathcal{O}_{3C,ijk}=\tr{Z_i \bar{Z_j}Z_k+\bar{Z_j}Z_i Z_k}-\frac{1}{4}\,\delta_{ij}\tr{Z_p\bar{Z_p}Z_k+\bar{Z_p} Z_p Z_k}\nonumber \\
-\frac{1}{4}\,\delta_{jk}\tr{Z_p\bar{Z_p}Z_i+\bar{Z_p}Z_p Z_i}, \,\,i,j,k=1,2,3
\end{eqnarray}
At length four we have the neutral BPS operators:\\
\begin{eqnarray}
   \label{neutral}
\mathcal{O}_{4G,neutral}=2\big[\tr{4 Z_2 Z_2 \bar{Z_2}\bar{Z_2}+2Z_2 \bar{Z_2}Z_2 \bar{Z_2}}-\frac{4}{5}\tr{Z_{(p}Z_2 \bar{Z}_{p)} \bar{Z}_2+\bar{Z}_{(p} Z_{p)} Z_{(2} \bar{Z}_{2)}}\big]\nonumber \\
-\big[\tr{4 Z_1 Z_1 \bar{Z_1}\bar{Z_1}+2Z_1 \bar{Z_1}Z_1 \bar{Z_1}}-\frac{4}{5}\tr{Z_{(p}Z_1 \bar{Z}_{p)} \bar{Z}_1+\bar{Z}_{(p} Z_{p)} Z_{(1} \bar{Z}_{1)}}\big]\nonumber \\
-\big[\tr{4 Z_3 Z_3 \bar{Z_3}\bar{Z_3}+2 Z_3 \bar{Z_3}Z_3 \bar{Z_3}}-\frac{4}{5}\tr{Z_{(p}Z_3 \bar{Z}_{p)} \bar{Z}_3+\bar{Z}_{(p} Z_{p)} Z_{(3} \bar{Z}_{3)}}\big],
\end{eqnarray}
The charge two BPS operator reads
\begin{eqnarray}
   \label{4G2}
\mathcal{O}_{4G,ijkl}=\tr{Z_i Z_j Z_k \bar{Z}_l+Z_i Z_j \bar{Z}_l Z_k+ Z_i Z_k Z_j\bar{Z}_l +Z_i Z_k \bar{Z}_l Z_j+Z_i \bar{Z}_l Z_j Z_k+Z_i \bar{Z}_l Z_k Z_j}\nonumber \\
-\frac{1}{5}\,\delta_{kl}\tr{Z_i Z_{(p}Z_j \bar{Z}_{p)} +Z_{(i} Z_{j)} Z_{(p}\bar{Z}_{p)}}
-\frac{1}{5}\,\delta_{jl}\tr{Z_k Z_{(p}Z_i \bar{Z}_{p)} +Z_{(k} Z_{i)} Z_{(p}\bar{Z}_{p)}}\nonumber \\
-\frac{1}{5}\,\delta_{il}\tr{Z_k Z_{(p}Z_j \bar{Z}_{p)} +Z_{(j} Z_{k)} Z_{(p}\bar{Z}_{p)}},
\,\,\, i,j,k,l=1,2,3\nonumber \\
\end{eqnarray}
where the repeated indices $p,q=1,2,3$ are summed over and the bracket in the indices means symmetrisation.
Namely,
\begin{eqnarray}
   \label{sym}
\tr{Z_{(p}Z_j \bar{Z}_{p)} \bar{Z}_l}=\sum_{p=1}^3\tr{Z_{p}Z_j \bar{Z}_{p} \bar{Z}_l+\bar{Z}_{p}Z_j Z_{p} \bar{Z}_l}\\
\tr{\bar{Z}_{(p} Z_{p)} Z_{(j} \bar{Z}_{l)}}=\sum_{p=1}^3\tr{Z_{p} \bar{Z}_{p} Z_{(j} \bar{Z}_{l)}+\bar{Z}_{p} Z_{p} Z_{(j} \bar{Z}_{l)}}=\nonumber \\
\sum_{p=1}^3\tr{Z_{p} \bar{Z}_{p} Z_{j} \bar{Z}_{l}+Z_{p} \bar{Z}_{p}  \bar{Z}_{l} Z_{j}+\bar{Z}_{p} Z_{p} Z_{j} \bar{Z}_{l}+\bar{Z}_{p} Z_{p}  \bar{Z}_{l}Z_{j}}.
\end{eqnarray}

\end{itemize}

\begin{table}[h]
\centering
 \begin{tabular}{|l|l|l|l|l|l|l|} \hline
Length & Class & $SU(4)^{\text{parity}}_{\text{length}}$ Rep. & Dim.  & $8\pi^{2}\,\gamma$ & Operator & Mixing  \\ \hline \hline
 \multirow{2}{*}{\raisebox{-6pt}{2}} & 2A &\ocx{$[0,0,0]^{+}_{2}$} & \ocx{1} &
  \ocx{$6$} & \ocx{$\mathcal{K}$} & no mixing
 \psx \cline{2-7}
& \ocx{2B} & \ocx{$[0,2,0]^{+}_{2}$} & \ocx{20} & \ocx{$0$} & \ocx{CPO}
& no mixing \psx \hline \hline
 \multirow{2}{*}{\raisebox{-10pt}{3}}
& \ocx{3B} & \ocx{$[0,1,0]^{-}_{3}$} & \ocx{6} & \ocx{$4$} & \ocx{$\mathcal{O}^{J=1}_{n=1}$} & resolved \psx \cline{2-7}
& \ocx{3C}& \ocx{$[0,3,0]^{-}_{3}$} & \ocx{50} & \ocx{$0$} & \ocx{CPO} & no mixing
 \psx \hline \hline
 \multirow{6}{*}{\raisebox{-26pt}{4}} & \ocx{4A} & \ocx{$[0,0,0]^{+}_{4}$} & \ocx{1} &
 \ocx{$\half \, (13 + \sqrt{41})$} & \ocx{$\cO_{4A}$} & resolved  \psx \cline{2-7}
& \ocx{4E} & \ocx{$[0,0,0]^{+}_{4}$} & \ocx{1} & \ocx{$\half \, (13 - \sqrt{41})$} &
\ocx{$\cO_{4E}$} & resolved \psx \cline{2-7}
   & \ocx{4B} & \ocx{$[0,2,0]^{+}_{4}$} & \ocx{20} & \ocx{$5 + \sqrt{5}$}
   &  \ocx{$\mathcal{O}^{J=2}_{n=2}$} & resolved \psx \cline{2-7}
 &\ocx{4F} & \ocx{$[0,2,0]^{+}_{4}$} & \ocx{20} & \ocx{$5 - \sqrt{5}$} &
 \ocx{$\mathcal{O}^{J=2}_{n=1}$} & resolved \psx \cline{2-7}
 & \ocx{\bf 4C} & \ocx{$[2,0,2]_{4}+[1,0,1]^{-}_{4}$} & \ocx{84 + 15} & \ocx{$6$} &
  & \psx \cline{2-7}
& \ocx{4G} &
 \ocx{$[0,4,0]^{+}_{4}$} & \ocx{105} & \ocx{$0$} & \ocx{CPO} & no mixing \psx \hline \hline
\multirow{8}{*}{\raisebox{-42pt}{5}}
& \ocx{\bf 5A} & \ocx{$[0,0,2]^{+}_{5}+[2,0,0]^{+}_{5}$} & \ocx{10 + $\bar{10}$} & \ocx{$7 + \sqrt{13}$} &  &  \psx \cline{2-7}
& \ocx{\bf 5H} & \ocx{$[0,0,2]^{+}_{5}+[2,0,0]^{+}_{5}$} & \ocx{10 + $\bar{10}$} & \ocx{$7 - \sqrt{13}$} &   &  \psx \cline{2-7}
& \ocx{\bf 5D} & \ocx{$[0,1,0]^{-}_{5}$ + {desc}} & \ocx{6 + 252} & \ocx{$5 + \sqrt{5}$} & \ocx{}   &  \psx \cline{2-7}
& \ocx{\bf 5I} & \ocx{$[0,1,0]^{-}_{5}$ + {desc}} & \ocx{6 + 252} & \ocx{$5 - \sqrt{5}$} & \ocx{}   &  \psx \cline{2-7}
& \ocx{\bf 5F} & \ocx{$[1,1,1]^{+}_{5}$ + $[1,1,1]^{-}_{5}$} &\ocx{64 + 64} & \ocx{$5$} &
  &  \psx \cline{2-7}
& \ocx{5J} & \ocx{$[0,3,0]^{-}_{5}$} &\ocx{50} & \ocx{$2$} & \ocx{$\cO^{J=3}_{n=1}$}
 & resolved \psx \cline{2-7}
& \ocx{\bf 5E} & \ocx{$[0,3,0]^{-}_{5}$ + {desc}} &\ocx{50 + 140} & \ocx{$6$} &
\ocx{$\mathcal{O}^{J=3}_{n=2}$} & resolved  \psx \cline{2-7}
& \ocx{5K} & \ocx{$[0,5,0]^{-}_{5}$} &\ocx{196} & \ocx{$0$} & \ocx{CPO} & no mixing
\psx
\cline{2-7}
& \ocx{\bf 5B} & \ocx{$[0,1,0]^{-}_{5}$} &\ocx{\bf 6+6} & \ocx{$10$} & &
\psx \cline{2-7} \hline
\end{tabular}
\caption{List of all scalar conformal primary operator up to length 5 with their one-loop
anomalous dimensions. Degenerate classes of
operators are printed in bold-face. $\mathcal{K}$ denotes the Konishi and
 $CPO$ chiral primary operators. The $\mathcal{O}^{J}_{n}$ refer to the BMN singlet
 operators in the nomenclature of \eqn{hws}.
 In the last column the resolved mixing problem with fermion, derivative and self-mixings
 of section \ref{sect:mixing} are displayed.}
\label{fig:sd}
\end{table}

\section{Results}
\label{sec:corresults}

The final result for the structure constants arises from two contributions: The
radiative one-loop corrections discussed in section \ref{three} as well as
the corrections arising from the operator mixing effects spelled out in section
\ref{sect:mixresults}, which in principle enable one to straightforwardly compute
three-point functions involving scalar operators up to length five by combinatorial means.

Let us begin with the dressing formulae to find the radiative corrections.
Clearly, due to the need to sum over
all permutations in these dressing formulae the complexity in the computations grows fast and needs
to be done on a computer. This has been implemented in a two step procedure. Starting with an
arbitrarily chosen basis of operators all two-point functions are computed and then diagonalized. All scalar operators up to length five are detailed in figure \ref{fig:sd}.
Similarly all three-point functions are computed in the original basis
and then projected to the diagonal basis where
the structure constants can be extracted. For operators up to length three this was done algebraically
with a {\sc Mathematica} program. Starting with length four the mixing matrix diagonalization could not
be performed algebraically any longer and we had to resort to numerics using Matlab.
Once the diagonal basis was constructed the numerically obtained structure constants could
in most cases be again fitted
to algebraic expressions derived by the algebraic form of the one-loop scaling dimensions.

Secondly the contribution from the structure constants from the mixing terms
of section \ref{sect:mixresults} were found as well. These arise from
tree-level contractions involving the operator corrections due to double bi-fermion,
bi-derivative and self-mixing, as well as bi-fermion corrections of one operator
and a Yukawa-interaction. It turns out that the relevant contibutions securing
conformal symmetry arise from suitable tree-level correlators only - the Yukawa contributions
always cancel.

Below  we list our main results sorted by correlator classes which are listed in the tables
\ref{244corr}, \ref{334corr}, \ref{444corr} and  \ref{255corr}\footnote{Here we have used everywhere $\lambda = g_{YM}^2 N$}. 
Note that only three-point functions which do not vanish at
tree-level are listed. We also stress that the majority of results for the
radiative corrections to the fractions
$C_{\alpha \beta \gamma}^{(1)}/C_{\alpha \beta \gamma}^{(0)}|_{\text{loop}}$ have been obtained numerically
and the quoted analytical results represents a biases fit allowing as non-rational factors only the
square root term appearing in the anomalous scaling dimensions of the operators involved in the
patricular three-point function. The numerical
precision in theses fits is typically of order $10^{-5}$ or better, for the raw data see
the appendix  A.2 of \cite{DiplomarbeitGrossardt}. Finally, the analytically obtained results are highlighted
in bold-face letters.

\subsection{$\vev{2|4|4}$ correlators}

Here a diagramatic analysis reveals that only the double bi-fermionic mixing
and the scalar self-mixings will contribute to the three-point correlator, whereas the
Yukawa-vertex insertion cancels against the bi-derivative mixing contributions.
This follows by considering the propagator dependances of these terms.

\newcommand{\gsbyandre}{0.75}
\newcommand{\smbyandre}{\footnotesize}
\begin{equation*}
\begin{array}{ccccccc}
\begin{overpic}[scale=\gsbyandre,tics=10]{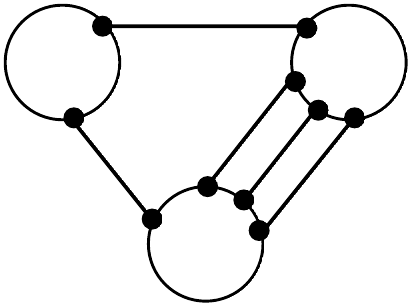}
\put(10,55){$1$}
\put(82,55){$3$}
\put(45,10){$2$}
\end{overpic}
& \raisebox{25pt}{+} &
\begin{overpic}[scale=\gsbyandre,tics=10]{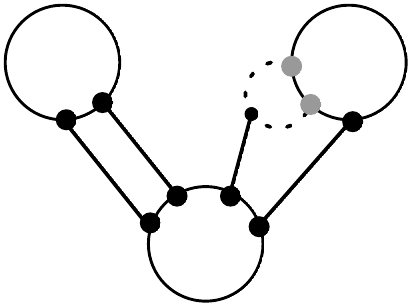}
\put(10,55){$1$}
\put(82,55){$3$}
\put(45,10){$2$}
\put(73,40){\smbyandre$\psi$}
\put(63,62){\smbyandre$\psi$}
\end{overpic}
& \raisebox{25pt}{+} &
\begin{overpic}[scale=\gsbyandre,tics=10]{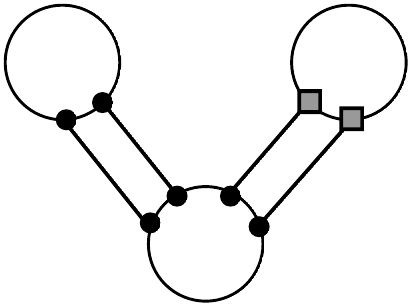}
\put(10,55){$1$}
\put(82,55){$3$}
\put(45,10){$2$}
\put(88,35){\smbyandre D}
\put(63,45){\smbyandre D}
\end{overpic}
& \raisebox{25pt}{+} &
\begin{overpic}[scale=\gsbyandre,tics=10]{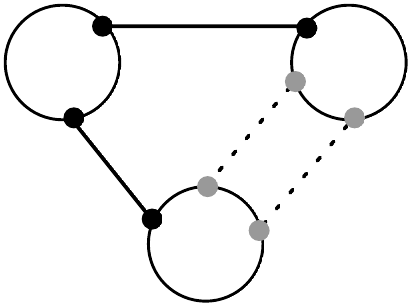}
\put(10,55){$1$}
\put(82,55){$3$}
\put(45,10){$2$}
\put(85,35){\smbyandre$\psi$}
\put(60,50){\smbyandre$\psi$}
\put(45,32){\smbyandre$\psi$}
\put(68,18){\smbyandre$\psi$}
\end{overpic} \\
\underline{\Delta_{12} \,\Delta_{13} \,\Delta_{23}^3}
&& \Delta_{12}^2 \,\Delta_{23}^3
&& \Delta_{12}^2 \,\Delta_{23}^3
&& \underline{\Delta_{12} \,\Delta_{13} \,\Delta_{23}^3}
\end{array}
\end{equation*}
The underlined terms contribute.

We hence only need to determine the ratio
\be
\label{eq:244-structure}
\frac{C^{(1)}_{\alpha\beta\gamma}}{C^{(0)}_{\alpha\beta\gamma}}\Bigr|_{\text{mixing}} =
\frac{\vev{2|4_{\psi\psi}|4}+\vev{2|4|4_{\psi\psi}}+ \vev{2|4_{\text{self}}|4}+\vev{2|4|4_{\text{self}}}}
{\vev{2|4|4}}\, .
\ee
The highly involved evaluation of these correlators was performed with the help of a
{\sc Mathematica} program. In addition one has the radiative corrections in the pure
$SO(6)$ sector whose form follows from the dressing procedure. We state the radiative contributions
and the mixing contributions seperately  and give
the complete result in the final column, see table \ref{244corr}.

\subsection{$\vev{3|3|4}$ correlators}

For the $\vev{3|3|4}$ the diagramatic analysis of

\begin{equation*}
\begin{array}{ccccc}
\begin{overpic}[scale=\gsbyandre,tics=10]{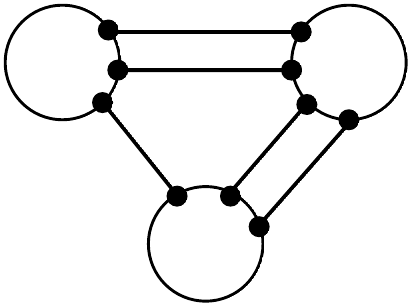}
\put(10,55){$1$}
\put(82,55){$3$}
\put(45,10){$2$}
\end{overpic}
& \raisebox{25pt}{+} &
\begin{overpic}[scale=\gsbyandre,tics=10]{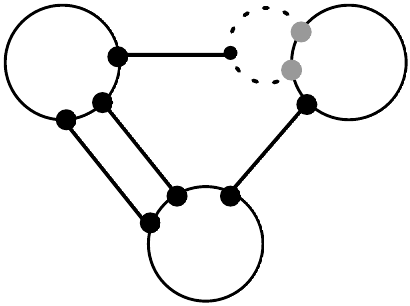}
\put(10,55){$1$}
\put(82,55){$3$}
\put(45,10){$2$}
\put(73,55){\smbyandre$\psi$}
\put(71,73){\smbyandre$\psi$}
\end{overpic}
& \raisebox{25pt}{+} &
\begin{overpic}[scale=\gsbyandre,tics=10]{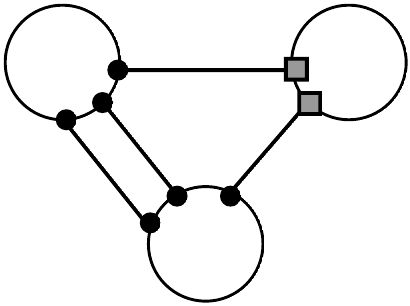}
\put(10,55){$1$}
\put(82,55){$3$}
\put(45,10){$2$}
\put(78,35){\smbyandre D}
\put(60,60){\smbyandre D}
\end{overpic} \\
\underline{\Delta_{12} \,\Delta_{23}^2 \,\Delta_{13}^2}
&& \Delta_{12}^2 \,\Delta_{13}^2 \,\Delta_{23} + 1 \leftrightarrow 2
&& \Delta_{12}^2 \,\left(\Delta_{13}^2 \,\Delta_{23} + \Delta_{13}
\,\Delta_{23}^2 - \underline{\Delta_{13}^2 \,\Delta_{23}^2
\,\Delta_{12}^{-1}}\right)
\end{array}
\end{equation*}

reveals that in the mixing sector we can only have corrections due to
self-mixing, bi-fermion and bi-derivative insertions for the operator
of engineering length four. Again, the Yukawa insertion cancels
against part of the bi-derivative term (see appendix \ref{sec:canc}),
and the surviving terms arise just from the contributions 
\be
\label{eq:344-structure}
\frac{C^{(1)}_{\alpha\beta\gamma}}{C^{(0)}_{\alpha\beta\gamma}}\Bigr|_{\text{mixing}} =
\frac{\vev{3|3|4_{\text{self}}}+\vev{3|4|4_{DD}}}{\vev{3|4|4}}\, .
\ee
The results are exposed in table \ref{334corr}.

\subsection{$\vev{4|4|4}$ correlators}

Turning to the $\vev{4|4|4}$ correlators a similar diagramatic analysis tells us that now
the Yukawa insertion  cancels part  of the double derivative corrections to the length four
operators, while the remaining term yields the surviving contribution respecting conformal
symmetry.

\begin{equation*}
\begin{array}{ccccc}
\begin{overpic}[scale=\gsbyandre,tics=10]{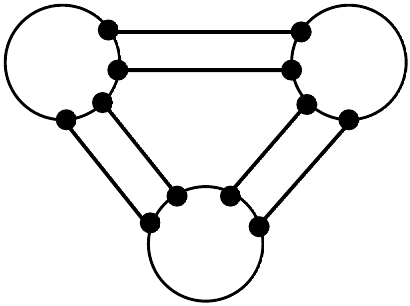}
\put(10,55){$1$}
\put(82,55){$3$}
\put(45,10){$2$}
\end{overpic}
& \raisebox{25pt}{+} &
\begin{overpic}[scale=\gsbyandre,tics=10]{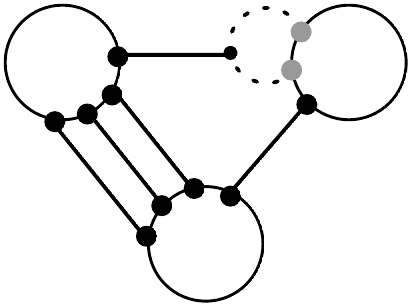}
\put(10,55){$1$}
\put(82,55){$3$}
\put(45,10){$2$}
\put(73,55){\smbyandre$\psi$}
\put(71,73){\smbyandre$\psi$}
\end{overpic}
& \raisebox{25pt}{+} &
\begin{overpic}[scale=\gsbyandre,tics=10]{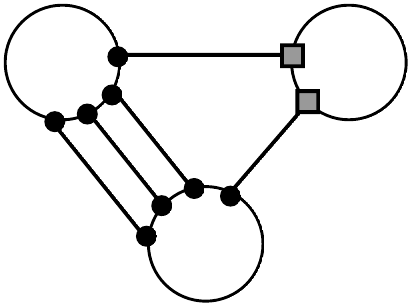}
\put(10,55){$1$}
\put(82,55){$3$}
\put(45,10){$2$}
\put(75,38){\smbyandre D}
\put(60,64){\smbyandre D}
\end{overpic} \\
\underline{\Delta_{12}^2 \,\Delta_{23}^2 \,\Delta_{13}^2}
&& \Delta_{12}^3 \,\Delta_{13}^2 \,\Delta_{23} + 1 \leftrightarrow 2
&& \Delta_{12}^3 \,\left(\Delta_{13}^2 \,\Delta_{23} + \Delta_{13}
\,\Delta_{23}^2 - \underline{\Delta_{13}^2 \,\Delta_{23}^2
\,\Delta_{12}^{-1}}\right)
\end{array}
\end{equation*}

Hence, one here needs to evaluate the mixing contributions
\be
\label{eq:444-structure}
\frac{C^{(1)}_{\alpha\beta\gamma}}{C^{(0)}_{\alpha\beta\gamma}}\Bigr|_{\text{mixing}} =
\frac{\vev{4|4|4_{DD}}+\vev{4|4_{DD}|4}+\vev{4_{DD}|4|4}+ \vev{4|4|4_{\text{self}}}
+\vev{4|4_{\text{self}}|4}+\vev{4_{\text{self}}|4|4}}
{\vev{4|4|4}}\, ,
\ee
which are summarized together with the radiative corrections in table \ref{444corr}.

\subsection{$\vev{2|5|5}$ correlators}

Finally the structure constants involving two length five operators and one length
two operator are similarly controlled by the bi-fermi and self-mixing insertions,
the Yukawa contribution cancels against the bi-derivative correction.

\begin{equation*}
\begin{array}{ccccccc}
\begin{overpic}[scale=\gsbyandre,tics=10]{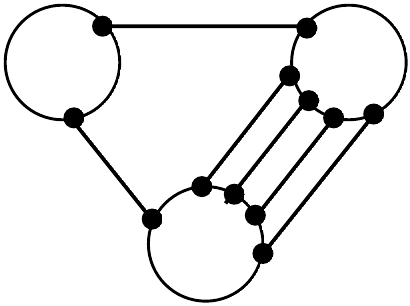}
\put(10,55){$1$}
\put(82,55){$3$}
\put(45,10){$2$}
\end{overpic}
& \raisebox{25pt}{+} &
\begin{overpic}[scale=\gsbyandre,tics=10]{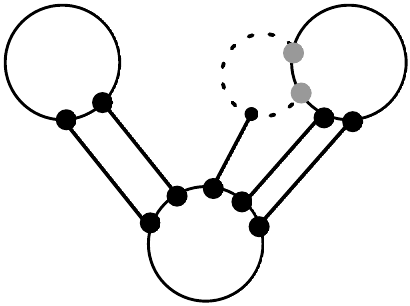}
\put(10,55){$1$}
\put(82,55){$3$}
\put(45,10){$2$}
\put(68,42){\smbyandre$\psi$}
\put(65,68){\smbyandre$\psi$}
\end{overpic}
& \raisebox{25pt}{+} &
\begin{overpic}[scale=\gsbyandre,tics=10]{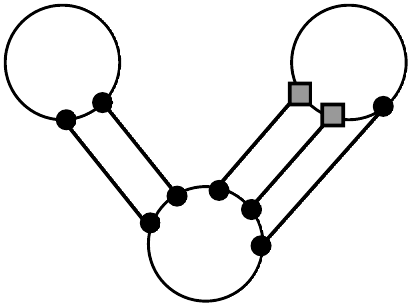}
\put(10,55){$1$}
\put(82,55){$3$}
\put(45,10){$2$}
\put(77,36){\smbyandre D}
\put(60,47){\smbyandre D}
\end{overpic}
& \raisebox{25pt}{+} &
\begin{overpic}[scale=\gsbyandre,tics=10]{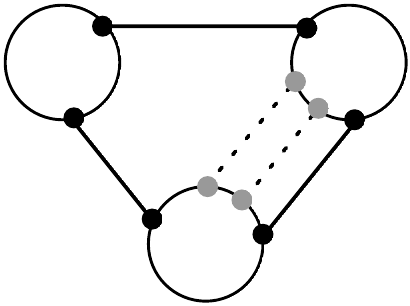}
\put(10,55){$1$}
\put(82,55){$3$}
\put(45,10){$2$}
\put(70,40){\smbyandre$\psi$} 
\put(60,50){\smbyandre$\psi$} 
\put(45,32){\smbyandre$\psi$} 
\put(58,30){\smbyandre$\psi$} 
\end{overpic} \\
\underline{\Delta_{12} \,\Delta_{13} \,\Delta_{23}^4}
&& \Delta_{12}^2 \,\Delta_{23}^4
&& \Delta_{12}^2 \,\Delta_{23}^4
&& \underline{\Delta_{12} \,\Delta_{13} \,\Delta_{23}^4}
\end{array}
\end{equation*}

Hence, we evaluate the contributions
\be
\label{eq:255-structure}
\frac{C^{(1)}_{\alpha\beta\gamma}}{C^{(0)}_{\alpha\beta\gamma}}\Bigr|_{\text{mixing}} =
\frac{\vev{2|5_{\psi\psi}|5_{\psi\psi}}+ \vev{2|5|5_{\text{self}}}
+\vev{2|5_{\text{self}}|5}}
{\vev{2|5|5}}\, ,
\ee
for the two cases in table \ref{255corr}.

\begin{landscape}
\begin{center}
\begin{longtable}{|l|l|l||l|l|l||l|l|l|}
 \hl
 $\Op_\alpha$ & $\Op_\beta$ & $\Op_\gamma$ & $8 \pi^2 \gamma_\alpha$ & $8 \pi^2 \gamma_\beta$ & $8 \pi^2 \gamma_\gamma$ & $\ft{-16 \pi^2 C_{\alpha \beta \gamma}^{(1)}}{C_{\alpha \beta \gamma}^{(0)}}|_{\text{loop}}$
 &$\ft{-16 \pi^2 C_{\alpha \beta \gamma}^{(1)}}{C_{\alpha \beta \gamma}^{(0)}}|_{\text{mixing}}$
  & Sum \ps\\ \hld \endhead

     \oc{2A} & \oc{4A} & \oc{4A} & \oc{6} & \oc{\gmVA} & \oc{\gmVA} & \oc{$\frac{1}{2}(25 + \sqrt{41})$}
 & \oc{$-\frac{3}{2}-\frac{7}{2\, \sqrt{41}}$} &
 \oc{$11+\frac{17}{\sqrt{41}}$} \ps\\* \hl

   \oc{2A} & \oc{4B} & \oc{4B} & \oc{6} & \oc{\gmVB} & \oc{\gmVB} & \oc{$11 + \sqrt{5}$}
 & \oc{$-\frac{3}{4}\, (3+\sqrt{5})$} &
 \oc{$\frac{1}{4}\, (35+\sqrt{5})$} \ps\\* \hl

    \oc{2A} & \oc{4E} & \oc{4E} & \oc{6} & \oc{\gmVE} & \oc{\gmVE} & \oc{$\frac{1}{2}(25 - \sqrt{41})$}
 & \oc{$-\frac{3}{2}+\frac{7}{2\, \sqrt{41}}$} &
 \oc{$11-\frac{17}{\sqrt{41}}$} \ps\\* \hl

   \oc{2A} & \oc{4F} & \oc{4F} & \oc{6} & \oc{\gmVF} & \oc{\gmVF} & \oc{$11 - \sqrt{5}$}
 & \oc{$-\frac{3}{4}\, (3-\sqrt{5})$} &
 \oc{$\frac{1}{4}\, (35-\sqrt{5})$} \ps\\* \hl

 \oc{2B} & \oc{4A} & \oc{4B} & \oc{\gmZB} & \oc{\gmVA} & \oc{\gmVB} & \oc{$5 + \sqrt{5}$}
 & \oc{$\frac{-3699+533 \sqrt{5}-651 \sqrt{41}-75 \sqrt{205}}{1640}$} &
 \oc{$\frac{4501+2173 \sqrt{5}-651 \sqrt{41}-75 \sqrt{205}}{1640}$} \ps\\* \hl
 \oc{2B} & \oc{4A} & \oc{4F} & \oc{\gmZB} & \oc{\gmVA} & \oc{\gmVF} & \oc{$5 - \sqrt{5}$}
  & \oc{$\frac{-3699-533 \sqrt{5}-651 \sqrt{41}+75 \sqrt{205}}{1640}$}
  & \oc{$\frac{4501-2173 \sqrt{5}-651 \sqrt{41}+75 \sqrt{205}}{1640}$} \ps\\* \hl
 \oc{2B} & \oc{4B} & \oc{4B} & \oc{\gmZB} & \oc{\gmVB} & \oc{\gmVB} & \oc{$\frac{2}{79} \, (115 + 14 \, \sqrt{5})$}
  & \oc{$-\frac{3}{395} \left(175+11 \sqrt{5}\right)$}
  & \oc{$\frac{1}{395} \left(615+107 \sqrt{5}\right)$} \ps\\* \hl
 \oc{2B} & \oc{4B} & \oc{4E} & \oc{\gmZB} & \oc{\gmVB} & \oc{\gmVE} & \oc{$5 + \sqrt{5}$}
  & \oc{$\frac{-3699+533 \sqrt{5}+651 \sqrt{41}+75 \sqrt{205}}{1640}$}
     & \oc{$\frac{4501+2173 \sqrt{5}+651 \sqrt{41}+75 \sqrt{205}}{1640}$} \ps\\* \hl
 \oc{2B} & \oc{4B} & \oc{4F} & \oc{\gmZB} & \oc{\gmVB} & \oc{\gmVF} & \oc{0}
  & \oc{$-\frac{35}{3}$} & \oc{$-\frac{35}{3}$} \ps\\* \hl
 \oc{2B} & \oc{4B} & \oc{4G} & \oc{\gmZB} & \oc{\gmVB} & \oc{\gmVG} & \oc{$5 + \sqrt{5}$}
  & \oc{$\frac{1}{10} \left(-25-7 \sqrt{5}\right)$}
  & \oc{$\frac{5}{2}+\frac{3}{2 \sqrt{5}}$} \ps\\* \hl
 \oc{2B} & \oc{4E} & \oc{4F} & \oc{\gmZB} & \oc{\gmVE} & \oc{\gmVF} & \oc{$5 - \sqrt{5}$}
  & \oc{$\frac{-3699-533 \sqrt{5}+651 \sqrt{41}-75 \sqrt{205}}{1640}$}
  & \oc{$\frac{4501-2173 \sqrt{5}+651 \sqrt{41}-75 \sqrt{205}}{1640}$} \ps\\* \hl
 \oc{2B} & \oc{4F} & \oc{4F} & \oc{\gmZB} & \oc{\gmVF} & \oc{\gmVF} & \oc{$\frac{2}{79} \, (115 - 14 \, \sqrt{5})$}
  & \oc{$\frac{3}{395} \left(11 \sqrt{5}-175\right)$}
  & \oc{$\frac{1}{395} \left(615-107 \sqrt{5}\right)$} \ps\\* \hl
 \oc{2B} & \oc{4F} & \oc{4G} & \oc{\gmZB} & \oc{\gmVF} & \oc{\gmVG} & \oc{$5 - \sqrt{5}$}
  & \oc{$\frac{1}{10} \left(7 \sqrt{5}-25\right)$} & \oc{$\frac{5}{2}-\frac{3}{2 \sqrt{5}}$}\ps\\* \hl
\caption{The evaluated $\vev{2|4|4}$ three-point correlators. \label{244corr}}
  \end{longtable}
  \begin{longtable}{|l|l|l||l|l|l||l|l|l|}
    \hl
    $\Op_\alpha$ & $\Op_\beta$ & $\Op_\gamma$ &
    $8 \pi^2 \gamma_\alpha$ & $8 \pi^2 \gamma_\beta$ & $8 \pi^2 \gamma_\gamma$ &
    $\ft{-16 \pi^2 C_{\alpha \beta \gamma}^{(1)}}{C_{\alpha \beta \gamma}^{(0)}}|_{\text{loop}}$
    &$\ft{-16 \pi^2 C_{\alpha \beta \gamma}^{(1)}}{C_{\alpha \beta \gamma}^{(0)}}|_{\text{mixing}}$
    & Sum \ps\\
    \hld \endhead
    \oc{3B} & \oc{3B} & \oc{4A} &
    \oc{\gmDB} & \oc{\gmDB} & \oc{\gmVA} &
     $\frac{1}{50}(261+9\sqrt{41})$ &
     $\frac{2}{205} \left(21 \sqrt{41}-121\right)$  &
     $\frac{8281+789 \sqrt{41}}{2050}$ \ps\\* \hl
    \oc{3B} & \oc{3B} & \oc{4E} &
    \oc{\gmDB} & \oc{\gmDB} & \oc{\gmVE} &
    $\frac{1}{50}(261-9\sqrt{41})$ &
    $-\frac{2}{205} \left(121+21 \sqrt{41}\right)$&
    $\frac{8281-789 \sqrt{41}}{2050}$ \ps\\* \hl
    \oc{3B} & \oc{3B} & \oc{4B} &
    \oc{\gmDB} & \oc{\gmDB} & \oc{\gmVB} &
    $\frac{1}{11}(87+3\sqrt{5})$ &
    $-\frac{1}{110} \left(175+3 \sqrt{5}\right)$  &
    $\frac{1}{110} \left(695+27 \sqrt{5}\right)$ \ps\\* \hl
    \oc{3B} & \oc{3B} & \oc{4F} &
    \oc{\gmDB} & \oc{\gmDB} & \oc{\gmVF} &
    $\frac{1}{11}(87-3\sqrt{5})$ &
    $\frac{1}{110} \left(123 \sqrt{5}-355\right)$ &
    $\frac{1}{110} \left(515+93 \sqrt{5}\right)$ \ps\\* \hl
    \oc{3B} & \oc{3C} & \oc{4B} &
    \oc{\gmDB} & \oc{\gmDC} & \oc{\gmVB} &
    $\frac{1}{11}(39+7\sqrt{5})$ &
    $-\frac{1}{110} \left(215+7 \sqrt{5}\right)$  &
    $\frac{7}{110} \left(25+9 \sqrt{5}\right)$ \ps\\* \hl
    \oc{3B} & \oc{3C} & \oc{4F} &
    \oc{\gmDB} & \oc{\gmDC} & \oc{\gmVF} &
    $\frac{1}{11}(39-7\sqrt{5})$ &
    $\frac{3}{110} \left(29 \sqrt{5}-105\right)$ &
    $\frac{1}{110} \left(75+17 \sqrt{5}\right)$ \ps\\* \hl
    \oc{3C} & \oc{3C} & \oc{4A} &
    \oc{\gmDC} & \oc{\gmDC} & \oc{\gmVA} &
    $\frac{1}{2}(13+\sqrt{41})$ &
    $\frac{2}{205} \left(371+89 \sqrt{41}\right)$ &
    $\frac{1}{210} \left(2849+461 \sqrt{41}\right)$ \ps\\* \hl
    \oc{3C} & \oc{3C} & \oc{4E} &
    \oc{\gmDC} & \oc{\gmDC} & \oc{\gmVE} &
    $\frac{1}{2}(13-\sqrt{41})$ &
    $-\frac{2}{205} \left(89 \sqrt{41}-371\right)$  &
    $\frac{1}{210} \left(2849-461 \sqrt{41}\right)$\ps\\* \hl
    \oc{3C} & \oc{3C} & \oc{4B} &
    \oc{\gmDC} & \oc{\gmDC} & \oc{\gmVB} &
    $5+\sqrt{5}$ &
    $-\frac{1}{10} \left(45+17 \sqrt{5}\right)$  &
    $\frac{1}{10} \left(5-7 \sqrt{5}\right)$ \ps\\* \hl
    \oc{3C} & \oc{3C} & \oc{4F} &
    \oc{\gmDC} & \oc{\gmDC} & \oc{\gmVF} &
    $5-\sqrt{5}$ &
    $\frac{1}{10} \left(55-23 \sqrt{5}\right) $  &
    $\frac{1}{10} \left(105-33\sqrt{5}\right)$\ps\\* \hl
    \oc{3B} & \oc{3C} & \oc{4G} &
    \oc{\gmDB} & \oc{\gmDC} & \oc{\gmVG} & 4  & 0  & 4 \ps\\* \hl
       \caption{The evaluated $\vev{3|3|4}$ three-point correlators.}
    \label{334corr}
  \end{longtable}
\begin{longtable}{|l|l|l||l|l|l||l|l|l|}
     \hl
 $\Op_\alpha$ & $\Op_\beta$ & $\Op_\gamma$ & $8 \pi^2 \gamma_\alpha$ & $8 \pi^2 \gamma_\beta$ & $8 \pi^2 \gamma_\gamma$ & $\ft{-16 \pi^2 C_{\alpha \beta \gamma}^{(1)}}{C_{\alpha \beta \gamma}^{(0)}}|_{\text{loop}}$
 &$\ft{-16 \pi^2 C_{\alpha \beta \gamma}^{(1)}}{C_{\alpha \beta \gamma}^{(0)}}|_{\text{mixing}}$
  & Sum \ps\\ \hld \endhead
 \oc{4A} & \oc{4A} & \oc{4A} & \oc{\gmVA} & \oc{\gmVA} & \oc{\gmVA} & \oc{${ \frac{1}{733} \, (7185 + 309 \,\sqrt{41})}$}
& \oc{$\frac{6 \left(11417 \sqrt{41}-105667\right)}{150265}$}
& \oc{$\frac{3 \left(279641+43949 \sqrt{41}\right)}{150265}$}
 \ps\\* \hl
 \oc{4A} & \oc{4A} & \oc{4E} & \oc{\gmVA} & \oc{\gmVA} & \oc{\gmVE} & \oc{${
\frac{1}{10} \, (21 - \sqrt{41})}$}
  & \oc{$-\frac{1}{820} \left(3847+383 \sqrt{41}\right)$}
  & \oc{$-\frac{1}{164} \left(425+93 \sqrt{41}\right)$}
   \ps\\* \hl
 \oc{4A} & \oc{4B} & \oc{4B} & \oc{\gmVA} & \oc{\gmVB} & \oc{\gmVB} &
\oc{$12.3279656$}
  & \oc{$\frac{\sqrt{41} \left(1263+527 \sqrt{5}\right)-88 \left(108+41
   \sqrt{5}\right)}{410} $}
  & \oc{$7.59846$}
 \ps\\* \hl
 \oc{4A} & \oc{4B} & \oc{4F} & \oc{\gmVA} & \oc{\gmVB} & \oc{\gmVF} &
\oc{$\frac{1}{2} \, (9 + \sqrt{41})$}
  & \oc{$\frac{-3149+205 \sqrt{5}-171 \sqrt{41}}{410}$}
  & \oc{$\frac{-1304+205 \sqrt{5}+34 \sqrt{41}}{410}$}
 \ps\\* \hl
 \oc{4A} & \oc{4E} & \oc{4E} & \oc{\gmVA} & \oc{\gmVE} & \oc{\gmVE} & \oc{${
\frac{1}{10} \, (21 + \sqrt{41})}$}
  & \oc{$\frac{1}{820} \left(383 \sqrt{41}-3847\right)$}
  & \oc{$\frac{1}{820} \left(301 \sqrt{41}-2781\right)$}
 \ps\\* \hl
 \oc{4A} & \oc{4F} & \oc{4F} & \oc{\gmVA} & \oc{\gmVF} & \oc{\gmVF} &
\oc{$4.865786$}
  & \oc{$\frac{-8479+3280 \sqrt{5}+1058 \sqrt{41}-445 \sqrt{205}}{410}$}
  & \oc{$3.05695$}
 \ps\\* \hl
 \oc{4A} & \oc{4G} & \oc{4G} & \oc{\gmVA} & \oc{\gmVG} & \oc{\gmVG} &
\oc{$\frac{1}{2} \, (13 + \sqrt{41})$}
  & \oc{$\frac{1}{410} \left(459 + 151 \sqrt{41}\right)$}
  & \oc{$\frac{2}{205} \left(781 + 89 \sqrt{41}\right)$}
 \ps\\* \hl
%
%
 \oc{4B} & \oc{4B} & \oc{4E} & \oc{\gmVB} & \oc{\gmVB} & \oc{\gmVE} &
\oc{$38.020253$}
  & \oc{$-\frac{4752}{205}-\frac{44}{5}\sqrt{5}-\frac{1263}{410} \sqrt{41}-\frac{527}{410} \sqrt{205}$}
  & \oc{$-42.9660$}
 \ps\\* \hl
%
%
 \oc{4B} & \oc{4B} & \oc{4G} & \oc{\gmVB} & \oc{\gmVB} & \oc{\gmVG} &
\oc{$\frac{4}{19} \, (25 + 7\, \sqrt{5})$}
  & \oc{$\frac{1}{190} \left(-785-121 \sqrt{5}\right)$}
  & \oc{$\frac{1}{190} \left(215+159 \sqrt{5}\right)$}
 \ps\\* \hl
 \oc{4B} & \oc{4E} & \oc{4F} & \oc{\gmVB} & \oc{\gmVE} & \oc{\gmVF} &
\oc{$\frac{1}{2} \, (9 - \sqrt{41})$}
  & \oc{$-\frac{3149}{410}+\frac{1}{2}\sqrt{5}+\frac{171}{410}\sqrt{41}$}
  & \oc{$-\frac{652}{205}+\frac{1}{2}\sqrt{5}-\frac{17}{205} \sqrt{41}$}
 \ps\\* \hl
%
%
 \oc{4E} & \oc{4E} & \oc{4E} & \oc{\gmVE} & \oc{\gmVE} & \oc{\gmVE} & \oc{${
\frac{1}{733} \, (7185  - 309 \,\sqrt{41})}$}
  & \oc{$-\frac{6 \left(105667+11417 \sqrt{41}\right)}{150265}$}
  & \oc{$-\frac{3 \left(43949 \sqrt{41}-279641\right)}{150265}$}
 \ps\\* \hl
%
%
 \oc{4E} & \oc{4F} & \oc{4F} & \oc{\gmVE} & \oc{\gmVF} & \oc{\gmVF} &
\oc{$4.785995$}
  & \oc{$-\frac{8479}{410}+\frac{89 \sqrt{\frac{5}{41}}}{2}+8 \sqrt{5}-\frac{529}{5 \sqrt{41}}$}
  & \oc{$1.01094$}
 \ps\\* \hl
 \oc{4E} & \oc{4G} & \oc{4G} & \oc{\gmVE} & \oc{\gmVG} & \oc{\gmVG} &
\oc{$\frac{1}{2} \, (13 - \sqrt{41})$}
  & \oc{$\frac{1}{410} \left(459-151 \sqrt{41}\right)$}
  & \oc{$-\frac{2}{205} \left(89 \sqrt{41}-781\right)$}
 \ps\\* \hl
%
\caption{The evaluated $\vev{4|4|4}$ three-point correlators. \label{444corr}}
  \end{longtable}

  \begin{longtable}{|l|l|l||l|l|l||l|l|l|}
     \hl
 $\Op_\alpha$ & $\Op_\beta$ & $\Op_\gamma$ & $8 \pi^2 \gamma_\alpha$ & $8 \pi^2 \gamma_\beta$ & $8 \pi^2 \gamma_\gamma$ & $\ft{-16 \pi^2 C_{\alpha \beta \gamma}^{(1)}}{C_{\alpha \beta \gamma}^{(0)}}|_{\text{loop}}$
 &$\ft{-16 \pi^2 C_{\alpha \beta \gamma}^{(1)}}{C_{\alpha \beta \gamma}^{(0)}}|_{\text{mixing}}$
  & Sum \ps\\ \hld \endhead
 \oc{2A} & \oc{5J} & \oc{5J}  & \oc{$6$} & \oc{$2$} & \oc{$2$}
& \oc{$\frac{38}{5}$}
& \oc{$-\frac{3}{5}$}
& \oc{$7$}
 \ps\\* \hl
 \oc{2B} & \oc{5J} & \oc{5J}  & \oc{$0$} & \oc{$2$} & \oc{$2$}
& \oc{$\frac{10}{7}$}
& \oc{$-\frac{15}{14}$}
& \oc{$\frac{5}{14}$}
 \ps\\* \hl
 \oc{2B} & \oc{5J} & \oc{5K} & \oc{$0$} & \oc{$2$} & \oc{$0$}
& \oc{$2$}
& \oc{$-\frac{5}{4}$}
& \oc{$\frac{3}{4}$}
   \ps\\* \hl
  \caption{The evaluated $\vev{2|5|5}$ three-point correlators. \label{255corr}}
  \end{longtable}
    \end{center}
\end{landscape}

As reported in the introduction we make the general observation, that
the radiative corrections to the three-point structure constants
for a three-point
function of two protected operators with one unprotected operator the structure constants
follow the simple pattern:
\be
\frac{C^{(1)}_{\alpha \beta \gamma}}{C^{(0)}_{\alpha \beta \gamma}}\Bigr |_{\text{loop}}=
-\frac{1}{2}\, \gamma_{\gamma}\, \qquad \text{if } \gamma_{\alpha}=\gamma_{\beta}=0\, .
\ee
This occurred in all applicable 17 cases we observed. Unfortunately this pattern
does not survive once the mixing contributions are included.

\subsection{Radiative contributions to $\vev{\mathcal{K}|\mathcal{O} | \mathcal{O}}$ correlators}

In this subsection we derive a compact result for the radiative contributions to the
three-point function of a Konishi operator with two arbitrary operators of same length from a diagonal basis. The three-point function then takes the general form
\begin{align}
 C^{(1)}_{\alpha \beta \K} \Bigr|_{\text{loop}} &= - \kla{\frac{\gamma_\alpha}{\Dn_\alpha} + \frac{\gamma_\beta}{\Dn_\beta} + \frac{\gamma_\K}{\Dn_\K}} C^{(0)}_{\alpha \beta \K} = -\frac{\delta_{\alpha \beta}}{4 \pi^2 \, \sqrt{3}} \kla{2 \gamma_\alpha + \frac{3}{8 \pi^2} \Dn_\alpha}\, ,
\end{align}
as already mentioned in the introduction.

This may be shown as follows.
Let $\K$ be the length two Konishi operator and the set $\{\Op_\alpha\}$ an arbitrary non-diagonal basis for the operators of length $\Delta^{(0)}$ that can be written in terms of attached vectors, namely
\begin{align}
 \K &= \frac{1}{\sqrt{12}} \sum_i \tr{\phi^i \phi^i} \\
 \Op_\alpha &= \tr{u^\alpha_1 \cdot \phi \cdots u^\alpha_{\oplength} \cdot \phi} & (\oplength > 2).
\end{align}
Let $Z_k \subset S_k$ denote the set of cyclic permutations of $(1,2,\dots,k)$.

We choose the renormalization scheme $\e \to e \e$ in which only the 2-gons hold finite contributions
\begin{equation}
 \skla{\begin{minipage}[h]{12mm}\begin{center}\footnotesize

\begin{fmffile}{2gon}
\begin{fmfgraph*}(5,14)
\fmfstraight
\fmfleft{dummy1,p1,a2,a1,p2,dummy2}
\fmfright{dummy3,p3,b1,b2,p4,dummy4}
\fmfpen{thin}
\fmf{plain}{dummy1,p1,a2,a1,p2,dummy2}
\fmf{plain}{dummy3,p3,b1,b2,p4,dummy4}
\fmffreeze
\fmf{plain,width=2}{a1,b2}
\fmf{plain,width=2}{a2,b1}
\fmf{dots}{p1,p3}
\fmf{dots}{p2,p4}
\fmfdot{a1,a2,b1,b2}
\fmfv{label=$u_1$,label.dist=1.5thick}{a1}
\fmfv{label=$u_2$,label.dist=1.5thick}{a2}
\fmfv{label=$v_1$,label.dist=1.5thick}{b1}
\fmfv{label=$v_2$,label.dist=1.5thick}{b2}
\fmflabel{$x_1$}{dummy1}
\fmflabel{$x_2$}{dummy3}
\end{fmfgraph*}
\end{fmffile}

\end{center}\end{minipage}}_\text{1-loop} = I_{12}^2 \, \frac{\lambda}{8 \pi^2} \kla{\ln \frac{\e^2}{x_{12}^2} + 1} \Bigg( \begin{minipage}[h]{8mm}\begin{center}\footnotesize

\begin{fmffile}{2gonBlank}
\begin{fmfgraph*}(5,14)
\fmfstraight
\fmfleft{dummy1,p1,a2,a1,p2,dummy2}
\fmfright{dummy3,p3,b1,b2,p4,dummy4}
\fmfpen{thin}
\fmf{plain}{dummy1,p1,a2,a1,p2,dummy2}
\fmf{plain}{dummy3,p3,b1,b2,p4,dummy4}
\fmffreeze
\fmf{plain,width=2}{a1,b2}
\fmf{plain,width=2}{a2,b1}
\fmf{dots}{p1,p3}
\fmf{dots}{p2,p4}
\fmfdot{a1,a2,b1,b2}
\end{fmfgraph*}
\end{fmffile}

\end{center}\end{minipage} - \begin{minipage}[h]{8mm}\begin{center}\footnotesize

\begin{fmffile}{2gonCrossed}
\begin{fmfgraph*}(5,14)
\fmfstraight
\fmfleft{dummy1,p1,a2,a1,p2,dummy2}
\fmfright{dummy3,p3,b1,b2,p4,dummy4}
\fmfpen{thin}
\fmf{plain}{dummy1,p1,a2,a1,p2,dummy2}
\fmf{plain}{dummy3,p3,b1,b2,p4,dummy4}
\fmffreeze
\fmf{plain,width=2}{a1,b1}
\fmf{plain,width=2}{a2,b2}
\fmf{dots}{p1,p3}
\fmf{dots}{p2,p4}
\fmfdot{a1,a2,b1,b2}
\end{fmfgraph*}
\end{fmffile}

\end{center}\end{minipage} + \frac{1}{2} \begin{minipage}[h]{8mm}\begin{center}\footnotesize

\begin{fmffile}{2gonSelf}
\begin{fmfgraph*}(5,14)
\fmfstraight
\fmfleft{dummy1,p1,a2,a1,p2,dummy2}
\fmfright{dummy3,p3,b1,b2,p4,dummy4}
\fmfpen{thin}
\fmf{plain}{dummy1,p1,a2,a1,p2,dummy2}
\fmf{plain}{dummy3,p3,b1,b2,p4,dummy4}
\fmfcurved
\fmf{plain,left,width=2,tension=0}{a1,a2}
\fmf{plain,left,width=2,tension=0}{b1,b2}
\fmf{dots}{p1,p3}
\fmf{dots}{p2,p4}
\fmfdot{a1,a2,b1,b2}
\end{fmfgraph*}
\end{fmffile}

\end{center}\end{minipage} \Bigg)\vspace{5pt}
\end{equation}
while the 3-gons only contribute to the logarithmic terms. For the two-point functions we get
\begin{align}
 \skla{\Op_\alpha(x_1) \, \Op_\beta(x_2)} &= I_{12}^{\oplength} \sum_{\sigma \in Z_\oplength} \Bigg[ \prod_{i=1}^{\oplength} u^\alpha_i \cdot u^\beta_{\sigma(i)} + \frac{\lambda}{8 \pi^2} \kla{\lnk{12}+1}\nnl
 &\bleq \times \sum_{\tau \in Z_\oplength} \Big( u^\alpha_{\tau(1)} \cdot u^\beta_{\tau \circ \sigma(1)} \, u^\alpha_{\tau(2)} \cdot u^\beta_{\tau \circ \sigma(2)} - u^\alpha_{\tau(1)} \cdot u^\beta_{\tau \circ \sigma(2)} \, \nnl
 &\bleq \times u^\alpha_{\tau(2)} \cdot u^\beta_{\tau \circ \sigma(1)} + \frac{1}{2} \, u^\alpha_{\tau(1)} \cdot u^\alpha_{\tau(2)} \, u^\beta_{\tau \circ \sigma(1)} \cdot u^\beta_{\tau \circ \sigma(2)} \Big) \nnl
 &\bleq \times \prod_{i=3}^{\oplength} u^\alpha_{\tau(i)} \cdot u^\beta_{\tau \circ \sigma(i)} \Bigg].
\end{align}

\noindent Now let $\D_\alpha = M_{\alpha \beta} \, \Op_\beta$ denote a diagonal basis of the length $\oplength$ subspace. Then
\begin{align}
 \skla{\D_\alpha(x_1) \, \D_\beta(x_2)} &= \frac{1}{x_{12}^{2 \oplength}} \kla{\delta_{\alpha \beta} + \lambda g_{\alpha \beta} + \lambda \gamma_\alpha \delta_{\alpha \beta} \lnk{12}}
 = M_{\alpha \gamma} \, M_{\beta \delta} \, \skla{\Op_\gamma(x_1) \, \Op_\delta(x_2)}
\end{align}
from which we immediately get the condition for tree-level diagonality
\begin{equation}
 \sum_{\sigma \in Z_\oplength} M_{\alpha \gamma} \, M_{\beta \delta} \,\prod_{i=1}^{\oplength} u^\gamma_i \cdot u^\delta_{\sigma(i)} = (2 \pi)^{2 \oplength} \, \delta_{\alpha \beta}. \label{eqn:konishi proof tree-level diagonality}
\end{equation}
Using this result we obtain
\begin{align}
 \skla{\D_\alpha(x_1) \, \D_\beta(x_2)} &= \frac{1}{x_{12}^{2 \oplength}} \Bigg( \delta_{\alpha \beta} + \frac{\lambda}{8 \pi^2} \kla{\lnk{12}+1} \Bigg[ \oplength \, \delta_{\alpha \beta} - \frac{1}{(2 \pi)^{2 \oplength}} \nnl
 &\bleq \times \sum_{\sigma \in Z_\oplength} \sum_{\tau \in Z_\oplength} M_{\alpha \gamma} \, M_{\beta \delta} \Big( u^\gamma_{\tau(1)} \cdot u^\delta_{\tau \circ \sigma(2)} \, u^\gamma_{\tau(2)} \cdot u^\delta_{\tau \circ \sigma(1)} \nnl
 &\bleq - \frac{1}{2} u^\gamma_{\tau(1)} \cdot u^\gamma_{\tau(2)} \, u^\delta_{\tau \circ \sigma(1)} \cdot u^\delta_{\tau \circ \sigma(2)} \Big) \times \prod_{i=3}^{\oplength} u^\gamma_{\tau(i)} \cdot u^\delta_{\tau \circ \sigma(i)} \Bigg] \Bigg)
\end{align}
and thus the condition for one-loop diagonality
\begin{align}
(2 \pi)^{2 \oplength} \, \delta_{\alpha \beta} \, \kla{\oplength - 8 \pi^2 \, \gamma_\alpha} &= \sum_{\sigma \in Z_\oplength} \sum_{\tau \in Z_\oplength} M_{\alpha \gamma} \, M_{\beta \delta} \Big( u^\gamma_{\tau(1)} \cdot u^\delta_{\tau \circ \sigma(2)} \, u^\gamma_{\tau(2)} \cdot u^\delta_{\tau \circ \sigma(1)} \nnl &
 \bleq - \frac{1}{2} u^\gamma_{\tau(1)} \cdot u^\gamma_{\tau(2)} \, u^\delta_{\tau \circ \sigma(1)} \cdot u^\delta_{\tau \circ \sigma(2)} \Big) \,  \prod_{i=3}^{\oplength} u^\gamma_{\tau(i)} \cdot u^\delta_{\tau \circ \sigma(i)} \label{eqn:konishi proof one-loop diagonality}
\intertext{and}
 g_\alpha &= \gamma_\alpha \, . \label{eqn:g alpha beta in eps to eeps renormalization}
\end{align}

\noindent The three-point functions are
\begin{align}
 &\hspace{-20pt}\skla{\D_\alpha(x_1) \, \D_\beta(x_2) \K(x_3)} = M_{\alpha \gamma} \, M_{\beta \delta} \, \skla{\Op_\alpha(x_1) \, \Op_\beta(x_2) \K(x_3)} \nnl
 &= \frac{1}{(2 \pi)^{2 \oplength +2} \, \sqrt{3} \, x_{12}^{2 \oplength -2} \, x_{13}^2 \, x_{23}^2} \, \sum_{\sigma \in Z_\oplength} \sum_{\tau \in Z_\oplength} M_{\alpha \gamma} \, M_{\beta \delta}  \times \Bigg[ \prod_{i=1}^{\oplength} u^\gamma_{\sigma(i)} \cdot u^\delta_{\tau(i)} \nnl
 &\bleq+ \frac{\lambda}{8 \pi^2} \sum_{\rho \in Z_{\oplength-2}} \Big( u^\gamma_{\sigma \circ \rho(1)} \cdot u^\delta_{\tau \circ \rho(1)} \, u^\gamma_{\sigma \circ \rho(2)} \cdot u^\delta_{\tau \circ \rho(2)} - u^\gamma_{\sigma \circ \rho(1)} \cdot u^\delta_{\tau \circ \rho(2)} \, u^\gamma_{\sigma \circ \rho(2)} \cdot u^\delta_{\tau \circ \rho(1)} \nnl
 &\bleq+ \frac{1}{2} u^\gamma_{\sigma \circ \rho(1)} \cdot u^\gamma_{\sigma \circ \rho(2)} \, u^\delta_{\tau \circ \rho(1)} \cdot u^\delta_{\tau \circ \rho(2)} \Big) \times \prod_{i=3}^{\oplength -2} \kla{u^\gamma_{\sigma \circ \rho(i)} \cdot u^\delta_{\tau \circ \rho(i)}} \nnl
 &\bleq\times u^\gamma_{\sigma(\oplength-1)} \cdot u^\delta_{\tau(\oplength-1)} \, u^\gamma_{\sigma(\oplength)} \cdot u^\delta_{\tau(\oplength)} + \lambda \times \text{logs} \Bigg]\nnl
 &\soll \frac{1}{x_{12}^{2 \oplength -2} \, x_{13}^2 \, x_{23}^2} \kla{C^{(0)}_{\alpha \beta \K} + \lambda \, \widetilde{C}^{(1)}_{\alpha \beta \K} + \lambda \times \text{logs}}
\end{align}
and we obtain the tree-level structure constant
{\allowdisplaybreaks \begin{align}
 C^{(0)}_{\alpha \beta \K} &= \frac{1}{(2 \pi)^{2 \oplength +2} \, \sqrt{3}} \sum_{\sigma \in Z_\oplength} \sum_{\tau \in Z_\oplength} M_{\alpha \gamma} \, M_{\beta \delta} \prod_{i=1}^{\oplength} u^\gamma_{\sigma(i)} \cdot u^\delta_{\tau(i)} \nnl
 &= \frac{\oplength}{(2 \pi)^{2 \oplength +2} \, \sqrt{3}} \sum_{\tau \in Z_\oplength} M_{\alpha \gamma} \, M_{\beta \delta} \prod_{i=1}^{\oplength} u^\gamma_{i} \cdot u^\delta_{\tau(i)},
\end{align}
}where we omitted one sum over all permutations in the second line because the first sum already delivers all possible contractions.

Using equation (\ref{eqn:konishi proof tree-level diagonality}) we get
\begin{equation}
C^{(0)}_{\alpha \beta \K} = \frac{\Dn}{4 \pi^2 \, \sqrt{3}} \, \delta_{\alpha \beta}\, .
\label{a1}
\end{equation}

\noindent The one-loop structure constant is
\begin{small}\begin{align}
 \widetilde{C}^{(1)}_{\alpha \beta \K} &= \frac{1}{(2 \pi)^{2 \oplength +4} \, \sqrt{12}} \sum_{\sigma \in Z_\oplength} \sum_{\tau \in Z_\oplength} \sum_{\rho \in Z_{\oplength-2}} M_{\alpha \gamma} \, M_{\beta \delta} \nnl
 &\bleq \times \Bigg[ \prod_{i=1}^{\oplength-2} \kla{u^\gamma_{\sigma \circ \rho(i)} \cdot u^\delta_{\tau \circ \rho(i)}} \times u^\gamma_{\sigma(\oplength-1)} \cdot u^\delta_{\tau(\oplength-1)} \, u^\gamma_{\sigma(\oplength)} \cdot u^\delta_{\tau(\oplength)} \nnl
 &\bleq - \Big( u^\gamma_{\sigma \circ \rho(1)} \cdot u^\delta_{\tau \circ \rho(2)} \, u^\gamma_{\sigma \circ \rho(2)} \cdot u^\delta_{\tau \circ \rho(1)} - \frac{1}{2} \, u^\gamma_{\sigma \circ \rho(1)} \cdot u^\gamma_{\sigma \circ \rho(2)} u^\delta_{\tau \circ \rho(1)} \cdot u^\delta_{\tau \circ \rho(2)} \Big) \nnl
 &\bleq \times \prod_{i=3}^{\oplength-2} \kla{u^\gamma_{\sigma \circ \rho(i)} \cdot u^\delta_{\tau \circ \rho(i)}} \times u^\gamma_{\sigma(\oplength-1)} \cdot u^\delta_{\tau(\oplength-1)} \, u^\gamma_{\sigma(\oplength)} \cdot u^\delta_{\tau(\oplength)} \Bigg] \nnl
 &= \frac{\delta_{\alpha \beta}}{(2 \pi)^{4} \, \sqrt{12}} \ekla{ (\oplength-2) \, \oplength  - (\oplength-2) \, (\oplength - 8 \pi^2 \, \gamma_\alpha) } \nnl
 &= \frac{(\oplength-2) \, \gamma_\alpha}{4 \pi^2 \, \sqrt{3}} \, \delta_{\alpha \beta},
 \label{a2}
\end{align}
\end{small}where the sum over the $\rho$-permutations gives only a factor of $(\oplength-2)$ and we made use of equations (\ref{eqn:konishi proof tree-level diagonality}) and (\ref{eqn:konishi proof one-loop diagonality}) in the second step.

The renormalization scheme independent structure constants
\be
 C^{(1)}_{\alpha \beta \gamma} = \widetilde{C}^{(1)}_{\alpha \beta \gamma} - \frac{1}{2} \, C^{(0)}_{\alpha \beta \gamma} \, \kla{g_\alpha  + g_\beta + g_\gamma}
\ee
may now be written down using \eqn{a1}, \eqn{a2} and
(\ref{eqn:g alpha beta in eps to eeps renormalization})
to find
\be
 C^{(1)}_{\alpha \beta \K} = \widetilde{C}^{(1)}_{\alpha \beta \K} - \frac{1}{2} \, C^{(0)}_{\alpha \beta \K} \, \kla{\gamma_\alpha + \gamma_\beta + \frac{3}{4 \pi^2}}
 =  - \kla{ \frac{\gamma_\alpha}{\oplength_\alpha} + \frac{\gamma_\beta}{\oplength_\beta} + \frac{\gamma_\K}{\oplength_\K} } \, C^{(0)}_{\alpha \beta \K} \, . \label{eqn:konishi beweis ergebnis}
\ee

\subsection*{Acknowledgements}

We thank Gleb Arutyunov, Niklas Beisert, Harald Dorn,
Johannes Henn, Charlotte Kristjansen and Rodolfo Russo
for helpful discussions. This work was supported by the Volkswagen Foundation.

\appendix

\section{Conventions}
\label{App:A}
In this Appendix we summarise our conventions.
The Lagrangian and super-symmetry transformations of the four dimensional ${\mathcal N}=4$ SYM
can be derived by dimensional reduction from the ten dimensional ${\mathcal N}=1$ SYM theory.
We adopt the mostly-minus metric $(+,-,-,-)$ and the following conventions for the $SU(N)$
gauge group generators:
\begin{equation}
\tr{T^a T^b}=\frac{1}{2} \delta^{ab},\,\,\,[T^a , T^b]=i f^{ab}_c T^c,\,\,\,
(T^a)^i_j (T^a)^k_l=\frac{1}{2} (\delta^i_l\delta^k_j- \frac{1}{N}\delta^i_j\delta^k_l)
\end{equation}

The Lagrangian reads
\begin{eqnarray}
L={\rm Tr}\Big[ -\frac{1}{2}F_{\mu\nu}F^{\mu\nu}+2 D_{\mu}\Phi_{AB} D^{\mu}\Phi^{AB} +2 i \psi^{\alpha A}\sigma^{\mu}_{\alpha \dot\alpha}
(D_{\mu}{\bar \psi}_A)^{\dot\alpha} + \nonumber \\
2 \,g_{YM}^2\,[\Phi^{AB},\Phi^{CD}][\Phi_{AB},\Phi_{CD}] -2\sqrt{2}\,g_{YM}([\psi^{\alpha A},\Phi_{AB}]\psi_{\alpha} ^B-
[{\bar \psi}_{\dot\alpha A},\Phi^{AB}]{\bar \psi}_B^{\dot\alpha})\Big],
\end{eqnarray}
where $\Phi_{AB}$ denote the six complex scalar fields of ${\mathcal N}=4$ SYM which satisfy
 $\Phi^{AB}=\frac{1}{2}\epsilon_{ABCD}\, \Phi^{CD}={\bar\Phi}_{AB}$.
Sometimes it is more convenient to work with three complex scalar fields $Z_1,\,Z_2, \,Z_3$ and their complex conjugates
defined as follows
\begin{align}
Z_{1}=2\,\Phi_{14}\, , \quad \bar Z_{1}=2\,\Phi_{23}=2\,\Phi^{14}\, \nn\\
Z_{2}=2\,\Phi_{24}\, , \quad \bar Z_{2}=2\, \Phi_{31}=2 \,\Phi^{24}\, \nn\\
Z_{3}=2\,\Phi_{34}\, , \quad \bar Z_{3}=2\,\Phi_{12}=2\,\Phi^{34}\, .
\end{align}
with $Z_{1}=\frac{1}{\sqrt{2}}(\phi_{1}+i\phi_{2})$, $Z_{2}=\frac{1}{\sqrt{2}}(\phi_{3}+i\phi_{4})$, $Z=Z_{3}=\frac{1}{\sqrt{2}}(\phi_{5}+i\phi_{6})$.

For the propagators we note
\begin{align}
\vev{Z_{i}(x)^{a}{}_{b}\, {\bar Z}_{j}(y)^{c}{}_{d}} & =  \frac{1}{2}\, \delta_{ij}\,
\delta^{c}_{b}\delta^{a}_{d}\, \Delta_{xy}\qquad
a,b,c,d=1,\ldots, N, \quad i,j=1,2,3\, , \nn\\
\vev{\psi_{\alpha}^{A}(x)^{a}{}_{b}\, {\bar \psi}_{\dot\alpha\, B}(y)^{c}{}_{d}}
& =  \frac{i}{2}\, \delta^{A}_{B}\, \sigma^{\mu}_{\alpha\dot\alpha}\,
\partial^{x}_{\mu} \Delta_{xy}\, ,
\end{align}
with $\Delta_{xy}=-\frac{1}{4\pi^{2}\, (x-y)^{2}}$. From these one deduces
\begin{align}
\vev{\partial_{\mu}Z_{i}(x)^{a}{}_{b}\, {\bar Z}_{j}(y)^{c}{}_{d}} & =  \frac{1}{2}\, \delta_{ij}\,
\delta^{c}_{b}\delta^{a}_{d}\, \partial^{x}_{\mu}\Delta_{xy}\, , \nn\\
\partial_{\mu}^{1}\Delta_{12}\, \partial^{1\, \mu}\Delta_{13} &=
-8\pi^{2}\, (\Delta_{12}\Delta_{13}^{2}+ \Delta_{12}^{2}\Delta_{13}-
\Delta_{12}^{2}\, \Delta_{13}^{2}\,\Delta_{23}^{-1}) \, .
\end{align}
Moreover, one may derive an effective spinor index free contraction of
the gluinos
\be \vev{\psi^{A}(x)^{a}_{b}\, {\bar
    \psi}_{B}(y)^{c}_{d}}_{\text{effective}} = i\,\sqrt{2}\, 2\pi\,
\Delta^{3/2}_{xy} \, \delta^{A}_{B}\, \delta^{a}_{d}\,
\delta^{c}_{b}\, ,
\ee
which appears in correlators involving only two
bi-fermion insertions of the form
\be \vev{\psi^{A_{1}\,
    \alpha}(x)^{a_{1}}_{b_{1}}\,
  \psi^{A_{2}}_{\alpha}(x)^{a_{2}}_{b_{2}}\, \bar\psi_{B_{1}\,
    \dot\alpha}(y)^{c_{1}}_{d_{1}}\,
  \bar\psi^{\dot\alpha}_{B_{2}}(y)^{c_{2}}_{d_{2}}}\, ,
\ee
which are
spinor-index singlets and are of relevance in the computations at
hand.

We report here the form of currents associated to the superconformal
transformations of $\mathcal{N}=4$ SYM (see Appendix A of
\cite{Georgiou:2008vk}):
\begin{subequations}
\label{supc}
\begin{multline}
\label{eq:bscc}
\bar{S}^{\mu \dot\alpha}_{\phantom{\mu} A}  =
  2  x_\tau (\bar\sigma^\tau)^{\dot\alpha \alpha}
\tr{(\sigma^{\rho \nu})_\alpha^\beta F_{\rho \nu}
      \sigma^{\mu}_{\beta\dot{\beta}}\bar{\psi}^{\dot{\beta}}_{A}
      + 2\sqrt{2} D_\rho \Phi_{AB} \sigma^\rho_{\alpha\dot{\alpha}}
      \bar{\sigma}^{\mu\,\dot{\alpha}\beta} \psi_{\beta}^B+\right.\\\left.
      - 4 \ii g [\Phi_{AC},\Phi^{CB}]\sigma^\mu_{\alpha\dot{\beta}}
      \bar{\psi}^{\dot{\beta}}_{B}}  + 8\sqrt{2}
     \tr{\phi_{AB} (\bar\sigma^{\mu})^{\dot\alpha \alpha} \psi^B_\alpha},
\end{multline}
\begin{multline}
\label{eq:scc}
{S^\mu}^A_\alpha =
    2  x_\tau \sigma^\tau_{\alpha \dot{\alpha}}
\tr{(\bar{\sigma}^{\rho \nu})^{\dot{\alpha}}_{\dot{\beta}} F_{\rho \nu}
      \bar{\sigma}^{\mu\,\dot{\beta}\beta}\psi^{A}_{\beta}
      -2\sqrt{2}D_\rho \Phi^{AB} \bar{\sigma}^{\rho\,\dot{\alpha}\alpha}
      \sigma^\mu_{\alpha\dot{\beta}} \bar{\psi}^{\dot{\beta}}_B +\right.\\\left.
     - 4 \ii g[\Phi^{AC},\Phi_{CB}]
      \bar{\sigma}^{\mu\dot{\alpha}\alpha}\psi_{\alpha}^B} - 8\sqrt{2}
      \tr{\phi^{AB} \sigma^{\mu}_{\alpha \dot{\alpha}} \bar{\psi}_B^{\dot{\alpha}}},
\end{multline}
\end{subequations}
from which one can derive the tree-level and order-$g_{YM}$
superconformal variation of the fields. In particular, we will use:
\begin{itemize}
\item the tree-level superconformal variation of a single
  fermion\cite{Georgiou:2008vk}:
  \begin{equation}
    \label{eq:sf}
    \bar{S}^{\dot{\alpha}}_A \bar{\psi}_{B\dot{\beta}} =
    4\sqrt{2} i \Phi_{AB}\delta^{\dot{\alpha}}_{\dot{\beta}};
  \end{equation}
\item the order-$g_{YM}$ superconformal variation of a pair of scalar
  fields\cite{Georgiou:2008vk}:
\begin{equation}
\label{eq:spp}
\bar{S}_A^{\dot\alpha} \Phi_{BC}
\Phi_{DE}(0)=  -\ii\, \frac{g N}{32 \pi^2} \left(
\epsilon_{ABC[D} \bar\psi_{E]}(0) -\epsilon_{ADE[B}
  \bar\psi^{\dot\alpha}_{C]}(0) \right)~,
\end{equation}
where $\epsilon_{ABC[D} \bar\psi_{E]}=\frac{1}{2}
(\epsilon_{ABCD} \bar\psi_{E}-\epsilon_{ABCE} \bar\psi_{D})$.
\end{itemize}

\section{Normalization of the states}
\label{App:norm}

In this section we compute explicitly the order-$g_{YM}^2$
contribution to the normalization of the non-BPS operators coming from
the mixing. We require that the two-point functions are canonically
normalized:
\begin{equation}
  \langle
\bar{\hat{\mathcal{O}}}(x)
\hat{\mathcal{O}}(y)
\rangle = \frac{(-1)^{L}}{[(x-y)^2]^{L}},
\end{equation}
where $L$ is the operator length.

Let us start from the Highest Weight State \eqref{hws}, focusing first
on the treelevel contribution coming from the leading term. Using:
\begin{equation}
 \sum_{p=0}^J \cos{\frac{\pi n (2p+3)}{J+3}} = - 2
\cos{\frac{\pi n}{J+3}}
\qquad\mbox{and}\qquad
\sum_{p=0}^J \cos^2{\frac{\pi n
    (2p+3)}{J+3}} = \frac{J+3}{2}- 2 \cos^2{\frac{\pi n}{J+3}},
\end{equation}
 it is
straightforward to show that
\begin{multline}
\sum_{p,q=0}^J
\cos{\frac{\pi n (2p+3)}{J+3}}\cos{\frac{\pi n (2q+3)}{J+3}} \times \\
\langle
\tr{\Phi_{AB} \bar{Z}^p \Phi^{AB} \bar{Z}^{J-p}}
\tr{\Phi_{AB} Z^q \Phi^{AB} Z^{J-q}}\rangle
\,=\,
(J+3)
\Big(
\frac{N}{8\pi^2}
\Big)^{J+2}
\frac{(-1)^{J+2}}{[(x-y)^2]^{J+2}},
\end{multline}
from which we get the leading term in \eqref{eq:norm}.

The fermionic terms contribute to the normalization at order $g^2$ via
tree-level contractions. So we shall compute
\begin{multline}
\sum_{p,q=0}^{J-1}
\sin{\frac{\pi n (2p+4)}{J+3}}\sin{\frac{\pi n (2q+4)}{J+3}} \times\\
\Big[
\langle
\tr{\bar{\psi}_{1\dot{\alpha}}\bar{Z}^p
\bar{\psi}_{2}^{\dot{\alpha}}\bar{Z}^{J-p-1}}(x)
\tr{\psi^{1 \alpha} Z^q \psi^2_\alpha Z^{J-q-1}}(y)
\rangle + \\
\langle
\tr{\psi^{3 \alpha} Z^p \psi^4_\alpha Z^{J-p-1}}(x)
\tr{\bar{\psi}_{3\dot{\alpha}}\bar{Z}^q
\bar{\psi}_{4}^{\dot{\alpha}}\bar{Z}^{J-q-1}}(y)
\rangle
\Big]
\end{multline}
The first term within squared brackets yields:
\begin{equation}
  \langle
\tr{\bar{\psi}_{1\dot{\alpha}}\bar{Z}^p
\bar{\psi}_{2}^{\dot{\alpha}}\bar{Z}^{J-p-1}}(x)
\tr{\psi^{1 \alpha} Z^q \psi^2_\alpha Z^{J-q-1}}(y)
\rangle =
\Big(
\frac{N}{2}
\Big)^{J+1}
32 \pi^2 \Delta_{xy}^{J+2} \delta_{q, J-p-1}
\end{equation}
while the second term just doubles this result.

So, if we includes the coefficients in front to the fermionic term in
\eqref{hws}, we get:
\begin{equation}
- \mathcal{N}^2\frac{N^2}{(8\sqrt{2}\pi^2)^2}\sin^2{\frac{\pi n}{J+3}}
\sum_{p=0}^{J-1}\sin^2{\frac{\pi n (2p+4)}{J+3}}
\Big(
\frac{N}{2}
\Big)^{J+1}
64 \pi^2 \Delta_{xy}^{J+2} \delta_{q, J-p-1}
\end{equation}
The sum yelds:
\begin{equation}
  \sum_{p=0}^{J-1}\sin^2{\frac{\pi n (2p+4)}{J+3}} =
\frac{J-1}{2} + 2 \cos^2\frac{2 \pi n}{J+3}
\end{equation}

Putting everything together, the finite part of the two point function
up to order $g^2$ is:
\begin{multline}
    \langle
\bar{\hat{\mathcal{O}}}^J_n(x)
\hat{\mathcal{O}}^J_n(y)
\rangle = \\
\mathcal{N}^2
(J+3)
\Big(
\frac{N}{8\pi^2}
\Big)^{J+2}
\Big[
1-\frac{g^2N}{\pi^2(J+3)}
\sin^2{\frac{\pi n}{J+3}}
\left(
\frac{J-1}{2} +
2 \cos^2\frac{2\pi n}{J+3}
\right)
\Big]
\frac{(-1)^{J+2}}{[(x-y)^2]^{J+2}}.
\end{multline}
Requiring this being canonically normalized:
\begin{equation}
  \langle
\bar{\hat{\mathcal{O}}}^J_n(x)
\hat{\mathcal{O}}^J_n(y)
\rangle = \frac{(-1)^{J+2}}{[(x-y)^2]^{J+2}},
\end{equation}
we get ($N_0 = \frac{N}{8\pi^2}$)
\begin{equation}
  \mathcal{N} = \sqrt{\frac{N_0^{-J-2}}{J+3}}
\left[
1-\frac{g^2N}{\pi^2(J+3)}
\sin^2{\frac{\pi n}{J+3}}
\left(
\frac{J-1}{2} +
2 \cos^2\frac{2\pi n}{J+3}
\right)
\right]^{-\frac{1}{2}}
\end{equation}
which, expanded for small $g$, gives the result in \eqref{eq:norm}.

Now let us focus on the operators $\mathcal{\hat{O}}_{4A}$ and $\mathcal{\hat{O}}_{4E}$. The tree level contribution can be rewritten as:
\begin{multline}
  \label{eq:tlnorm}
\Big(
  \frac{8\pi^2}{N}
\Big)^4 \mathcal{N}_{A/E}^2
\left[
\langle
\tr{\Phi_{AB} \Phi^{AB} \Phi_{CD} \Phi^{CD}}(x)
\tr{\Phi_{A'B'} \Phi^{A'B'} \Phi_{C'D'} \Phi^{C'D'}}(y)
\rangle \right.\\\left.
+ 2 \alpha_{A/E}
\langle
\tr{\Phi_{AB} \Phi^{AB} \Phi_{CD} \Phi^{CD}}(x)
\tr{\Phi_{A'B'}  \Phi_{C'D'} \Phi^{A'B'} \Phi^{C'D'}}(y)
\rangle \right.\\\left.
+ \alpha_{A/E}^2
\langle
\tr{\Phi_{AB} \Phi_{CD} \Phi^{AB}  \Phi^{CD}}(x)
\tr{\Phi_{A'B'}  \Phi_{C'D'} \Phi^{A'B'} \Phi^{C'D'}}(y)
\rangle\right]
\end{multline}

We recall that in the $SU(4)$ notation the correlator between the
scalar fields reads:
\begin{equation}
  \langle \Phi_{AB}(x)^a \Phi_{CD}(y)^b \rangle =
\delta^{ab} \epsilon_{ABCD}\Delta_{xy}
\qquad\text{and}\qquad
\Phi^{AB} = \frac{1}{2} \epsilon^{ABCD} \Phi_{CD}
\end{equation}

The computation of the different terms in \eqref{eq:tlnorm} is then straigthforward and it yields:
\begin{equation}
 \mathcal{N}_{A/E}^2
\left[
\frac{21}{4} + 3\,\alpha_{A/E} + 9\,\alpha_{A/E}^2
\right] \frac{(-1)^4}{[(x-y)^2]^4} =
\mathcal{N}_{A/E}^2\frac{738 \mp 102\sqrt{41}}{16}
\frac{(-1)^4}{[(x-y)^2]^4}
\end{equation}
where we have replaced $\alpha_{A/E} =  \frac{5\mp\sqrt{41}}{4}$.

The contribution of the fermionic subleading terms is:
\begin{multline}
  \Big(
  \frac{8\pi^2}{N}
\Big)^4 \mathcal{N}_{A/E}^2
\frac{g^2N^2}{(16\sqrt{2}\pi^2)^2}
\frac{(3\mp\sqrt{41})^2}{16}
\left[
\langle
\tr{\Phi^{AB}\bar{\psi}_{A\dot{\alpha}} \bar{\psi}_B^{\dot{\alpha}}}(x)
\tr{\Phi_{CD}\psi^{C \alpha} \psi^D_\alpha}(y)
\rangle
+\right.\\\left
\langle
\tr{\Phi_{AB}\psi^{A \alpha} \psi^B_\alpha}(x)
\tr{\Phi^{CD}\bar{\psi}_{C\dot{\alpha}} \bar{\psi}_D^{\dot{\alpha}}}(y)
\rangle
\right]
\end{multline}
Each  term in the last formula gives the same result, namely:
\begin{equation}
  \langle
\tr{\Phi^{AB}\bar{\psi}_{A\dot{\alpha}} \bar{\psi}_B^{\dot{\alpha}}}(x)
\tr{\Phi_{CD}\psi^{C \alpha} \psi^D_\alpha}(y)
\rangle = 3 \Big(\frac{N}{2}\Big)^3
\Delta_{xy}
\bar{\sigma}^{\mu \dot{\alpha} \alpha}
\sigma^\nu_{\alpha \dot{\alpha}}
\partial_\mu^{(x)}\Delta_{xy}\partial_\nu^{(x)}\Delta_{xy}
\end{equation}
But $\bar{\sigma}^{\mu \dot{\alpha} \alpha} \sigma^\nu_{\alpha
  \dot{\alpha}} = 2 \eta^{\mu\nu} + [\mu,\nu]$, where $[\mu,\nu]$
denotes a term antisymmetric in $\mu$ and $\nu$ which does not
contribute to the final result, and
$\partial_\mu^{(x)}\Delta_{xy}\partial^{\mu\,(x)}\Delta_{xy} =
-16\pi^2 \Delta_{xy}^3$, so all togheter the contribution of the
fermionic subleading mixing term is:
\begin{equation}
 - \mathcal{N}^2_{A/E}\frac{g_{YM}^2 N}{(8\pi)^2} 3 (3\mp\sqrt{41})^2
\frac{(-1)^4}{[(x-y)^2]^4}
\end{equation}
and then the finite part of the two point function up to order $g_{YM}^2$ is:
\begin{equation}
 \langle\bar{\mathcal{\hat{O}}}_{4A/E}(x) \mathcal{\hat{O}}_{4A/E}\rangle =
\mathcal{N}^2_{A/E}\Big[\frac{738 -102\sqrt{41}}{16} - \frac{g_{YM}^2 N}{(8 \pi)^2} 3 (50 \mp 6 \sqrt{41})
\Big]
\frac{(-1)^4}{[(x-y)^2]^4}
\end{equation}
Requiring that the above correlator is canonically normalized we get:
\begin{equation}
  \mathcal{N}_{A/E} = \frac{4}{\sqrt{738 - 102\sqrt{41}}}
\Big[
1+ \frac{g_{YM}^2 N}{4 \pi^2}
\frac{25 \mp 3\sqrt{41}}{246 \mp 34\sqrt{41}}
\Big]
\end{equation}

\section{Cancellation of the terms violating conformal invariance}
\label{sec:canc}

In the correlators we computed in section \ref{sec:corresults}, the
terms coming from a Yukawa insertion which do not respect conformal
invariance cancel out agagainst similar contributions from the
bi-derivative mixing terms. This behavour has already been shown in
\cite{Georgiou:2009tp}, where the contribution to the one loop
structure constant has been computed for a class of correlators
involving the highest weight state in \eqref{hws}.

In this section, we are going to show explicitly the same cancellation
occurring in the classes of correlators $\vev{3B|3B|4A}$ and $\vev{3B|3B|4E}$.

Thus let us consider
\begin{equation}
  \label{eq:3b3b4ae}
  \langle
  \hat{\mathcal{O}}^1_1(x_1)
  \bar{\hat{\mathcal{O}}}^1_1(x_2)
  \hat{\mathcal{O}}_{4A/E}(x_3)
  \rangle,
\end{equation}
and drop the overall normalization, which does not play any role in this computation. So we are going to take:
\begin{align}
  \hat{\mathcal{O}}^1_1 &= \tr{\Phi_{AB} \Phi^{AB} Z}\\
\intertext{and write}
  \hat{\mathcal{O}}_{4A/E} & = \mathcal{O}_{scal} +
\mathcal{O}_{\psi\psi} + \mathcal{O}_{DD},\notag\\
\intertext{where}
  \mathcal{O}_{\psi\psi} & = - g \,\mathcal{N} \frac{3\mp\sqrt{41}}{4}
\Big[\tr{\Phi_{AB}\psi^{A\alpha}\psi^B_\alpha}
- \tr{\Phi^{AB} \bar{\psi}_{A \dot{\alpha}} \bar{\psi}_B^{\dot{\alpha}}}
\Big]\\
  \mathcal{O}_{DD} & =
g^2 \mathcal{N}^2 (3\mp\sqrt{41})
\tr{D_\mu \Phi_{AB} D^\mu \Phi^{AB}}
\end{align}
with $\mathcal{N} = \frac{N}{16\sqrt{2}\pi^2}$.
We want to show that in the sum $\langle \hat{\mathcal{O}}^1_1
\bar{\hat{\mathcal{O}}}^1_1 \mathcal{O}_{\psi\psi}\rangle +
\langle\hat{\mathcal{O}}^1_1 \bar{\hat{\mathcal{O}}}^1_1 \mathcal{O}_{DD}\rangle$
all the terms which do not respect conformal invariance cancel out.

Let us start from $\langle\hat{\mathcal{O}}^1_1 \bar{\hat{\mathcal{O}}}^1_1
\mathcal{O}_{\psi\psi}\rangle$. It is sufficient to focus on the term with
unbarred fermions, because the other term will just double this
result. The Yukawa coupling relevant for this computation is:
\begin{equation}
  \ii 4 \sqrt{2} g
  \int d^4w
  \tr{\Phi^{XY} \bar{\psi}_X^{\dot{\alpha}}\bar{\psi}_Y^{\dot{\beta}}}
  \epsilon_{\dot{\alpha}\dot{\beta}}
\end{equation}
We can contract the scalar in the Yukawa with any scalar field in the
operator in $x_1$, and this forces the contraction of the remaining
scalar in longer operator with a scalar in $x_2$. Obviously, one must
also consider the opposite situation, where the scalar of the Yukawa
is contracted with a scalar in $x_2$, and hence the remaining scalar in
$\mathcal{O}_{DD}$ with one in $x_1$. However, this just exchange the
role of $x_1$ and $x_2$ in the result, so, by now, it is sufficient to
focus on the first case, and then add the same result with $x_1$ and $x_2$
swapped.

For each of the three diagrams we get contracting the Yukawa scalar
with a specific scalar in $x_1$, we can further contract the remaining
six scalar in three possible ways. Taking into account all
the diagrams, and then summing the result with $x_1\leftrightarrow
x_2$ one gets:
\begin{multline}
\langle\hat{\mathcal{O}}^1_1 \bar{\hat{\mathcal{O}}}^1_1
\mathcal{O}_{\psi\psi}\rangle =
- \ii g^2 \frac{N^4}{8\sqrt{2}} \mathcal{N} 3 (3\mp\sqrt{41})
\epsilon_{\dot{\alpha} \dot{\beta}} \epsilon_{\alpha\beta}
\bar{\sigma}^{\mu \dot{\beta}\alpha}\bar{\sigma}^{\nu \dot{\alpha}\beta} \\
\Delta_{12}^2
\int d^4w
\partial_\mu^{(w)}\Delta_{w3} \partial_\nu^{(w)}\Delta_{w3}
\big[
\Delta_{2w}\Delta_{13}+ \Delta_{1w}\Delta_{23}
\big].
\end{multline}

However $\epsilon_{\dot{\alpha} \dot{\beta}} \epsilon_{\alpha\beta}
\bar{\sigma}^{\mu \dot{\beta}\alpha}\bar{\sigma}^{\nu
  \dot{\alpha}\beta} = - 2 \eta^{\mu\nu}$ plus terms antisymmetric in $\mu$ and $\nu$, and then the integrals become:
\begin{align}
\int d^4w
\Delta_{2w}\partial_\mu^{(w)}\Delta_{w3} \partial_\nu^{(w)}\Delta_{w3}
= &-\frac{\ii}{2} \Delta_{23}^2,\\
\int d^4w
\Delta_{1w}\partial_\mu^{(w)}\Delta_{w3} \partial_\nu^{(w)}\Delta_{w3}
= &-\frac{\ii}{2} \Delta_{13}^2.
\end{align}
Substituting $\mathcal{N} = \frac{N}{16\sqrt{2}\pi^2}$
we get:
\begin{equation}
\label{psipsi}
 \langle\hat{\mathcal{O}}^1_1 \bar{\hat{\mathcal{O}}}^1_1
\mathcal{O}_{\psi\psi}\rangle =
g_{YM}^2\frac{N^5}{(16\sqrt{2})^2\pi^2} 3(3\mp\sqrt{41})
\Delta_{12}^2\big[ \Delta_{13} \Delta_{23}^2 +  \Delta_{13}^2 \Delta_{23} \big].
\end{equation}

Moving to the computation of the derivative terms, one must contract
one $D_\mu\Phi_{AB}$ with any of the scalars in $x_1$ and the other
one with any of the scalar in $x_2$. This can be done in nine
independent ways. Then one must exchange the role of $x_1$ and
$x_2$. However this would just double the result of the former
case. Thus, summing all the diagrams one gets:
\begin{equation}
  \langle\hat{\mathcal{O}}^1_1 \bar{\hat{\mathcal{O}}}^1_1 \mathcal{O}_{DD}\rangle =
g^2 \frac{N^3}{8}\mathcal{N}^2 3 (3 \mp \sqrt{41}) \Delta_{12}^2
\partial_\mu^{(x_3)}\Delta_{23} \partial^{\mu\,(x_3)}\Delta_{13}
\end{equation}
Since $\partial_\mu^{(x_3)}\Delta_{23} \partial^{\mu\,(x_3)}\Delta_{13} =
- 8 \pi^2 (\Delta_{13}\Delta_{23}^2+\Delta_{13}^2\Delta_{23} -
\Delta_{13}^2\Delta_{23}^2\Delta_{12}^{-1})$, and replacing
$\mathcal{N}= \frac{N}{16\sqrt{2}\pi^2}$, one finally gets:
\begin{equation}
\label{DD}
  \langle\hat{\mathcal{O}}^1_1 \bar{\hat{\mathcal{O}}}^1_1 \mathcal{O}_{DD}\rangle =
- g^2 \frac{N^5}{(16\sqrt{2})^2\pi^2} 3 (3 \mp \sqrt{41}) \Delta_{12}^2
\big[
\Delta_{13}\Delta_{23}^2+\Delta_{13}^2\Delta_{23} -
\Delta_{13}^2\Delta_{23}^2\Delta_{12}^{-1}
\big].
\end{equation}

Comparing \eqref{DD} with \eqref{psipsi}, one notice that the first
and the second term in \eqref{DD} cancel out against \eqref{psipsi},
and that, up to the overall normalization, the contribution of the
bi-fermion and bi-derivative mixing terms to the correlator reduces
to:
\begin{equation}
  g^2 \frac{N^5}{(16\sqrt{2})^2\pi^2} 3 (3 \mp \sqrt{41})
\Delta_{12} \Delta_{13}^2 \Delta_{23}^2,
\end{equation}
in agreement with conformal invariance prescriptions.

\bibliographystyle{nb}
\bibliography{botany}

\end{document}